\documentclass[fleqn,useAMS]{Papers}
\usepackage{newtxtext,newtxmath}
\usepackage[T1]{fontenc}
\usepackage{subfloat}
\usepackage{makecell}
\usepackage{comment}

\usepackage[square,sort,numbers]{natbib}
\usepackage{graphicx}	% Including figure files
\usepackage{amsmath}	% Advanced maths commands
\usepackage{lipsum}
\usepackage{caption}
\usepackage{textgreek} 
\usepackage{xcolor}  
\newcommand{\rev}[1]{{\color{black} #1}} 
\title[High order statistics for dark matter flow]{On the statistical theory of self-gravitating collisionless dark matter flow: high order kinematic and dynamic relations}
\author[Z. Xu]{Zhijie (Jay) Xu,$^{1}$\thanks{E-mail: \href{mailto:zhijie.xu@pnnl.gov}{zhijie.xu@pnnl.gov}; \href{mailto:zhijiexu@hotmail.com}{zhijiexu@hotmail.com}}
\\
$^{1}$Physical and Computational Sciences Directorate, Pacific Northwest National Laboratory; Richland, WA 99354, USA\\
}

\date{Accepted XXX. Received YYY; in original form ZZZ}

\pubyear{2024}

\begin{document}
\label{firstpage}
\pagerange{\pageref{firstpage}--\pageref{lastpage}}
\maketitle

\begin{abstract}
Dark matter, if it exists, accounts for five times as much as ordinary baryonic matter. %The flow of dark matter has a wide presence in our universe. 
To better understand the self-gravitating collisionless dark matter flow on different scales, statistical theory involving kinematic and dynamic relations must be developed for different types of flow, e.g. incompressible, constant divergence, and irrotational flow. This is mathematically challenging because of the intrinsic complexity of dark matter flow %(constant divergence flow on small scales and irrotational flow on large scales) 
and the lack of a self-closed description of flow velocity. This paper extends our previous work on second-order statistics (Phys. Fluids 35, 077105 \citep{Xu:2023-On-the-statistical-theory-of-self-gravitating}) to kinematic relations of any order for any type of flow. %The self-closed velocity evolution was established on small and large scales, from which 
Dynamic relations were also developed to relate statistical measures of different orders. The results were validated by N-body simulations. On large scales, we found i) third-order velocity correlations can be related to density correlation or pairwise velocity; ii) the $p$th-order velocity correlations follow $\propto a^{(p+2)/2}$ for odd $p$ and $\propto a^{p/2}$ for even $p$, where $a$ is the scale factor; iii) the overdensity $\delta$ is proportional to density correlation on the same scale, $\langle\delta\rangle\propto\langle\delta\delta'\rangle$; iv) velocity dispersion on a given scale $r$ is proportional to the overdensity on the same scale. On small scales, i) a self-closed velocity evolution is developed by decomposing the velocity into motion in haloes and motion of haloes; ii) the evolution of vorticity and enstrophy are derived from the evolution of velocity; iii) dynamic relations are derived to relate second- and third-order correlations; iv) while the first moment of pairwise velocity follows $\langle\Delta u_L\rangle=-Har$ ($H$ is the Hubble parameter), the third moment follows $\langle(\Delta u_L)^3\rangle\propto\varepsilon_uar$ that can be directly compared with simulations and observations, where $\varepsilon_u\approx10^{-7}$m$^2$/s$^3$ is the constant rate for energy cascade; v) the $p$th order velocity correlations follow $\propto a^{(3p-5)/4}$ for odd $p$ and $\propto a^{3p/4}$ for even $p$. Finally, the combined kinematic and dynamic relations lead to exponential and one-fourth power-law velocity correlations on large and small scales, respectively. 
\end{abstract}

\begin{keywords}
\vspace*{-20pt}
Dark matter flow; Collisionless; Self-gravitating; Velocity correlations; Kinematic relations; Dynamic relations;
\end{keywords}

%%%%%%%%%%%%%%%%%%%%%%%%%%%%%%%%%%%%%%%%%%%%%%%%%
%%%%%%%%%%%%%%%%% BODY OF PAPER %%%%%%%%%%%%%%%%%%
\begingroup
\let\clearpage\relax
\tableofcontents
\endgroup
\vspace*{-15pt}
%\newpage

%\begin{multicols}{1}
%\begingroup
%\let\clearpage\relax
%\tableofcontents
%\endgroup
%\vspace*{-10pt}
%\newpage
%\end{multicols}

\section{Introduction}
\label{sec:1}
\begin{figure}
\includegraphics*[width=\columnwidth]{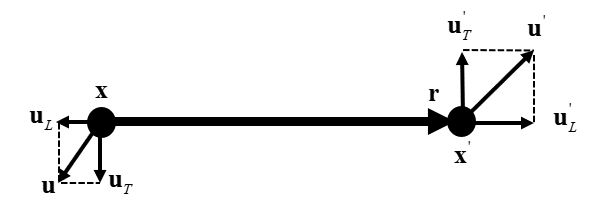}
\caption{A schematic plot for the longitudinal ($u_{L} $ and $u_{L}^{'} $) and transverse ($u_{T} $ and $u_{T}^{'} $) velocities for a particle pair separated by a distance of \textit{r}.} 
\label{fig:1}
\end{figure}

The existence of dark matter is supported by many astronomical observations \citep{Rubin:1970-Rotation-of-Andromeda-Nebula-f,Rubin:1980-Rotational-Properties-of-21-Sc}. In the standard $\Lambda$CDM paradigm (cold dark matter), the amount of dark matter is about five times that of baryonic matter \citep{Spergel:2003-First-Year-Wilkinson-Microwave-Anisotropy,Komatsu:Seven-year-Wilkinson-Microwave-Anisotropy-Probe,Aghanim:2021-Planck-2018-results--VI--Cosmo}. Therefore, the flow of collisionless dark matter is the flow with the largest presence in our universe. The self-gravitating collisionless fluid dynamics (SG-CFD) concerns the dynamic evolution and statistics of dark matter velocity and density fields, which are valuable for probing large-scale structure formation and density fluctuations and for constraining cosmological parameters \citep{Ma:2015-Constraining-cosmology-with-pa}. Statistical analysis of dark matter velocity is applied primarily to pairwise velocity, that is, the difference in the velocity of a pair of masses along the direction of separation \citep{Xu:2023-On-the-statistical-theory-of-self-gravitating} (Fig. \ref{fig:1}). The pairwise velocity was originally introduced to describe the evolution of a system of self-gravitating particles \citep{Davis:1977-Integration-of-Bbgky-Equations} and was later applied to probe the cosmological density parameter \citep{Ferreira:1999-Streaming-velocities-as-a-dyna,Juszkiewicz:2000-Evidence-for-a-low-density-uni}. Lower-order moments of pairwise velocity were also proposed as a diagnostic tool for laws of gravity on large scales \citep{Hellwing:2014-Clear-and-Measurable-Signature}. Another common statistical quantity is the two-point second-order velocity correlation functions that were introduced in the 1980s to quantify the cosmic velocity field \citep{Gorski:1988-On-the-Pattern-of-Perturbation} and were extensively studied in our previous work \citep{Xu:2023-On-the-statistical-theory-of-self-gravitating}. Later, it was applied to a real data set of local supercluster samples \citep{Gorski:1989-Cosmological-Velocity-Correlat} and the Spiral Field I-Band (SFI) catalog of peculiar velocities \citep{Borgani:2000-Correlation-analysis-of-SFI-pe}. 

Directly measuring velocity statistics from real samples is still very challenging in practice, since only the radial velocity component can be directly observed. On the other hand, N-body simulation is an invaluable tool for studying the dynamics of collisionless dark matter flow (SG-CFD) on different scales and capturing very complex gravitational collapse \citep{Angulo:2012-Scaling-relations-for-galaxy-c,Springel:2005-The-cosmological-simulation-co,Peebles:1989-A-Model-for-the-Formation-of-t,Efstathiou:1985-Numerical-Techniques-for-Large}. A complete review of N-body simulations for large-scale structure and galaxy formation can be found in \citet{Angulo:2022-Large-scale-dark-matter-simulations} and \citet{Vogelsberger:2020-Cosmological-simulations-of-galaxy}. The peculiar velocity from N-body simulations is relatively accessible, with a large amount of simulation data for the development of statistical theory. Tremendous amounts of knowledge on the nature of self-gravitating collisionless dark matter flow (SG-CFD) can be obtained from this practice. Two approaches are often applied to study the velocity field from the N-body simulations. 

The first approach is a halo-based method, where all haloes of different sizes are identified based on a given algorithm (for example, friend of friend) \citep{Xu:2023-Maximum-entropy-distributions-of-dark-matter}. All particles in the entire N-body system are then divided into halo particles and out-of-halo particles. The statistics of density \citep{Xu:2022-Two-thirds-law-for-pairwise-ve}, velocity \citep{Xu:2023-On-the-statistical-theory-of-self-gravitating}, and acceleration \citep{Xu:2022-The-origin-of-MOND-acceleratio} evolve differently for halo particles and out-of-halo particles, which can be studied separately. In this approach, the largest haloes in the system introduce a critical scale $r_t$. Below this scale, the dominant flow is the flow in haloes, i.e. the dynamics of in-halo particles. Above that scale, the dynamics of haloes and out-of-halo particles becomes dominant. Therefore, the flow of dark matter exhibits distinct behavior on small and large scales, i.e. a constant divergence nature on small scales and an irrotational nature on large scales, as well as a clear transition between two regimes \citep{Xu:2023-On-the-statistical-theory-of-self-gravitating} (Fig. \ref{fig:19}).

%\citep{Xu:2022-Two-thirds-law-for-pairwise-ve,Xu:2022-The-origin-of-MOND-acceleratio}. Our previous study falls primarily into this category, for which the inverse mass and energy cascade and halo mass functions can be developed rigorously \citep{Xu:2021-Inverse-mass-cascade-mass-function,Xu:2021-Inverse-mass-cascade-halo-density,Xu:2021-Inverse-and-direct-cascade-of-,Xu:2022-The-evolution-of-energy--momen}. Maximum entropy distributions of dark matter particle velocity and energy were also formulated \citep{Xu:2021-The-maximum-entropy-distributi,Xu:2021-Mass-functions-of-dark-matter-}. Relevant applications of the mass/energy cascade theory in dark matter flow are also presented for dark matter particle mass and properties \citep{Xu:2022-Postulating-dark-matter-partic}, MOND (modified Newtonian dynamics) theory \citep{Xu:2022-The-origin-of-MOND-acceleratio}, and baryonic-to-halo mass relation \citep{Xu:2022-The-baryonic-to-halo-mass-rela}.  

An alternative correlation-based approach focuses on various statistical measures of the velocity field. Correlation-based statistical analysis was originally developed for incompressible homogeneous turbulence \citep{Batchelor:1953-The-Theory-of-Homogeneous-Turb}. The velocity correlation functions, first introduced in the 1930s \citep{Taylor:1935-Statistical-theory-of-turbulan,Taylor:1932-The-transport-of-vorticity-and}, play a central role in the statistical theory of turbulence. Other statistical measures can be related to correlation functions and used to describe how energy and enstrophy are distributed on different scales. For SG-CFD, the two-point correlation, structure, dispersion, and spectrum functions are among the most commonly used statistical measures to quantify the peculiar velocity field \citep{Xu:2023-On-the-statistical-theory-of-self-gravitating}. Real-space two-point statistics are calculated by a pairwise averaging over all pairs of particles on the same scale (or separated by the same distance $r$). Therefore, the scale and redshift variation of these statistical measures can be studied in detail \citep{Xu:2022-Two-thirds-law-for-pairwise-ve,Kitaura:2016-Bayesian-redshift-space-distor,Pueblas:2009-Generation-of-vorticity-and-ve}. In this approach, haloes are not explicitly identified. However, on small scales, most pairs of particles with a small separation $r$ should come from the same halo. On large scales, pairs of particles with a large separation $r$ often come from different haloes. Therefore, the effect of haloes on real-space statistics involving the critical scale $r_t$ can be clearly demonstrated by the scale-dependence (or r-dependence) of these statistical measures (Fig. \ref{fig:20}) \citep{Xu:2022-Two-thirds-law-for-pairwise-ve}. 

Statistical analysis in Fourier space by projecting a velocity field onto structured grids (e.g. cloud-in-cell) may involve additional information loss due to the projection and the conversion between real and Fourier space. In this paper, we mainly apply a correlation-based approach to formulate the statistical theory of SG-CFD, where real-space statistics of different orders are directly calculated without field projection. In this way, we are able to obtain the most complete information on N-body systems at different scales and redshifts. 

The statistical theory of stochastic flow is mostly concerned about two types of relations: i) the kinematic relations between statistical measures of the same order; and ii) the dynamic relations between statistical measures of different orders. Kinematic relations can be developed for a given type of flow (incompressible, constant divergence, or irrotational) under the assumption of translational and rotational symmetry. However, the dynamic relations can only be developed from self-closed dynamic equations for the evolution of the velocity field, such as the Burgers or Navier-Stokes equations. In fact, the celebrated Kolmogorov four-fifth law is an exact result of the dynamic relations derived from the Navier-Stokes equation \citep{de_Karman:1938-On-the-statistical-theory-of-i,Kolmogoroff:1941-The-local-structure-of-turbule,Kolmogoroff:1941-Dissipation-of-energy-in-the-l}. By contrast, kinematic and dynamic relations for self-gravitating collisionless dark matter flow (SG-CFD) are not completely developed and far from satisfactory due to several reasons:
\begin{enumerate}
\item \noindent SG-CFD flow is intrinsically complex with different nature of flow on different scales, i.e. a constant divergence flow on small scales and irrotational flow on large scales \citep{Xu:2023-On-the-statistical-theory-of-self-gravitating}. Kinematic and dynamic relations need to be developed for both types of flow. 
%\item \noindent Dynamic equations for velocity evolution (Jeans equation) are not self-closed (closure problem). No dynamic relations can be derived without self-closed equations for velocity evolution. 
\item \noindent Existing work mostly focuses on the first- and second-order velocity statistics, while the peculiar velocity field contains much richer information beyond the second order. %An example is the third-order velocity correlations that are intimately related to the energy cascade and transfer across sales (Eqs. \eqref{ZEqnNum445071} and \eqref{eq:258}). 
However, it is very challenging to explore high-order statistics that inherently involve vector calculus and tensors of great complexity. 
\end{enumerate}

Our previous work focuses mainly on second-order statistics including the kinematic relations of second order \citep{Xu:2023-On-the-statistical-theory-of-self-gravitating}. This paper will extend to high-order statistics, kinematic relations, and dynamic relations. The main objective of this paper is to establish the formal language of statistical theory for collisionless dark matter flow by introducing basic notations, putting in place necessary equations, and laying down the fundamental rules. The theory itself is intrinsically complex because of the stochasticity, nonlinearity, and multiscale nature. However, we are still able to appreciate the beauty of nature, as evidenced by the hidden symmetry in kinematic and dynamic relations. From this practice, we are able to demonstrate that on large scales of irrotational flow:  
\begin{enumerate}
\item \noindent The third-order velocity correlations can be closely related to density correlation and pairwise velocity (Eqs. \eqref{ZEqnNum846484} and \eqref{ZEqnNum216206});
\item \noindent The effective viscosity originates from the inverse energy cascade from small to large scales (Eq. \eqref{ZEqnNum133069}); 
\item \noindent The $p$th order velocity correlations scale as $\propto a^{{\left(p+2\right)/2}}$ for odd order \textit{p} and scale as $\propto a^{p/2} $ for even order \textit{p} (Eq. \eqref{ZEqnNum252855}); 
\item \noindent The overdensity on a given scale \textit{r} is proportional to the density correlation on the same scale, i.e. $\langle \delta \rangle \propto \langle \delta \delta ^{'} \rangle$ (Eq. \eqref{ZEqnNum711863}). Therefore, the void region where $\langle \delta \rangle<0$ can be related to the negative density correlation $\langle\delta\delta{'}\rangle<0$ on the same scale where $r \approx$30Mpc/h;
\item \noindent The excess velocity dispersion on a given scale is proportional to the overdensity on the same scale, i.e. $\langle u^{2} \rangle -3u^2 \propto \langle \delta \rangle$ (Eq. \eqref{ZEqnNum864804}). Low density void region has a negative excess velocity dispersion due to $\langle \delta \rangle<0$, i.e. the local velocity dispersion $\langle u^{2} \rangle$ is less than the asymptotic dispersion $3u^2$; 
\item \noindent The exponential velocity correlations originates from the kinematic and dynamic relations on large scales (Eq. \eqref{eq:196});
\end{enumerate}
\noindent While on small scales of constant divergence flow:
\begin{enumerate}
\item \noindent A self-closed velocity evolution is developed by decomposing the velocity into motion in haloes and motion of haloes (Eq. \eqref{ZEqnNum598457});
\item \noindent Vorticity and enstrophy evolution can be derived from the self-closed equation of velocity (Eqs. \eqref{ZEqnNum968516} and \eqref{ZEqnNum517069}); 
\item \noindent Dynamic relations are derived to relate second- and third-order correlations (Eq. \eqref{ZEqnNum168148}); 
\item \noindent The third moment of the pairwise velocity is determined by the rate $\varepsilon_u$ of the energy cascade or $\langle(\Delta u_L)^3\rangle\propto\varepsilon_uar$ (Eq. \eqref{eq:258});
\item \noindent The power-law velocity correlations originates from the kinematic and dynamic relations on small scales (Eq. \eqref{eq:189});
\item \noindent The $p$th order velocity correlations follow $\propto a^{(3p-5)/4}$ for odd order $p$ and $\propto a^{3p/4}$ for even order $p$.
\end{enumerate}

There is a tremendous amount of knowledge that can be learned from this practice, much more than what we present here. With the second-order statistics presented in \citep{Xu:2023-On-the-statistical-theory-of-self-gravitating}, this paper is organized as follows: Section \ref{sec:2} introduces the N-body simulation data, followed by the third-order statistical measures in Section \ref{sec:3}. The general kinematic relations are presented in Section \ref{sec:4} with results from N-body simulations for comparison in Section \ref{sec:5}. The dynamic relations are formulated on large and small scales in Sections \ref{sec:6} and \ref{sec:7}, respectively. The origin of exponential correlations on large scales and power-law correlations on small scales are presented in Sections \ref{sec:6.3} and \ref{sec:6.3-2}.

\section{N-body simulations and numerical data}
\label{sec:2}
The simulation data are publicly available and are generated from N-body simulations carried out by the Virgo consortium. A comprehensive description of the data can be found in \citep{Frenk:2000-Public-Release-of-N-body-simul,Jenkins:1998-Evolution-of-structure-in-cold}. As a first step, the current study uses simulation runs with $\Omega$ = 1 and the standard CDM power spectrum (SCDM) to focus on matter-dominant gravitational flow. Similar analysis can be extended to other models with different assumptions and parameters in the future. \rev{The current simulation includes 17 million particles with a particle mass of $2.27\times 10^{11}$ ${M_{\odot}/h}$. The gravitational softening length is $36kpc/h$.
The simulation box is about 239.5 Mpc/h, where \textit{h=0.5} is the reduced Hubble constant. The same set of data has been widely used in several studies ranging from clustering statistics \citep{Jenkins:1998-Evolution-of-structure-in-cold} to halo formation in a large-scale environment \citep{Colberg:1999-Linking-cluster-formation-to-l}, and in the testing of models for halo abundances and mass functions \citep{Sheth:2001-Ellipsoidal-collapse-and-an-im}. Some key parameters of the N-body simulations are listed in Table \ref{tab:1}}. 

\begin{table}
\caption{Numerical parameters of N-body simulation}
\begin{tabular}{p{0.25in}p{0.1in}p{0.05in}p{0.05in}p{0.05in}p{0.05in}p{0.37in}p{0.1in}p{0.4in}p{0.35in}} 
\hline 
Run & $\Omega_{0}$ & $\Lambda$ & $h$ & $\Gamma$ & $\sigma _{8}$ & \makecell{L\\(Mpc/h)} & $N$ & \makecell{$m_{p}$\\($M_{\odot}/h$)} & \makecell{$l_{soft}$\\(Kpc/h)} \\ 
\hline 
SCDM1 & 1.0 & 0.0 & 0.5 & 0.5 & 0.51 & \centering 239.5 & $256^{3}$ & 2.27$\times 10^{11}$ & \makecell{\centering 36} \\ 
\hline 
\end{tabular}
\label{tab:1}
\end{table}

In this paper, a correlation-based approach is used to directly calculate the real-space statistics by pairwise averaging over all particle pairs on a given scale. This approach can maximally extract the information contained in N-body simulations and generate complete statistics on all scales without involving any projection kernels (e.g. cloud-in-cell). However, it is also computationally intensive to identify all pairs of particles on all scales. The selected N-body simulation has a relatively low resolution with fewer particles compared with some recent large-scale cosmological simulations. This allows computationally affordable calculations of various real-space two-point statistics. With increasing computing power, the same approach can be applied similarly to other simulations with higher resolution and more particles. Two relevant datasets from this N-body simulation, that is, halo-based and correlation-based statistics of dark matter flow, can be found at Zenodo.org \citep{Xu:2022-Dark_matter-flow-dataset-part1, Xu:2022-Dark_matter-flow-dataset-part2}. %along with the accompanying presentation slides, "A comparative study of dark matter flow \& hydrodynamic turbulence and its applications" \citep{Xu:2022-Dark_matter-flow-and-hydrodynamic-turbulence-presentation}. All data files are also available on GitHub \citep{Xu:Dark_matter_flow_dataset_2022_all_files}.

\begin{comment}
\begin{table}
\caption{Numerical parameters of N-body simulation}
\begin{tabular}{p{0.25in}p{0.05in}p{0.05in}p{0.05in}p{0.05in}p{0.05in}p{0.4in}p{0.1in}p{0.4in}p{0.4in}} 
\hline 
Run & $\Omega_{0}$ & $\Lambda$ & $h$ & $\Gamma$ & $\sigma _{8}$ & \makecell{L\\(Mpc/h)} & $N$ & \makecell{$m_{p}$\\$M_{\odot}/h$} & \makecell{$l_{soft}$\\(Kpc/h)} \\ 
\hline 
SCDM1 & 1.0 & 0.0 & 0.5 & 0.5 & 0.51 & \centering 239.5 & $256^{3}$ & 2.27$\times 10^{11}$ & \makecell{\centering 36} \\ 
\hline 
\end{tabular}
\label{tab:1}
\end{table}
\end{comment}

%\section{Second order velocity statistics}
%\label{sec:3-1}
%Various statistical measures were introduced to characterize the velocity field of dark matter flow. This section provides a brief overview of the kinematic relations.

\section{Two-point third order velocity statistics}
\label{sec:3}
The real-space correlation functions are one of the most fundamental statistical measures that can be used to describe the random and multiscale nature of any stochastic flow. As shown in Fig. \ref{fig:1}, for velocity $\boldsymbol{\mathrm{u}}$ and $\boldsymbol{\mathrm{u}}^{'}$ of two DM particles at two different locations separated by a distance \textit{r}, where $\boldsymbol{\mathrm{x}}^{'} =\boldsymbol{\mathrm{x}}+\boldsymbol{\mathrm{r}}$, the longitudinal and transverse velocities at location $\boldsymbol{\mathrm{x}}$ can be written as \citep{Xu:2023-On-the-statistical-theory-of-self-gravitating}
\begin{equation} 
\label{ZEqnNum286963} 
u_{L} =\boldsymbol{\mathrm{u}}\cdot \hat{\boldsymbol{\mathrm{r}}}=u_{i} \hat{r}_{i}  \quad \textrm{and} \quad \boldsymbol{\mathrm{u}}_{T} =-\left(\boldsymbol{\mathrm{u}}\times \hat{\boldsymbol{\mathrm{r}}}\times \hat{\boldsymbol{\mathrm{r}}}\right)=\boldsymbol{\mathrm{u}}-\left(\boldsymbol{\mathrm{u}}\cdot \hat{\boldsymbol{\mathrm{r}}}\right)\hat{\boldsymbol{\mathrm{r}}} ,        
\end{equation} 
where $\hat{\boldsymbol{\mathrm{r}}}$ is the unit vector and $\hat{r}_{i} ={r_{i} /r}$ is the Cartesian component. 
The two-point second order velocity correlation tensor quantifies the degree to which the velocities at two locations are statistically correlated. For a homogeneous and isotropic tensor, it can be mathematically defined as \citep{Xu:2023-On-the-statistical-theory-of-self-gravitating}
\begin{equation} 
\label{ZEqnNum195930} 
Q_{ij} \left(r\right)=\left\langle u_{i} \left(\boldsymbol{\mathrm{x}}\right)u_{j} \left(\boldsymbol{\mathrm{x}}^{'} \right)\right\rangle \equiv \left\langle u_{i} u_{j}^{'} \right\rangle=A_{2} \left(r\right)r_{i} r_{j} +B_{2} \left(r\right)\delta _{ij}, 
\end{equation} 
where $A_{2}$ and $B_{2}$ are two scalar functions of distance $r$. In \citep{Xu:2023-On-the-statistical-theory-of-self-gravitating}, three velocity correlation functions were introduced by index contraction of $Q_{ij} \left(r\right)$ in Eq. \eqref{ZEqnNum195930}, i.e. a total correlation $R_{2} \left(r\right)$
\begin{equation} 
\label{ZEqnNum935612} 
R_{2} \left(r\right)=Q_{ij} \delta _{ij} =\left\langle \boldsymbol{\mathrm{u}}\cdot \boldsymbol{\mathrm{u}}_{}^{'} \right\rangle =\left\langle u_{i} u_{i}^{'} \right\rangle \equiv L_2(r)+2T_2(r), \end{equation} 
a longitudinal correlation function $L_{2} \left(r\right)$
\begin{equation}
\label{ZEqnNum950112} 
L_{2} \left(r\right)=Q_{ij} {r_{i} r_{j} /r^{2} } =\left\langle u_{L} u_{L}^{'} \right\rangle,       
\end{equation} 
and a transverse (lateral) correlation function
\begin{equation} 
\label{ZEqnNum991811} 
T_{2} \left(r\right)=Q_{ij} n_{i} n_{j} ={\left\langle \boldsymbol{\mathrm{u}}_{T} \cdot \boldsymbol{\mathrm{u}}_{T}^{'} \right\rangle /2},   
\end{equation} 
where \textbf{\textit{n}} is a normal vector perpendicular to \textbf{r }(i.e. $\boldsymbol{\mathrm{n}}\cdot \hat{\boldsymbol{\mathrm{r}}}=0$).

\begin{figure}
\includegraphics*[width=\columnwidth]{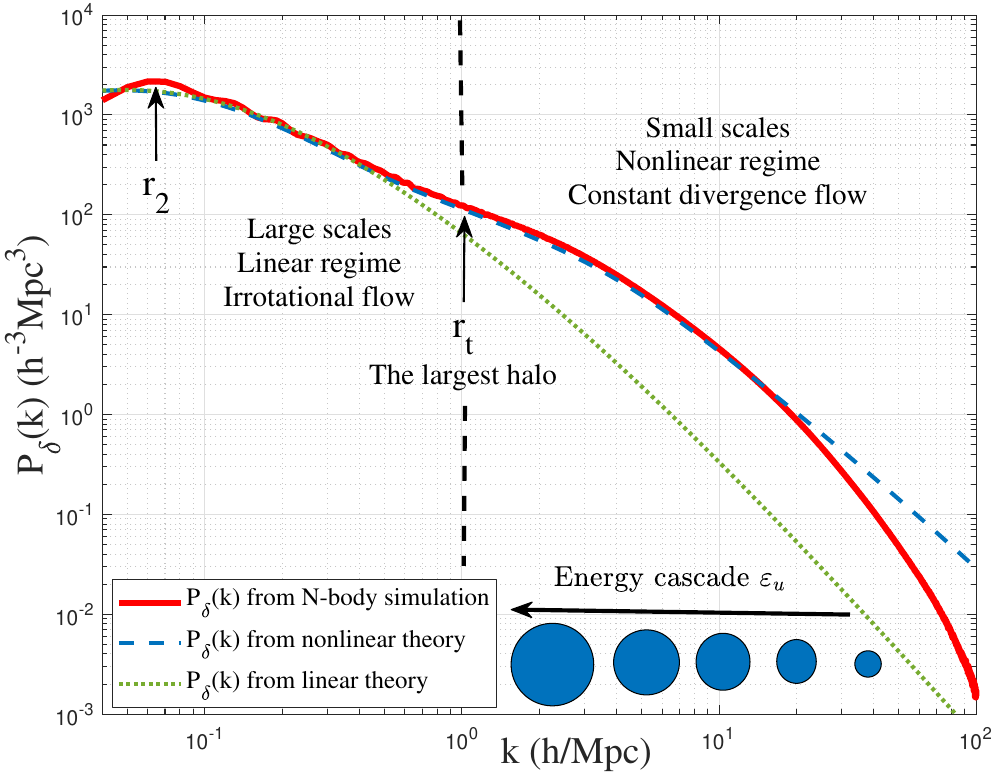}
\caption{The variation of density power spectrum $P_{\delta}(k)$ with comoving wavenumber $k$ at \textit{z}=0. Linear and non-linear theory predictions are also presented for comparison. The pivot wavenumber $k_{\max}$ (or scale $r_2$=21.3 Mpc/h \citep{Xu:2023-On-the-statistical-theory-of-self-gravitating}) denotes the size of the horizon at the matter-radiation equality. The scale $r_t\approx$1Mpc/h is roughly the size of the largest halo (see Fig. \ref{fig:20}), where the longitudinal velocity correlation equals the transverse velocity correlation ($L_2(r_t)=T_2(r_t)$ in Eqs. \eqref{ZEqnNum950112} and \eqref{ZEqnNum991811}). Dark matter flow is irrotational on large scales $r>r_t$ (linear regime) and constant divergence on small scales $r<r_t$ (nonlinear regime). More details are presented in \citep{Xu:2023-On-the-statistical-theory-of-self-gravitating}. There also exists a halo-mediated energy cascade with a constant rate of $\varepsilon_u$ on small scales $r<r_t$ (nonlinear regime), which can be used to derive halo mass functions, density profiles \citep{Xu:2023-Dark-matter-halo-mass-functions-and}, and dark matter particle properties \citep{Xu:2022-Postulating-dark-matter-partic}. The effective viscosity $\nu$ on large scales $r>r_t$ can be related to the energy cascade $\varepsilon_u$ on small scales $r<r_t$ (Eq. \eqref{ZEqnNum133069}). This paper focuses on the kinematic and dynamic relations, and two-point high order statistics on small and large scales, respectively.}
\label{fig:19}
\end{figure}

The kinematic relations between these second order correlation functions have been developed for different types of flow \citep{Xu:2023-On-the-statistical-theory-of-self-gravitating}. Since cosmic peculiar velocity exhibits a different nature on small and large scales, the kinematic relations are also different on different scales. To illustrate this, Figure \ref{fig:19} presents the density power spectrum $P_{\delta}(k)$ as a function of the wavenumber $k$ from the N-body simulation (solid red line). The corresponding predictions from linear theory (dotted green line) and nonlinear theory (dashed blue line) are also presented for comparison \citep{Jenkins:1998-Evolution-of-structure-in-cold}. The comoving length scale $r_2$ or the pivot wavenumber $k_{max}$ ($k_{max}r_2=\sqrt{2}$ \citep{Xu:2023-On-the-statistical-theory-of-self-gravitating}) is related to the horizon size at the matter-radiation equality, where $k_{max}\propto \Omega_m h^2$ is proportional to the matter content $\Omega_m$. The length scale $r_t\approx 1Mpc/h$ is roughly the size of the largest halo at $z=0$, which can be defined as the scale when the longitudinal correlation equals the transverse correlation, i.e. $L_2(r_t)\equiv T_2(r_t)$ \citep{Xu:2023-On-the-statistical-theory-of-self-gravitating}. On small scales, the longitudinal correlation is dominant over the transverse correlation due to stronger gravity. On large scales, the transverse correlation $T_2$ is dominant over the longitudinal correlation $L_2$. Here, $r_t$ is the transition scale.

On small scales ($r<r_t$) in the highly nonlinear regime, the virialized haloes have a vanishing (proper) radial flow, and the peculiar velocity $\textbf{u}$ satisfies a constant divergence or $\nabla \cdot \textbf{u}=-3Ha$, where $H$ is the Hubble parameter and $a$ is the scale factor \citep{Xu:2022-The-mean-flow--velocity-disper}. On large scales ($r>r_t$) in the linear regime, the flow becomes irrotational or $\nabla \times \textbf{u}=0$. The different nature of flow on different scales is a unique feature of the self-gravitating collisionless dark matter flow (SG-CFD) that is different from conventional hydrodynamic turbulence \citep{Xu:2023-On-the-statistical-theory-of-self-gravitating}. In the same work, we show that the incompressible flow and constant divergence flow share the same kinematic relations for even order correlation functions. Therefore, for dark matter flow on small scales, three second order correlation functions satisfy kinematic relations
\begin{equation}
T_{2} =\frac{1}{2r} \left(r^{2} L_{2} \right)_{,r} \quad \textrm{and} \quad R_{2} =\frac{1}{r^{2} } \left(r^{3} L_{2} \right)_{,r}.      
\label{ZEqnNum314105}
\end{equation}
On large scales of irrotational nature, the kinematic relations read 
\begin{equation}
R_{2} =\frac{1}{r^{2} } \left(r^{3} T_{2} \right)_{,r} \quad \textrm{and} \quad L_{2} =\left(rT_{2} \right)_{,r}. 
\label{ZEqnNum320035}
\end{equation}

\begin{table}
\caption{One-dimensional velocity dispersion $u(z)$ (Km/s)}
%\begin{tabular}{p{0.1in}p{0.1in}p{0.1in}p{0.1in}p{0.1in}p{0.1in}p{0.1in}}
\begin{tabular}{lccccc}
\hline 
Redshift $z$  & 10   & 5      & 3      & 2      & 1.5    \\
Dispersion $u$  & 86.47& 118.50 & 148.61 & 177.15 & 199.76  \\ \hline
Redshift $z$  & 1.0    & 0.5    & 0.3    & 0.1    & 0.0   \\
Dispersion $u$  & 231.29 & 277.67 & 303.37 & 335.42 & 354.61 \\ \hline
\end{tabular}
\label{tab:2}
\end{table}

The functional form of correlation functions can also be obtained based on these kinematic relations. The N-body simulation strongly suggests an exponential form for the transverse velocity correlation function on large scales:
\begin{equation} 
\label{ZEqnNum971850} 
T_{2} \left(r,a\right)=a_{0} u^{2} \exp \left(-\frac{r}{r_{2} } \right),         
\end{equation} 
which should be an intrinsic property of large-scale dynamics (see Section \ref{sec:6.3}). Here $u^2$ is the one-dimensional velocity dispersion of all dark matter particles. We show that $a_{0} u^{2} \propto a$ and $a_{0} \left({u/u_{0} } \right)^{2} =0.45a$, where $u_{0}^{2} =u^{2} \left(a=1\right)$ is the dispersion in the present epoch of $z=0$. \rev{Table \ref{tab:2} lists the values of velocity dispersion u(z) at different redshifts $z$. The correlation functions at different redshifts can conveniently be normalized by the velocity dispersion $u(z)$ at that redshift.}

On large scales, the kinetic energy (or second order correlation functions) increases as $\propto a_{0} u^{2} \propto a$, consistent with the energy evolution in the N-body system \citep{Xu:2022-Postulating-dark-matter-partic}. The dynamic origin of the exponential correlations is discussed in Section \ref{sec:6.3} of this paper. With kinematic relations in Eq. \eqref{ZEqnNum320035}, the longitudinal and total correlation functions read:
\begin{equation} 
\label{ZEqnNum344034} 
\begin{split}
&L_{2} \left(r,a\right)=a_{0} u^{2} \exp \left(-\frac{r}{r_{2} } \right)\left(1-\frac{r}{r_{2} } \right),\\
&R_{2} \left(r,a\right)=a_{0} u^{2} \exp \left(-\frac{r}{r_{2} } \right)\left(3-\frac{r}{r_{2} } \right),  
\end{split}
\end{equation} 
where $L_{2} =0$ at $r=r_{2}$. Here $r_{2}$ = 21.3 Mpc/h is a redshift independent comoving length scale (for SCDM simulation of matter-dominant) that could be related to the size of the horizon at the matter-radiation equality (Fig. \ref{fig:19}).

On small scales in the nonlinear regime of constant divergence, the longitudinal velocity correlation $L_{2} \left(r\right)$ has the form of
\begin{equation} 
\label{ZEqnNum178092} 
L_{2} \left(r\right)=u^{2} \left[1-\left(\frac{r}{r_{1} } \right)^{n} \right],          
\end{equation} 
where $n=1/4$ (see Eq. \eqref{eq:189} for derivation) and $r_1$ is another length scale \citep{Xu:2023-On-the-statistical-theory-of-self-gravitating}. Similarly, the kinematic relations in Eq. \eqref{ZEqnNum314105} can be used to derive the total and transverse velocity correlation functions 
\begin{equation} 
\label{ZEqnNum955991} 
\begin{split}
&R_{2} =u^{2} \left[3-\left(3+n\right)\left(\frac{r}{r_{1} } \right)^{n} \right],  \\ 
&T_{2} =u^{2} \left[1-\frac{2+n}{2} \left(\frac{r}{r_{1} } \right)^{n} \right].     
\end{split}
\end{equation} 
Complete velocity correlation functions can be constructed on all scales on the basis of solutions on small and large scales \citep{Xu:2023-On-the-statistical-theory-of-self-gravitating}.

All results discussed so far focus on the statistical theory of two-point second order statistics. In this paper, the statistical theory of two-point high order statistics is developed that includes both kinematic relations for correlations of the same order and the dynamic relations for correlations of different orders. This is a challenging task involving high order tensors and vector calculus of great complexity. In this section, we start our journey with two-point third-order statistics. As usual, we restrict our discussion to homogeneous and isotropic flow, which will significantly simplify the velocity correlation tensors and the development of theory. The SG-CFD flow is of constant divergence nature on small scales and is irrotational on large scales. Constant divergence flow and incompressible flow share the same kinematic relations of even order, whereas they can be different for odd-order statistics. Therefore, we will need to discuss two-point third order statistics for different types of flow, i.e., incompressible, constant divergence, or irrotational flow, respectively.    

\subsection{Third order velocity correlation tensor}
\label{sec:3.1}
Due to homogeneous and isotropic symmetry, the two-point third order velocity correlation tensor can be generally defined as
\begin{equation} 
\label{ZEqnNum883006}
\begin{split}
Q_{ijk} \left(\boldsymbol{\mathrm{x}},\boldsymbol{\mathrm{r}}\right)&=Q_{ijk} \left(\boldsymbol{\mathrm{r}}\right)=Q_{ijk} \left(r\right)\\
&=\left\langle u_{i} \left(\boldsymbol{\mathrm{x}}\right)u_{j} \left(\boldsymbol{\mathrm{x}}\right)u_{k} \left(\boldsymbol{\mathrm{x}}^{'} \right)\right\rangle =\left\langle u_{i} u_{j} u_{k}^{'} \right\rangle 
\end{split}
\end{equation} 
for velocity \textbf{u} at two different locations separated by a distance \textit{r} (Fig. \ref{fig:1}). The prime notation indicates the field evaluated at the location $\boldsymbol{\mathrm{x}}^{'}$. 
For third order isotropic tensor $Q_{ijk}(r)$, symmetry requires
\begin{equation} 
\label{ZEqnNum264952} 
Q_{ijk} \left(-r\right)=\left\langle u_{i}^{'} u_{j}^{'} u_{k}^{} \right\rangle =-\left\langle u_{i} u_{j} u_{k}^{'} \right\rangle =-Q_{ijk} \left(r\right).       
\end{equation} 
The most general form of the isotropic third order correlation tensor can be written as,
\begin{equation} 
\label{eq:3}
\begin{split}
Q_{ijk} \left(r\right)=A_{3} \left(r\right)r_{i} r_{j} r_{k} &+B_{3} \left(r\right)r_{i} \delta _{jk}\\
&+C_{3} \left(r\right)r_{j} \delta _{ki} +D_{3} \left(r\right)r_{k} \delta _{ij},
\end{split}
\end{equation} 
where $A_{3} \left(r\right)$, $B_{3} \left(r\right)$, $C_{3} \left(r\right)$ and $D_{3} \left(r\right)$ are all symmetric regular functions of scale \textit{r}. Because of the symmetry about indexes \textit{i} and \textit{j} (using definition in Eq. \eqref{ZEqnNum883006}), 
\begin{equation}
Q_{ijk} \left(r\right)=Q_{jik} \left(r\right) \quad \textrm{leads to} \quad B\left(r\right)=C\left(r\right). 
\label{ZEqnNum859159}
\end{equation}
\noindent The final form of the third order correlation tensor simply reads
\begin{equation} 
\label{ZEqnNum953203} 
Q_{ijk} \left(r\right)=A_{3} r_{i} r_{j} r_{k} +B_{3} \left(r_{i} \delta _{jk} +r_{j} \delta _{ki} \right)+D_{3} r_{k} \delta _{ij} ,     
\end{equation} 
which is fully determined by three scalar functions $A_{3} $, $B_{3} $ and $D_{3}$. 

Using contraction in index, the longitudinal triple (third order) correlation function is defined as
\begin{equation} 
\label{ZEqnNum550536} 
L_{3} \left(r\right)=Q_{ijk} \hat{r}_{i} \hat{r}_{j} \hat{r}_{k} =\left\langle u_{L}^{2} u_{L}^{'} \right\rangle =A_{3} r^{3} +\left(2B_{3} +D_{3} \right)r,        
\end{equation} 
where $\hat{r}_{i} ={r_{i}/r} $ is the normalized Cartesian components of vector \textbf{\textit{r}} satisfying $\hat{r}_{i} \hat{r}_{i} =1$. Einstein summation is employed. Here, $u_{L}$ is the longitudinal velocity in Fig. \ref{fig:1}. Two total third order correlation functions can be defined as,
\begin{equation} 
\label{ZEqnNum811607}
\begin{split}
R_{3} \left(r\right)&=\frac{1}{2} Q_{ijk} \left(\delta _{ik} \hat{r}_{j} +\delta _{jk} \hat{r}_{i} \right)\\
&=\left\langle u_{L}^{} \boldsymbol{\mathrm{u}}\cdot \boldsymbol{\mathrm{u}}_{}^{'} \right\rangle =A_{3}^{} r^{3} +\left(4B_{3} +D_{3} \right)r,
\end{split}
\end{equation} 
and
\begin{equation} 
\label{ZEqnNum819453} 
R_{31} \left(r\right)=Q_{ijk} \delta _{ij} \hat{r}_{k} =\left\langle \boldsymbol{\mathrm{u}}\cdot \boldsymbol{\mathrm{u}}u_{L}^{'} \right\rangle =A_{3}^{} r^{3} +\left(2B_{3} +3D_{3} \right)r.      
\end{equation} 
The transverse third-order correlation function can be defined as
\begin{equation} 
\label{ZEqnNum796638} 
T_{3} \left(r\right)={\left\langle u_{L}^{} \boldsymbol{\mathrm{u}}_{T} \cdot \boldsymbol{\mathrm{u}}_{T}^{'} \right\rangle/2} ={\left(R_{3} -L_{3} \right)/2} =B_{3} r.       
\end{equation} 
All correlations satisfy the odd symmetry $f\left(-r\right)=-f\left(r\right)$. 

Next, the divergence and curl of the third-order correlation tensor are derived. Some tensor/vector algebras are involved and only the final results are presented here. The divergence reads
\begin{equation}
\begin{split}
Q_{ijk,k}&=\frac{\partial Q_{ijk} \left(r\right)}{\partial r_{k} } =\frac{\partial \left\langle u_{i} u_{j} u_{k}^{'} \right\rangle }{\partial r_{k} }\\&=\left(5A_{3} +\frac{\partial A_{3} }{\partial r} r+\frac{2}{r} \frac{\partial B_{3} }{\partial r} \right)r_{i} r_{j} +\left(2B_{3} +\frac{\partial D_{3} }{\partial r} r+3D_{3} \right)\delta _{ij},\\
Q_{iik,k}&=Q_{ijk,k} \delta _{ij}\\&=\left(5A_{3} +\frac{\partial A_{3} }{\partial r} r+\frac{2}{r} \frac{\partial B_{3} }{\partial r} \right)r^{2}+3\left(2B_{3} +\frac{\partial D_{3} }{\partial r} r+3D_{3} \right),
\end{split}
\label{ZEqnNum405014} 
\end{equation} 
and
\begin{equation} 
\label{ZEqnNum699561} 
\begin{split}
Q_{ijk,i} =\frac{\partial Q_{ijk} \left(r\right)}{\partial r_{i} } &=\left(5A_{3} +\frac{\partial A_{3} }{\partial r} r+\frac{1}{r} \frac{\partial B_{3} }{\partial r} +\frac{1}{r} \frac{\partial D_{3} }{\partial r} \right)r_{j} r_{k}\\
&+\left(4B_{3} +\frac{\partial B_{3} }{\partial r} r+D_{3} \right)\delta _{jk}.
\end{split}
\end{equation} 
Other derivatives can be derived from Eq. \eqref{ZEqnNum699561} for later use,
\begin{equation} 
\label{ZEqnNum813280} 
\begin{split}
Q_{ijk,ij}&=\frac{\partial Q_{ijk} \left(r\right)}{\partial r_{i} \partial r_{j} } =\left(r^{2} \frac{\partial ^{2} A_{3} }{\partial r^{2} } +10r\frac{\partial A_{3} }{\partial r}\right.\\
&\left.+20A_{3} +2\frac{\partial ^{2} B_{3} }{\partial r^{2} } +\frac{8}{r} \frac{\partial B_{3} }{\partial r} +\frac{\partial ^{2} D_{3} }{\partial r^{2} } +\frac{4}{r} \frac{\partial D_{3} }{\partial r} \right)r_{k},
\end{split}
\end{equation} 
\begin{equation} 
\label{ZEqnNum415504} 
\begin{split}
Q_{iki,k}&=Q_{ikk,i} =Q_{ijk,i} \delta _{jk}\\&=5A_{3} r^{2} +\frac{\partial A_{3} }{\partial r} r^{3} +12B_{3} +4r\frac{\partial B_{3} }{\partial r} +3D_{3} +r\frac{\partial D_{3} }{\partial r},
\end{split}
\end{equation} 
where symmetry condition (Eq. \eqref{ZEqnNum859159}) is used for deriving Eq. \eqref{ZEqnNum415504}. 

With the definition of correlation functions $R_{3} $ and $R_{31} $ in Eqs. \eqref{ZEqnNum811607} and \eqref{ZEqnNum819453}, Eqs. \eqref{ZEqnNum405014} and \eqref{ZEqnNum415504} can be concisely written as
\begin{equation}
\begin{split}
&Q_{iki,k} =Q_{ijk,i} \delta _{jk} =Q_{ikk,i} =\frac{1}{r^{2} } \left(r^{2} R_{3} \right)_{,r}\\
&\textrm{and}\\
&Q_{iik,k} =\frac{1}{r^{2} } \left(r^{2} R_{31} \right)_{,r}.
\end{split}
\label{ZEqnNum952051}
\end{equation}

\noindent Similarly, the curl of third order velocity correlation tensor reads
\begin{equation} 
\label{ZEqnNum285705} 
\begin{split}
\nabla \times Q_{mni} \left(r\right)&=\varepsilon _{ijk} Q_{mnk,j}\\ &=\left(A_{3} -\frac{1}{r} \frac{\partial B_{3} }{\partial r} \right)\left(\varepsilon _{imk} r_{n} r_{k} +\varepsilon _{ink} r_{m} r_{k} \right).
\end{split}
\end{equation} 
where $\varepsilon_{ijk}$ is the standard Levi-Civita symbol.  

\subsection{Kinematic relations for third order correlations}
\subsubsection{Kinematic relations for incompressible flow}
\label{sec:3.1.1}
This section formulates the kinematic relations for incompressible flow. A more compact formulation is also presented in the Appendix \ref{appendix:A}, which facilitates generalization to arbitrary order. Incompressible flow requires divergence-free ($\nabla \cdot \textbf{u}=0$) that leads to (Eq. \eqref{ZEqnNum405014}):
\begin{equation} 
\label{ZEqnNum280236} 
5A_{3} r+\frac{\partial A_{3} }{\partial r} r^{2} +2\frac{\partial B_{3} }{\partial r} =0,         
\end{equation} 
\begin{equation} 
\label{ZEqnNum113602} 
2B_{3} +\frac{\partial D_{3} }{\partial r} r+3D_{3} =0.          
\end{equation} 
Differentiating Eq. \eqref{ZEqnNum113602} with $r$ and subtracting Eq. \eqref{ZEqnNum280236} leads to
\begin{equation} 
\label{ZEqnNum193922} 
\left(\frac{\partial D_{3} }{\partial r} r\right)_{,r} +3\frac{\partial D_{3} }{\partial r} =r^{2} \frac{\partial A_{3} }{\partial r} +5A_{3} r=\left(A_{3} r^{2} \right)_{,r} +3A_{3} r.      
\end{equation} 
From Eqs. \eqref{ZEqnNum113602} and \eqref{ZEqnNum193922}), we should have
\begin{equation}
A_{3}=\frac{1}{r} \frac{\partial D_{3}^{} }{\partial r} \quad \textrm{and} \quad A_{3} r^{2} +2B_{3} +3D_{3} =0.
\label{ZEqnNum934928}
\end{equation}
\noindent Equation \eqref{ZEqnNum113602} can be analytically solved (using Eqs. \eqref{ZEqnNum550536} and \eqref{ZEqnNum934928}),
\begin{equation}
B_{3} =-\frac{r}{2} \frac{\partial D_{3} }{\partial r} -\frac{3}{2} D_3 \quad \textrm{and} \quad D_{3} =-\frac{L_{3}}{2r}.     
\label{eq:21}
\end{equation}
\noindent With all scalar functions $A_{3}$ ,$B_{3}$, and $D_{3}$ expressed in terms of the longitudinal correlation $L_{3}(r)$, the third order correlation tensor can be expressed as (prime denotes the derivative with respect to \textit{r})
\begin{equation} 
\label{ZEqnNum235209} 
\begin{split}
Q_{ijk} \left(r\right)&=\frac{L_{3} -rL_{3}^{'} }{2} \hat{r}_{i} \hat{r}_{j} \hat{r}_{k} \\
&+\frac{2L_{3} +rL_{3}^{'} }{4} \left(\hat{r}_{i} \delta _{jk} +\hat{r}_{j} \delta _{ki} \right)-\frac{L_{3} }{2} \hat{r}_{k} \delta _{ij}.
\end{split}
\end{equation} 

From Eqs. \eqref{ZEqnNum813280}, \eqref{ZEqnNum280236} and \eqref{ZEqnNum113602}, $Q_{ijk,ij} =0$. Since the first-order isotropic tensor for incompressible flow must vanish \citep{Xu:2023-On-the-statistical-theory-of-self-gravitating}, we have
\begin{equation} 
\label{eq:23} 
Q_{ijk,ij} =Q_{ijk,ik} =Q_{ijk,jk} =0.         
\end{equation} 
Multiplying Eq. \eqref{ZEqnNum235209} by $\delta _{jk}$ and taking the divergence, 
\begin{equation} 
\label{ZEqnNum589090} 
\begin{split}
Q_{iki,k}&=Q_{ijk,i} \delta _{jk} =Q_{ikk,i}\\&=\frac{1}{r^{2} } \left(R_{3} r^{2} \right)_{,r}=\frac{1}{2r^{2} } \frac{\partial }{\partial r} \left(\frac{1}{r} \frac{\partial }{\partial r} \left(r^{4} L_{3} \right)\right),
\end{split}
\end{equation} 
with the following identity used
\begin{equation}
\left(\hat{r}_{i} \right)_{,j} =\frac{1}{r} \left(\delta _{ij} -\hat{r}_{i} \hat{r}_{j} \right)=\left(\hat{r}_{j} \right)_{,i} \quad \textrm{and} \quad \left(\hat{r}_{i} \right)_{,i} =\frac{2}{r}.     
\label{ZEqnNum679713}
\end{equation}

Finally, the kinematic relations between third-order correlations can be easily obtained by contraction in the index notation of Eq. \eqref{ZEqnNum235209} (using definitions in Eqs. \eqref{ZEqnNum811607} and \eqref{ZEqnNum796638}) such that
\begin{equation}
\begin{split}
&R_{3} =\frac{1}{2r^{3} } \left(r^{4} L_{3} \right)_{,r},\quad T_{3} =\frac{1}{4r} \left(r^{2} L_{3} \right)_{,r},\\
&\textrm{and}\\
&r^{2} \left(r^{2} R_{3} \right)_{,r} =2\left(r^{4} T_{3} \right)_{,r}.
\end{split}
\label{ZEqnNum374800}
\end{equation}

\noindent These kinematic relations will be generalized to arbitrary higher order in Section \ref{sec:4} with a new method of derivation presented in the appendix. From Eqs. \eqref{ZEqnNum934928} and \eqref{ZEqnNum819453}, the correlation function $R_{31}$ vanishes for incompressible flow,
\begin{equation}
\label{ZEqnNum218978} 
R_{31} \left(r\right)=\left\langle \boldsymbol{\mathrm{u}}\cdot \boldsymbol{\mathrm{u}}u_{L}^{'} \right\rangle =0. \end{equation} 
The same correlation $R_{31}$ does not vanish for dark matter flow with a constant divergence on small scales, showing that two types of flow are different in odd order correlations (Fig. \ref{fig:8}).

\subsubsection{Kinematic relations for constant divergence flow}
\label{sec:3.1.2}
Dark matter flow is of constant divergence on small scales. The kinematic relations for constant divergence flow are different from the incompressible flow for odd order correlations. The peculiar radial flow ${u}_r$ in virialized haloes satisfies $\boldsymbol{\mathrm{u}_r}=-Ha\boldsymbol{\mathrm{r}}$ from the stable clustering hypothesis \citep{Xu:2021-A-non-radial-two-body-collapse}. The divergence of peculiar velocity $\nabla \cdot \boldsymbol{\mathrm{u}}=-3Ha$ (in local spherical coordinate), that is, a spatially constant divergence \citep{Xu:2022-The-mean-flow--velocity-disper}. Without loss of generality, assume that $\nabla \cdot \boldsymbol{\mathrm{u}}=\theta \left(t\right)$, where $\theta=-3Ha$ is a constant in space. The divergence of the third-order correlation tensor simply reads
\begin{equation} 
\label{eq:28}
\begin{split}
Q_{ijk,k} &=\nabla ^{'} \cdot \left\langle u_{i} \left(\boldsymbol{\mathrm{x}}\right)u_{j} \left(\boldsymbol{\mathrm{x}}\right)u_{k} \left(\boldsymbol{\mathrm{x}}^{'} \right)\right\rangle\\
&=\left\langle u_{i} \left(\boldsymbol{\mathrm{x}}\right)u_{j} \left(\boldsymbol{\mathrm{x}}\right)\frac{\partial u_{k} \left(\boldsymbol{\mathrm{x}}^{'} \right)}{\partial x_{k}^{'} } \right\rangle =\theta \left\langle u_{i} u_{j} \right\rangle,
\end{split}
\end{equation} 
where the prime stands for taking derivative at location $\boldsymbol{\mathrm{x}}^{'}$. The constant divergence requires (Eq. \eqref{ZEqnNum405014}),
\begin{equation} 
\label{ZEqnNum139076}
\begin{split}
Q_{ijk,k}=\left\langle u_{i} u_{j} \right\rangle \theta=&\left(5A_{3} +\frac{\partial A_{3} }{\partial r} r+\frac{2}{r} \frac{\partial B_{3} }{\partial r} \right)r_{i} r_{j} \\
&+\left(2B_{3} +\frac{\partial D_{3} }{\partial r} r+3D_{3} \right)\delta _{ij}.
\end{split}
\end{equation} 
Multiplying $\hat{r}_{i} \hat{r}_{j} $ on both sides of Eq. \eqref{ZEqnNum139076} gives
\begin{equation} 
\label{eq:30} 
\left\langle u_{L}^{2} \right\rangle \theta =\left(5A_{3} +\frac{\partial A_{3} }{\partial r} r+\frac{2}{r} \frac{\partial B_{3} }{\partial r} \right)r^{2} +\left(2B_{3} +\frac{\partial D_{3} }{\partial r} r+3D_{3} \right).     
\end{equation} 

Using definitions of correlation functions in Eqs. \eqref{ZEqnNum550536} and \eqref{ZEqnNum811607}, an exact kinematic relation can be obtained between correlation functions, longitudinal dispersion $\langle u_{L}^{2} \rangle $ and the divergence $\theta$, 
\begin{equation}
\label{ZEqnNum647632} 
R_{3} +\frac{1}{2} \left\langle u_{L}^{2} \right\rangle \theta r=\frac{1}{2r^{3} } \left(r^{4} L_{3} \right)_{,r}.  
\end{equation} 

Similarly, multiply $\delta _{ij}$ on both sides of Eq. \eqref{ZEqnNum139076} leads to
\begin{equation} 
\label{eq:32}
\begin{split}
\left\langle u^{2} \right\rangle \theta=\left(5A_{3} +\frac{\partial A_{3} }{\partial r} r+\frac{2}{r} \frac{\partial B_{3} }{\partial r} \right)r^{2}+3\frac{\partial D_{3}}{\partial r}r+6B_{3}+9D_{3}. \end{split}    
\end{equation} 
An exact kinematic relation for $R_{31} $ and the total velocity dispersion $\langle u^{2} \rangle$ on the scale $r$ can be obtained from Eqs. \eqref{ZEqnNum405014} and \eqref{ZEqnNum952051},
\begin{equation} 
\label{ZEqnNum487892} 
\left\langle u^{2} \right\rangle \theta =\frac{1}{r^{2} } \left(r^{2} R_{31} \right)_{,r} .          
\end{equation} 
With $\langle u^{2} \rangle \approx 3 \langle u_{L}^{2} \rangle$ %\citep[see][Fig. 20]{Xu:2022-Two-thirds-law-for-pairwise-ve} 
that is exact on both small and large scales (\citep [see] [Fig. (20)]{Xu:2022-Two-thirds-law-for-pairwise-ve}, the kinematic relation between three third order correlation functions finally reads,
\begin{equation} 
\label{ZEqnNum552000} 
R_{3} +\frac{1}{6r} \left(r^{2} R_{31} \right)_{,r} =\frac{1}{2r^{3} } \left(r^{4} L_{3} \right)_{,r} .        
\end{equation} 
For small \textit{r} with velocity dispersion $\langle u_{L}^{2} \rangle$ independent of \textit{r} , solution of $R_{31} $ from Eq. \eqref{ZEqnNum487892} is 
\begin{equation}
\label{ZEqnNum126019} 
R_{31} =\theta \left\langle u_{L}^{2} \right\rangle r,           
\end{equation} 
i.e. a linear relation $R_{31}\propto r$ on small scale that can be directly tested by N-body simulations (Fig. \ref{fig:3}). In particular, with $\theta=0$, kinematic relations Eqs. \eqref{ZEqnNum647632}, \eqref{ZEqnNum487892}, and \eqref{ZEqnNum552000} are reduced to Eqs. \eqref{ZEqnNum374800} and \eqref{ZEqnNum218978}, as expected. 

\subsubsection{Kinematic relations for irrotational flow}
\label{sec:3.1.3}
Dark matter flow is irrotational on large scales. The curl-free condition ($\nabla \times \boldsymbol{\mathrm{u}}=0$) leads to (using Eq. \eqref{ZEqnNum285705}),
\begin{equation} 
\label{eq:36} 
\frac{1}{r} \frac{\partial B_{3} }{\partial r} =A_{3} .            
\end{equation} 
The kinematic relations between velocity correlations can be similarly found from Eqs. \eqref{ZEqnNum550536}-\eqref{ZEqnNum796638},
\begin{equation}
\begin{split}
&\left(rR_{3} \right)_{,r} +R_{31} =\frac{1}{r^{3} } \left(r^{4} L_{3} \right)_{,r}, \quad 3R_{3} -R_{31} =\frac{2}{r^{3} } \left(r^{4} T_{3} \right)_{,r},\\ 
&\textrm{and}\\
&3L_{3} -R_{31} =2\left(rT_{3} \right)_{,r}.
\end{split}
\label{ZEqnNum337667}
\end{equation}

\noindent All kinematic relations developed for a constant divergence flow on small scales and irrotational flow on large scales can be validated by N-body simulations and presented in Section \ref{sec:5}.

\section{Two-point high order velocity statistics}
\label{sec:4}
In this section, we formulate the general kinematic relations for correlation functions of any order (beyond the second and third orders) for different types of flow. This is a challenging task that involves a significant amount of tensor/vector algebra and calculus. The reader can simply skip the details of the derivation in the appendix \ref{appendix:A} and jump directly to the final results, that is, Eqs. \eqref{M_ZEqnNum234089}-\eqref{M_ZEqnNum533624} for incompressible flow, Eq. \eqref{M_ZEqnNum591192} for constant divergence on small scales, and Eqs. \eqref{M_ZEqnNum900611}-\eqref{M_ZEqnNum433592} for irrotational flow on large scales.

\subsection{High order velocity correlation functions}
\label{sec:4.1}
Similarly to the second and third order tensors (Eqs. \eqref{ZEqnNum195930} and \eqref{ZEqnNum883006}), the correlation tensor $Q$ of arbitrary order \textit{p} can be defined as
\begin{equation}
\label{eq:M_92} 
\left({}_{\left(p\right)} Q_{ij} {}_{k..mn} \right)=\left\langle u_{i} u_{j} u_{k} ...u_{m} u_{n}^{'} \right\rangle .         
\end{equation} 
Scalar correlation functions are defined by tensor contraction of $Q$ (similar to Eqs. \eqref{ZEqnNum550536}, \eqref{ZEqnNum811607}, and \eqref{ZEqnNum819453}). For an even number \textit{q}, the total correlation function of order $(p,q+1)$ is defined as
\begin{equation} 
\label{M_ZEqnNum442788} 
R_{\left(p,q+1\right)} =\left\langle u^{q} u_{L}^{p-q-2} u_{i} u_{i}^{'} \right\rangle =\left\langle u^{q} u_{L}^{p-q-2} \boldsymbol{\mathrm{u}}\cdot \boldsymbol{\mathrm{u}}_{}^{'} \right\rangle .       
\end{equation} 
For even number \textit{q}, the longitudinal and transverse correlation functions of order $(p,q)$ are defined as: 
\begin{equation} 
\label{eq:M_96} 
L_{\left(p,q\right)} =\left\langle u^{q} u_{L}^{p-q-1} u_{L}^{'} \right\rangle  
\end{equation} 
and
\begin{equation} 
\label{M_ZEqnNum340773} 
T_{\left(p,q\right)} ={\left(R_{\left(p,q+1\right)} -L_{\left(p,q\right)} \right)/2} .         
\end{equation} 

\begin{figure*}
\includegraphics*[width=\textwidth]{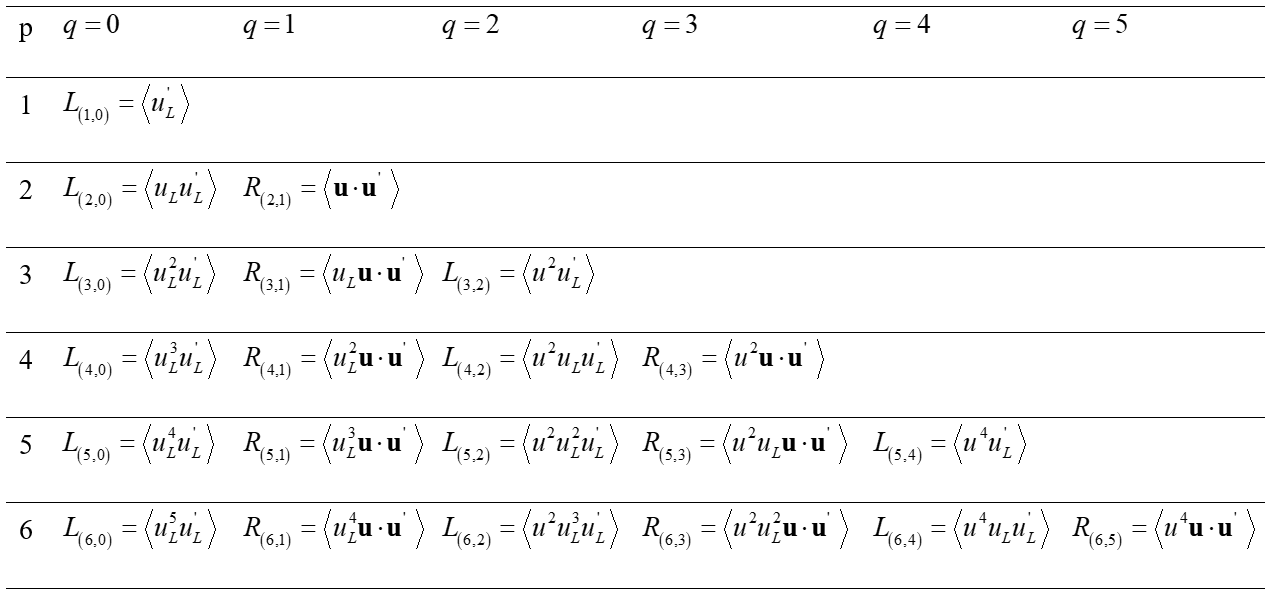}
\caption{Velocity correlation functions of different order}
\label{fig:2}
\end{figure*}

Figure \ref{fig:2} lists the velocity correlation functions up to the sixth order ($p=6$). In particular, for third order correlations, $R_{3}=R_{(3,1)}$ and $R_{31}=L_{(3,2)}$. Like second order correlations, all these correlation functions can be calculated similarly from N-body simulations. We first identify all pairs of particles with the same separation $r$, followed by a pairwise average to calculate these correlations on a given scale $r$. The results are presented in Section \ref{sec:5}.

\subsection{Correlation functions in the limit \texorpdfstring{$r\to 0$ and $r\to \infty $}{}}
\label{sec:4.2}
We first identify the limiting ratio for odd order \textit{p} (see Appendix \ref{appendix:A}),
\begin{equation}
{\mathop{\lim }\limits_{r\to 0}} \frac{\left\langle u^{q} u_{L}^{p-q-1} \right\rangle }{\left\langle u_{L}^{p-1} \right\rangle } =\frac{p}{p-q} \quad \textrm{with q=0...p-1}.     
\label{M_ZEqnNum747023}
\end{equation}
\noindent Equation \eqref{M_ZEqnNum747023} is also valid for $r\to \infty$ where velocity distributions are independent of scale \textit{r}, i.e. 
\begin{equation}
{\mathop{\lim }\limits_{r\to \infty }} \frac{\left\langle u^{q} u_{L}^{p-q-1} \right\rangle }{\left\langle u_{L}^{p-1} \right\rangle } =\frac{p}{p-q} \quad \textrm{with q=0...p-1}.     
\label{M_ZEqnNum300527}
\end{equation}
\noindent Next, using the definition of correlation functions (Eqs. \eqref{M_ZEqnNum442788}-\eqref{M_ZEqnNum340773}), for correlation functions of odd order \textit{p},
\begin{equation} 
\label{eq:M_102}
\begin{split}
{\mathop{\lim }\limits_{r\to 0,\infty }} \frac{L_{\left(p,q\right)} }{L_{\left(p,0\right)} }={\mathop{\lim }\limits_{r\to 0,\infty }} \frac{\left\langle u^{q} u_{L}^{p-q-1} \right\rangle }{\left\langle u_{L}^{p-1} \right\rangle } =\frac{p}{p-q}. \end{split} 
\end{equation} 
Similar relations can be obtained for correlation functions of even order \textit{p} (from Eq. \eqref{M_ZEqnNum300527}),
\begin{equation} 
\label{eq:M_103} 
{\mathop{\lim }\limits_{r\to 0}} \frac{R_{\left(p,q+1\right)} }{L_{\left(p,0\right)} } ={\mathop{\lim }\limits_{r\to 0}} \frac{\left\langle u^{q} u_{L}^{p-q-2} \boldsymbol{\mathrm{u}}\cdot \boldsymbol{\mathrm{u}}_{}^{'} \right\rangle }{\left\langle u_{L}^{p-1} u_{L}^{'} \right\rangle } =\frac{p+1}{p-q-1} ,       
\end{equation} 
and
\begin{equation} 
\label{M_ZEqnNum950434} 
{\mathop{\lim }\limits_{r\to 0,\infty }} \frac{L_{\left(p,q\right)} }{L_{\left(p,0\right)} } ={\mathop{\lim }\limits_{r\to 0,\infty }} \frac{\left\langle u^{q} u_{L}^{p-q-1} u_{L}^{'} \right\rangle }{\left\langle u_{L}^{p-1} u_{L}^{'} \right\rangle } =\frac{p+1}{p+1-q} .       
\end{equation} 

\subsection{Kinematic relations for high order correlations}
\label{sec:4.3}
\subsubsection{Kinematic relations for incompressible flow}
\label{sec:4.3.1}
For incompressible flow, the general kinematic relations for correlations of any order \textit{p} are obtained as (see Appendix \ref{appendix:A4.3}):
\begin{equation}
\left(p-q-1\right)R_{\left(p,q+1\right)} =\frac{1}{r^{p-q} } \left(r^{p-q+1} L_{\left(p,q\right)} \right)_{,r},     \label{M_ZEqnNum234089}
\end{equation}
\begin{equation} 
\label{eq:M_110} 
2\left(p-q-1\right)T_{\left(p,q\right)} =\frac{1}{r} \left(r^{2} L_{\left(p,q\right)} \right)_{,r} ,        
\end{equation} 
\begin{equation} 
\label{M_ZEqnNum533624} 
\left(r^{2} R_{\left(p,q+1\right)} \right)_{,r} =\frac{2}{r^{p-q-1} } \left(r^{p-q+1} T_{\left(p,q\right)} \right)_{,r}.  
\end{equation} 
These kinematic relations are new and exact and should be valid for homogeneous isotropic incompressible turbulence.

\subsubsection{Kinematic relations for constant divergence flow}
\label{sec:4.3.2}
For constant divergence flow on small scales, a general kinematic relation is obtained as (see Appendix \ref{appendix:A4.4}):
\begin{equation} 
\label{M_ZEqnNum365670} 
\begin{split}
\left(p-q-1\right)R_{\left(p,q+1\right)} &+\left\langle u^{q} u_{L}^{p-q-1} \right\rangle \theta r \\
&=\frac{1}{r^{p-q} } \left(r^{p-q+1} L_{\left(p,q\right)} \right)_{,r}.
\end{split}
\end{equation} 
Equation \eqref{M_ZEqnNum365670} reduces to Eq. \eqref{M_ZEqnNum234089} for incompressible flow with $\theta=0$. 

For correlation functions of even order \textit{p} (\textit{q} is always an even number, see Eq. \eqref{M_ZEqnNum365670} and Fig. \ref{fig:2}), we should have
\begin{equation}
\label{M_ZEqnNum136987} 
{\mathop{\lim }\limits_{r\to 0}} \left\langle u^{q} u_{L}^{p-q-1} \right\rangle =0.          
\end{equation} 
Therefore, here we can demonstrate that the kinematic relations for even order correlations of constant divergence flow should always be the same as that of incompressible flow. Equations \eqref{M_ZEqnNum234089}-\eqref{M_ZEqnNum533624} are still valid for correlations of even order \textit{p} in constant divergence flow. 

For odd order \textit{p}, two special cases are considered here. With $q=p-1$ and $q=0$ from Eq. \eqref{M_ZEqnNum365670}, we should have relations
\begin{equation} 
\label{M_ZEqnNum383689} 
\left\langle u^{p-1} \right\rangle \theta r=\frac{1}{r} \left(r^{2} L_{\left(p,p-1\right)} \right)_{,r}  
\end{equation} 
and
\begin{equation} 
\label{M_ZEqnNum213847} 
\left(p-1\right)R_{\left(p,1\right)} +\left\langle u_{L}^{p-1} \right\rangle \theta r=\frac{1}{r^{p} } \left(r^{p+1} L_{\left(p,0\right)} \right)_{,r} .       
\end{equation} 
For $r\to 0$, the correlation function $L_{(p,p-1)} $ can be directly solved from Eq. \eqref{M_ZEqnNum383689} that is proportional to $r$ on small scales (Eq. \eqref{M_ZEqnNum747023}),
\begin{equation}
\label{M_ZEqnNum719510} 
L_{\left(p,p-1\right)} =\frac{p}{3} \theta \left\langle u_{L}^{p-1} \right\rangle r=\frac{1}{3} \theta \left\langle u_{}^{p-1} \right\rangle r.        
\end{equation} 

Especially for $p=1$ and $q=0$ in Eq. \eqref{M_ZEqnNum213847}, the mean pairwise velocity (or first order structure function) $S_{1}^{lp} (r)=\langle \Delta u_{L} \rangle =\langle u_{L}^{'} -u_{L} \rangle =2\langle u_{L}^{'} \rangle $ can be directly related to the divergence $\theta$ as
\begin{equation} 
\label{M_ZEqnNum643394} 
\theta =\frac{1}{2r^{2} } \left(r^{2} \left\langle \Delta u_{L} \right\rangle \right)_{,r} .          
\end{equation} 
With $\langle\Delta u_{L}\rangle =-Har$ on small scales from stable clustering hypothesis \citep{Xu:2021-A-non-radial-two-body-collapse}, the divergence on small scales $\theta={-3Ha/2}$ can be obtained. Here $r$ is the distance between two particles. Equation \eqref{M_ZEqnNum643394} is an important kinematic relation for the constant divergence flow on small scales. However, it is actually valid for an entire range of scales (see Eq. \eqref{ZEqnNum732704}), and will be used repeatedly in Section \ref{sec:6}. On large scales, we will show that $\theta=-Ha\xi(r)$ (Eq. \eqref{ZEqnNum344666}), where $\xi(r)=\langle\delta \delta^{'} \rangle$ is the density correlation function.

With Eq. \eqref{M_ZEqnNum747023}, Eqs. \eqref{M_ZEqnNum365670} and \eqref{M_ZEqnNum213847}, the general kinematic relations for odd order \textit{p} should finally read
\begin{equation} 
\label{M_ZEqnNum591192} 
\begin{split}
\left(p-q-1\right)R_{\left(p,q+1\right)} +\frac{1}{p-q} \frac{1}{r} &\left(r^{2} L_{\left(p,p-1\right)} \right)_{,r}\\
&=\frac{1}{r^{p-q} } \left(r^{p-q+1} L_{\left(p,q\right)} \right)_{,r} .     
\end{split}
\end{equation} 

\subsubsection{Kinematic relations for irrotational flow}
\label{sec:4.3.3}
For irrotational flow on large scales, kinematic relations of any order \textit{p} and even number $0\le q\le p-1$ are obtained (Appendix \ref{appendix:A4.5})
\begin{equation} 
\label{M_ZEqnNum900611} 
\begin{split}
\left(R_{\left(p,q+1\right)} r\right)_{,r} +\left(p-q-2\right)&L_{\left(p,q+2\right)}\\
&=\frac{1}{r^{p-q} } \left(r^{p-q+1} L_{\left(p,q\right)} \right)_{,r},
\end{split}
\end{equation} 
\begin{equation} 
\label{M_ZEqnNum602698} 
\begin{split}
\left(p-q\right)R_{\left(p,q+1\right)} -\left(p-q-2\right)&L_{\left(p,q+2\right)}\\
&=\frac{2}{r^{p-q} } \left(r^{p-q+1} T_{\left(p,q\right)} \right)_{,r},
\end{split}
\end{equation} 
\begin{equation} 
\label{M_ZEqnNum433592} 
\left(p-q\right)L_{\left(p,q\right)} -\left(p-q-2\right)L_{\left(p,q+2\right)} =2\left(rT_{\left(p,q\right)} \right)_{,r} .      
\end{equation} 
In Eqs. \eqref{M_ZEqnNum900611}-\eqref{M_ZEqnNum433592}, terms involving the correlation function $L_{(p,q+2)}$ should disappear if $q\ge p-2$.

\section{Results from N-body simulations }
\label{sec:5}
This section presents the two-point velocity statistics from N-body simulations. All statistical measures are computed in the same way as second-order measures \citep{Xu:2023-On-the-statistical-theory-of-self-gravitating}. In an N-body system, all particle pairs with a given separation \textit{r} are identified first with particle velocities and locations recorded. Any statistical quantity on the scale \textit{ r} is calculated as the average of that quantity for all particle pairs with the same separation. The kinematic relations developed in Section \ref{sec:4} can be systematically verified by N-body simulations. 

\begin{figure}
\includegraphics*[width=\columnwidth]{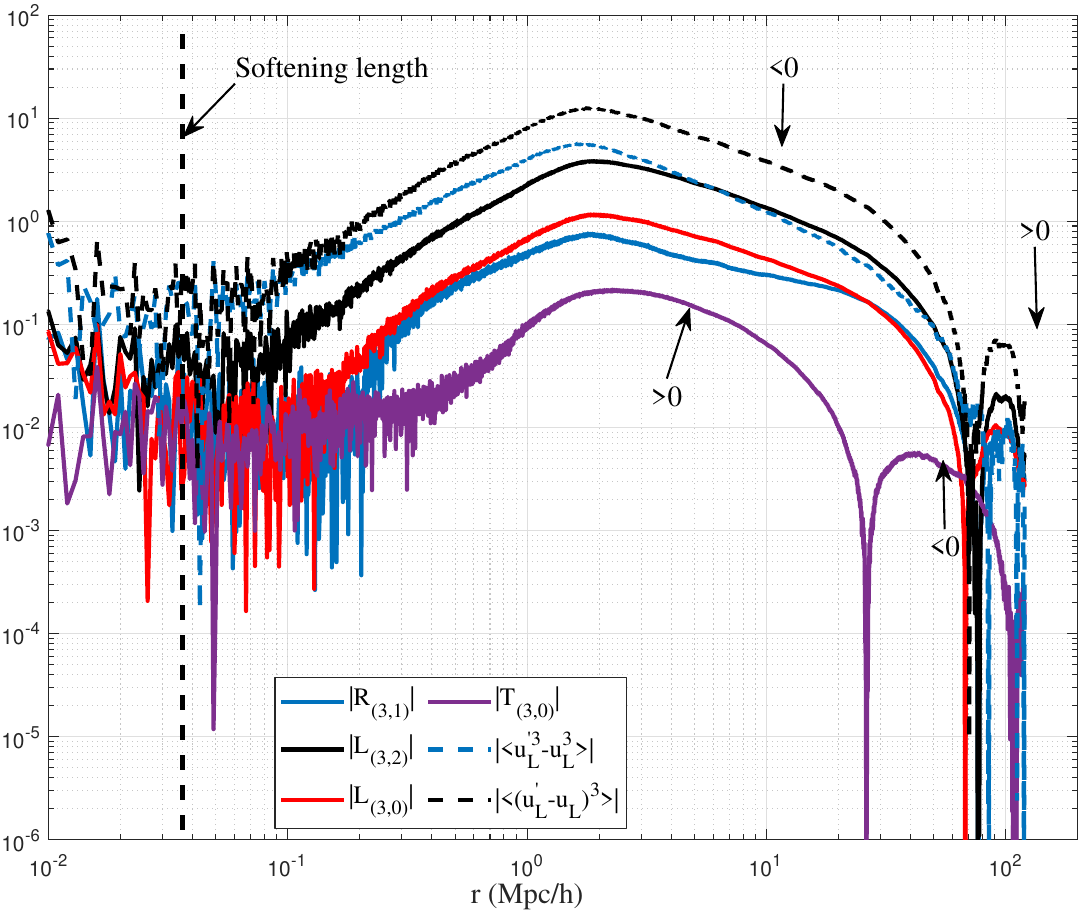}
\caption{Two-point third order velocity correlation functions $L_{(3,0)} $, $R_{(3,1)} $, $L_{(3,2)} $, $T_{(3,0)} $, the third moment of longitudinal velocity $-\langle u_{L}^{'3} \rangle =\langle u_{L}^{3} \rangle$, and the third order structure function $\langle (u_{L}^{'} -u_{L} )^{3} \rangle $ at $z=0$. All statistical measures are normalized by $u_0^{3}$, where $u_0^2$ is the one-dimensional velocity dispersion of entire system. Only the transverse correlation $T_{(3,0)} >0$ is greater than zero on small scales. The vertical dashed line indicates the softening length of 36kpc/h. Odd order correlations have large fluctuations when approaching the softening length because odd correlations become vanishing small on small scales. While even order correlations approach constant values with smaller fluctuations on small scales (Fig. \ref{fig:4}).} 
\label{fig:3}
\end{figure}

Figure \ref{fig:3} plots the variation of all third order correlation/structure functions and the moment of longitudinal velocity with scale \textit{r} at $z=0$. All correlation functions are normalized by $u_0^{3}$, where $u_0$ is the one-dimensional velocity dispersion of the entire system. All third order statistical measures vanish on both small and large scales and are negative on small scales. The only exception is the transverse correlation $T_{(3,0)} >0$. 

\begin{figure}
\includegraphics*[width=\columnwidth]{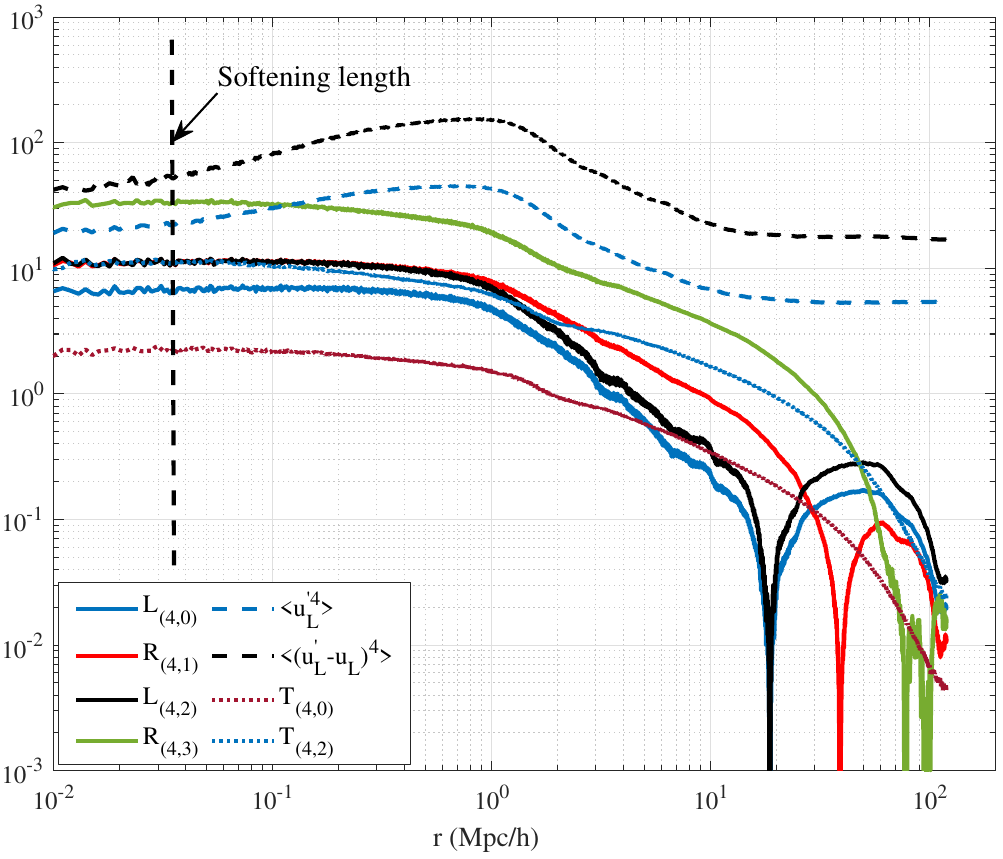}
\caption{Two-point fourth order velocity correlation functions $L_{(4,0)} $, $R_{(4,1)} $, $L_{(4,2)}$, $R_{(4,3)} $,  $T_{(4,0)}$, $T_{(4,2)}$, the fourth  moment of longitudinal velocity $\langle u_{L}^{'4} \rangle $, and fourth order structure function $\langle (u_{L}^{'} -u_{L} )^{4} \rangle $ at $z=0$. All statistical measures are normalized by $u_0^{4}$, where $u_0^2$ is the one-dimensional velocity dispersion of entire system. The transverse correlation function $T_{(4,2)}>T_{(4,0)} >0$ on small scales. All even order correlation functions approach constant values on small scales. The vertical dashed line indicates the softening length of 36kpc/h.} 
\label{fig:4}
\end{figure}

Similarly, Figure \ref{fig:4} presents the fourth order correlation/structure functions and moment at $z=0$ (normalized by $u_0^{4}$). All fourth statistical measures are positive on all scales and approach constant values in the limit of $r\to 0$ (small scales). 

The general solution of the correlation function $L_{(p,p-1)}$ of odd order \textit{p} on small scales (from Eq. \eqref{M_ZEqnNum719510}) is
\begin{equation} 
\label{ZEqnNum676855} 
\frac{L_{\left(p,p-1\right)} }{u^{p} } =\frac{p}{3} \theta r\frac{\left\langle u_{L}^{p-1} \right\rangle }{u^{p} } =-2^{{\left(p-3\right)/2} } pK_{p-1} \left(u_{L} ,0\right)\frac{Har}{u} ,     
\end{equation} 
where divergence $\theta =-{3Ha/2} $ and ${\mathop{\lim }\limits_{r\to 0}} {\langle u_{L}^{2} \rangle/ u^{2} } =2$ (Fig. \ref{fig:20}). Generalized kurtosis $K_{p-1} (u_{L} ,r)$ of longitudinal velocity $u_{L}$ on small scales $r$ is defined as
\begin{equation}
\label{eq:37} 
K_{n} \left(u_{L} ,r\to 0 \right)=\frac{\langle \left( u_{L}\right)^{n} \rangle }{\left \langle  u_{L}^2 \right\rangle ^{{n/2}}}. 
\end{equation} 
The value of $K_n$ can be found for the maximum entropy distribution of velocity \citep{Xu:2023-Maximum-entropy-distributions-of-dark-matter} and is presented in a separate article \citep{Xu:2022-Two-thirds-law-for-pairwise-ve}, where the first three values are $K_{0}=1$, $K_{2}=1$, and $K_{4}=4.8$. 

\begin{figure}
\includegraphics*[width=\columnwidth]{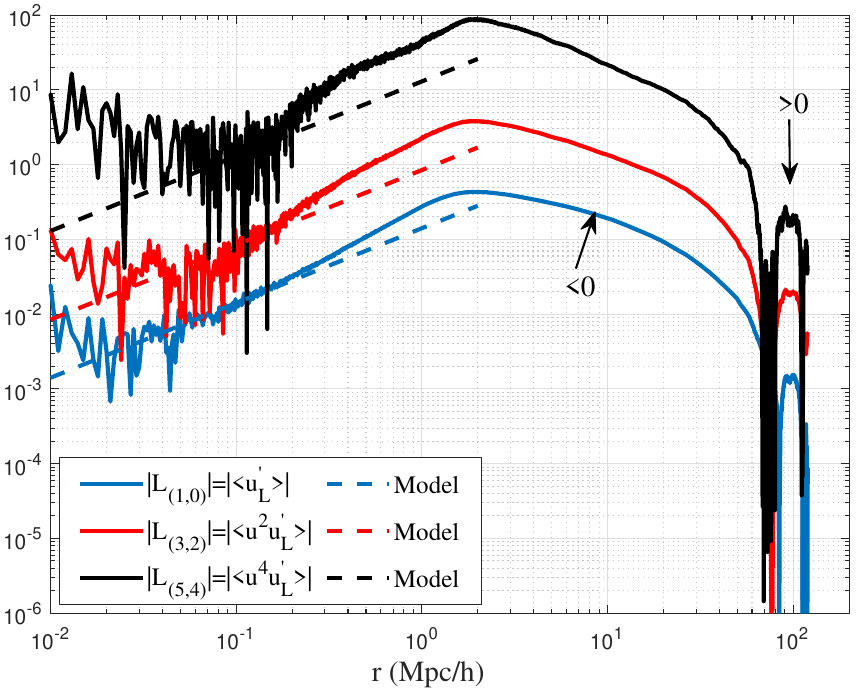}
\caption{The variation of correlation functions $L_{(1,0)}$ (mean longitudinal velocity), $L_{(3,2)}$ and $L_{(5,4)}$ at \textit{z}=0 (normalized by $u$, $u^{3}$, and $u^{5}$ in Table \ref{tab:2}). Dashed line shows the model of Eq. \eqref{ZEqnNum676855} on small scales for comparison.} 
\label{fig:5}
\end{figure}

The solution in Eq. \eqref{ZEqnNum676855} can be compared with the results of N-body simulations. Figure \ref{fig:5} presents the correlation functions $L_{(1,0)} $, $L_{(3,2)} $ and $L_{(5,4)} $ at $z=0$ (normalized by $u_0$, $u_0^{3}$ and $u_0^{5}$, respectively). The dashed lines show the solution of Eq. \eqref{ZEqnNum676855} that agrees with N-body simulations on small scales. 

For correlation functions of odd order \textit{p} and even numbers \textit{q}, kinematic relations for a constant divergence flow on small scales in Eq. \eqref{M_ZEqnNum591192} can be transformed to (integrating both sides)
\begin{equation} 
\label{ZEqnNum167557} 
\begin{split}
&H_{\left(p,q\right)}^{S}\left(r\right)=\frac{1}{\left(p-q\right)} \cdot \frac{L_{\left(p,p-1\right)} }{L_{\left(p,q\right)} }\\
&+\frac{\left(p-q-1\right)}{r^{p-q+1} L_{\left(p,q\right)} } \int _{0}^{r}\left(R_{\left(p,q+1\right)} -\frac{L_{\left(p,p-1\right)} }{p-q} \right) r^{p-q} dr=1.
\end{split}
\end{equation} 

For correlation functions of even order \textit{p} for constant divergence flow on small scales, the kinematic relations are the same as those of an incompressible flow (Eq. \eqref{M_ZEqnNum365670}) and can be transformed to
\begin{equation}
\label{ZEqnNum729182} 
H_{\left(p,q\right)}^{S} \left(r\right)=\frac{\left(p-q-1\right)}{r^{p-q+1} L_{\left(p,q\right)} } \int _{0}^{r}R_{\left(p,q+1\right)}  r^{p-q} dr=1.       
\end{equation} 

Similarly, kinematic relations of any order \textit{p} and even \textit{q} for irrotational flow on large scales (Eq. \eqref{M_ZEqnNum602698}) can be transformed to
\begin{equation} 
\label{ZEqnNum467352} 
\begin{split}
H_{\left(p,q\right)}^{L} \left(r\right)&=\frac{1}{2r^{p-q+1} T_{\left(p,q\right)} } \int _{0}^{r}\left[\left(p-q\right)R_{\left(p,q+1\right)}\right.\\ &\left.-\left(p-q-2\right)L_{\left(p,q+2\right)} \right]r^{p-q} dr =1.
\end{split}
\end{equation} 
The functions $H_{(p,q)}^{S}$ and $H_{(p,q)}^{L}$ are introduced to verify the general kinematic relations and can be directly computed from N-body simulations based on the velocity correlation functions. They are expected to be 1 on both small and large scales. 

\begin{figure}
\includegraphics*[width=\columnwidth]{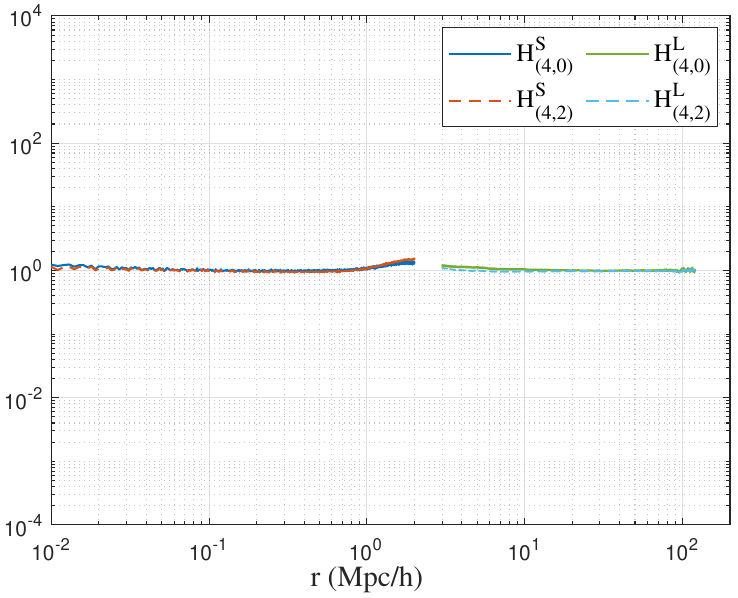}
\caption{The variation of functions $H_{(4,0)}^{S} $ and $H_{(4,2)}^{S} $ (Eq. \eqref{ZEqnNum729182} ) on small scales and $H_{(4,0)}^{L} $ and $H_{(4,2)}^{L} $ (Eq. \eqref{ZEqnNum467352}) on large scales from N-body simulation at $z=0$. Both functions are expected to be 1 from the kinematic relations for fourth order correlation functions. Results from N-body simulations confirms this and validates the kinematic relations obtained.}
\label{fig:6}
\end{figure}

\begin{figure}
\includegraphics*[width=\columnwidth]{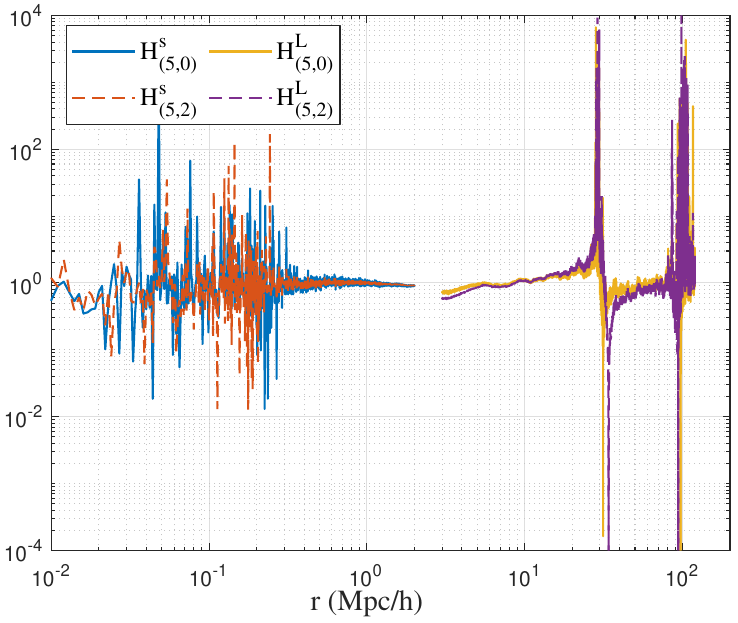}
\caption{The variation of functions $H_{(5,0)}^{S} $ and $H_{(5,2)}^{S} $ (Eq. \eqref{ZEqnNum167557}) on small scales and $H_{(5,0)}^{L} $ and $H_{(5,2)}^{L} $ (Eq. \eqref{ZEqnNum467352}) on large scales from N-body simulation at \textit{z}=0. Both functions are expected to be 1, as predicted by the kinematic relations for fifth order correlations. N-body simulation confirms this result and validates the kinematic relations obtained. The large deviations of functions $H_{(5,0)}^{L} $ and $H_{(5,2)}^{L}$ at r=30 and 100 Mpc/h are due to the transverse correlation function $T_{(p,q)}$ approaching zero at these scales (Eq. \eqref{ZEqnNum467352}).}
\label{fig:7}
\end{figure}

Figures \ref{fig:6} and \ref{fig:7} plot the variation of functions $H^{S}_{(p,q)}$ on small scales and $H_{(p,q)}^{L}$ on large scales computed based on the fourth and fifth order correlation functions from N-body simulations. Both functions are expected to be 1 from kinematic relations for fourth and fifth order correlations, as shown in both figures. These results validate our derivation for high order kinematic relations.

\section{Dynamic relations on large scales}
\label{sec:6}
So far, we have considered only kinematic relations, that is, the relations between correlation functions of the same order (the same order \textit{p} in Fig. \ref{fig:2}). However, the dynamic relations between correlation functions of different orders (different \textit{p} in Fig. \ref{fig:2}) contain key information on the dynamics of dark matter flow and the connections between velocity and density fields. In this section, we will focus on the dynamic relations that can only be determined from the dynamic evolution of the velocity field. 

The basic dynamics of self-gravitating collisionless flow (SG-CFD) follows from the collisionless Boltzmann equations (CBE), where the Jeans equations of different order can be systematically constructed \citep{Mo:2010-Galaxy-formation-and-evolution}. However, the closure problem is well known for Jeans' equations, which are not self-closed. The self-closed system of dynamic equations must be introduced on small and large scales, respectively. These dynamic equations are subsequently converted into the dynamic relations between correlation functions of different orders. This section focuses on the dynamics on large scales. Section \ref{sec:7} focuses on the dynamics on small scales.

\subsection{Dynamic relations and effective viscosity}
\label{sec:6.1}
On large scales ($r>r_t$ in Fig. \ref{fig:19}), the peculiar velocity is irrotational that can be written as the gradient of velocity potential ($\boldsymbol{\mathrm{v}}=\nabla \Phi_v$). The dynamic equation for the peculiar velocity $\boldsymbol{\mathrm{v}}$ is usually modeled via an empirical "adhesion approximation" \citep{Gurbatov:1989-The-Large-Scale-Structure-of-t,Buchert:2005-Adhesive-gravitational-cluster},  
\begin{equation} 
\label{ZEqnNum709847} 
\frac{\partial \boldsymbol{\mathrm{v}}}{\partial t} +\frac{1}{a} \boldsymbol{\mathrm{v}}\cdot \nabla \boldsymbol{\mathrm{v}}=c\left(a\right)\boldsymbol{\mathrm{v}}+\nu \left(a\right)\nabla ^{2} \boldsymbol{\mathrm{v}},         
\end{equation} 
where $c(a)$ is a damping coefficient on large scales and $\nu(a)$ is an "artificial" viscosity in adhesion model. This section will attempt to elucidate the dynamic origin of the "adhesion approximation" and the origin of the "artificial" viscosity $\nu(a)$. By neglecting the second order inertial and viscosity terms, we have the Zeldovich approximation on large scales from Eq. \eqref{ZEqnNum709847},
\begin{equation}
\label{ZEqnNum911265} 
\frac{\partial \boldsymbol{\mathrm{v}}}{\partial t} =c\left(a\right)\boldsymbol{\mathrm{v}}.            
\end{equation} 
For matter-dominant cosmology with $H^{2} ={8\pi G\rho _{0}/3}$,
\begin{equation} 
\label{eq:138} 
c\left(a\right)=\left(\frac{4\pi G\rho _{0} }{Hf\left(\Omega _{m} \right)} -H\right)=\frac{1}{2} H,        
\end{equation} 
where $\rho _{0} \left(a\right)$ is the mean matter density. The function $f(\Omega_m)\approx\Omega_m^{0.6}$ depends on the content of matter $\Omega_m$. The adhesion approximation (a phenomenological model) extends the range of the Zeldovich approximation into the weakly nonlinear regime. 

Artificial ("effective") viscosity has its origin in the inverse energy cascade (Eq. \eqref{ZEqnNum133069}). On small scales ($r<r_t$ in Fig. \ref{fig:19}), there exists a continuous halo-mediated inverse cascade of kinetic energy from small to large scales until the largest halo scale $r_t$ \citep{Xu:2023-Dark-matter-halo-mass-functions-and}. The effective viscosity $\nu(a)$ on the scale $r_t$ is required to continuously dissipate the energy cascaded from all scales below $r_t$. At this time, to develop dynamic relations, we adopt the adhesion approximation with a time-varying viscosity $\nu \left(a\right)$ as the dominant dynamics on large scales. Two identities can be introduced for an arbitrary velocity field $\boldsymbol{\mathrm{u}}$,
\begin{equation} 
\label{ZEqnNum683197} 
\boldsymbol{\mathrm{u}}\cdot \nabla \boldsymbol{\mathrm{u}}=\frac{1}{2} \nabla \left(\boldsymbol{\mathrm{u}}\cdot \boldsymbol{\mathrm{u}}\right)+\left(\nabla \times \boldsymbol{\mathrm{u}}\right)\times \boldsymbol{\mathrm{u}},  \end{equation} 
and
\begin{equation} 
\label{ZEqnNum248930} 
\nabla ^{2} \boldsymbol{\mathrm{u}}=\nabla \left(\nabla \cdot \boldsymbol{\mathrm{u}}\right)-\nabla \times \left(\nabla \times \boldsymbol{\mathrm{u}}\right).         
\end{equation} 
For irrotational flow ($\nabla \times \boldsymbol{\mathrm{v}}=0$) on large scales, this leads to
\begin{equation}
\boldsymbol{\mathrm{v}}\cdot \nabla \boldsymbol{\mathrm{v}}=\frac{1}{2} \nabla \left(\boldsymbol{\mathrm{v}}\cdot \boldsymbol{\mathrm{v}}\right) \quad \textrm{and} \quad \nabla ^{2} \boldsymbol{\mathrm{v}}=\nabla \left(\nabla \cdot \boldsymbol{\mathrm{v}}\right),     
\label{ZEqnNum307676}
\end{equation}

\noindent such that the dynamic Eq. \eqref{ZEqnNum709847} can be rewritten as
\begin{equation}
\label{ZEqnNum284543} 
\frac{\partial \boldsymbol{\mathrm{v}}}{\partial t} +\frac{1}{2a} \nabla \left(\boldsymbol{\mathrm{v}}\cdot \boldsymbol{\mathrm{v}}\right)=c\left(a\right)\boldsymbol{\mathrm{v}}+\nu \left(a\right)\nabla ^{2} \boldsymbol{\mathrm{v}}.       
\end{equation} 

Index notation of Eq. \eqref{ZEqnNum284543} at two different locations $\boldsymbol{\mathrm{x}}$ and $\boldsymbol{\mathrm{x}}^{'} $ with a separation of \textit{r} reads
\begin{equation}
\label{ZEqnNum373540} 
\frac{\partial v_{j} }{\partial t} +\frac{1}{2a} \frac{\partial \left(v_{i} v_{i} \right)}{\partial x_{j} } =cv_{j} +\nu \nabla ^{2} v_{j} ,         
\end{equation} 
and
\begin{equation} 
\label{ZEqnNum198491} 
\frac{\partial v_{i}^{'} }{\partial t} +\frac{1}{2a} \frac{\partial \left(v_{j}^{'} v_{j}^{'} \right)}{\partial x_{i}^{'} } =cv_{i}^{'} +\nu \nabla ^{'} {}^{2} v_{i}^{'}.       
\end{equation} 
Multiplying Eqs. \eqref{ZEqnNum373540} and \eqref{ZEqnNum198491} by $v_{i}^{'} $ and $v_{j} $, respectively, adding two equations together and taking the average over all pairs ($\boldsymbol{\mathrm{x}}$ and $\boldsymbol{\mathrm{x}}^{'}$) with the same separation \textit{r} lead to  
\begin{equation} 
\label{ZEqnNum527474} 
\begin{split}
\frac{\partial \left\langle v_{j} v_{i}^{'} \right\rangle }{\partial t} &+\frac{1}{2a} \left\langle v_{i}^{'} \frac{\partial \left(v_{k} v_{k} \right)}{\partial x_{j} } +v_{j}^{} \frac{\partial \left(v_{k}^{'} v_{k}^{'} \right)}{\partial x_{i}^{'} } \right\rangle\\
&=c\left\langle v_{j} v_{i}^{'} +v_{i}^{'} v_{j} \right\rangle +\nu \nabla ^{2} \left\langle v_{j} v_{i}^{'} +v_{i}^{'} v_{j} \right\rangle.
\end{split}
\end{equation} 
The following facts were used for this equation,
\begin{enumerate}
\item \noindent Velocity $v_{i}^{'} $ is independent of $\boldsymbol{\mathrm{x}}$ and $v_{j} $ is independent of $\boldsymbol{\mathrm{x}}^{'} $.
\item \noindent Partial derivatives ${\partial/\partial x_{j}}$ and ${\partial/\partial x_{i}^{'}}$ can be replace by ${-\partial/\partial r_{j}}$ and ${\partial/\partial r_{i} } $, respectively, 
where ${\partial/\partial } x_{i}^{'} ={\partial/\partial } r\cdot \hat{r}_{i} $ and ${\partial /\partial } x_{j}^{} ={-\partial/\partial } r\cdot \hat{r}_{j} $, with unit vectors $\hat{r}_{j} ={r_{j}/r} $ and $\hat{r}_{i} ={r_{i}/r} $. 
\item \noindent $\langle v_{k} v_{k} v_{j}^{'} \rangle (\boldsymbol{\mathrm{r}})=\langle v_{k}^{'} v_{k}^{'} v_{j}^{} \rangle (-\boldsymbol{\mathrm{r}})=-\langle v_{k}^{'} v_{k}^{'} v_{j}^{} \rangle (\boldsymbol{\mathrm{r}})$ from the symmetry of the third order tensor (see Eq. \eqref{ZEqnNum264952}).
\end{enumerate}

From Eq. \eqref{ZEqnNum527474}, the time evolution of second order velocity correlation tensor $Q_{ij}$ can be related to the third order tensor as
\begin{equation} 
\label{ZEqnNum674282} 
\frac{\partial Q_{ij} }{\partial t} =\frac{1}{2a} \left(\frac{\partial Q_{kki} }{\partial r_{j} } +\frac{\partial Q_{kkj} }{\partial r_{i}^{} } \right)+2cQ_{ij} +2\nu \nabla ^{2} Q_{ij}  .       
\end{equation} 
The dynamics of second order correlation tensor $Q_{ij} $ is dependent on the third order correlation tensor $Q_{kki} $, which is dependent on the fourth order correlation tensor. Here we hit the closure problem. 

Since the second order correlation functions on large scales can be modeled explicitly (Eqs. \eqref{ZEqnNum971850} and \eqref{ZEqnNum344034}) \citep{Xu:2023-On-the-statistical-theory-of-self-gravitating}, third and higher order correlations can be obtained through the dynamic relations (e.g. Eq. \eqref{ZEqnNum674282}) on large scales. Multiplying both sides of Eq. \eqref{ZEqnNum674282} by $\delta _{ij} $ leads to the evolution of second order scalar correlation $R_2$, 
\begin{equation} 
\label{ZEqnNum923677} 
\frac{\partial R_{2}^{} }{\partial t} =2\Gamma \left(r\right)+2cR_{2}^{} +2\nu \left(\frac{1}{r^{2} } \frac{\partial }{\partial r} \left(r^{2} \frac{\partial R_{2}^{} }{\partial r} \right)\right),       
\end{equation} 
where the third order function $\Gamma (r)$ represents the energy transfer across scales due to the nonlinear advection term in Eq. \eqref{ZEqnNum709847}. %The energy transfer function $\Gamma (r)$ mimics the mass transfer function $T_m(m_h,a)$ for the inverse mass cascade in the halo mass space \citep[see][Eq. (18)]{Xu:2021-Inverse-mass-cascade-mass-function}. 
The energy transfer function $\Gamma (r)$ describes the removal of kinetic energy from a small scale ($\Gamma (r)<0$) and the deposition of kinetic energy at a larger scale ($\Gamma (r)>0$). From Eq. \eqref{ZEqnNum674282}, we should have 
\begin{equation} 
\label{eq:148} 
\Gamma \left(r\right)=\frac{1}{2a} \frac{\partial Q_{kki} }{\partial r_{i} } .          
\end{equation} 
From the definition of third order tensor (Eqs. \eqref{ZEqnNum883006} and \eqref{ZEqnNum953203}),
\begin{equation} 
\label{eq:149} 
Q_{kki} =Q_{jki} \delta _{jk} =\left(A_{3} r^{2} +2B_{3} +3D_{3} \right) \hat{r}_{i} =R_{31} \hat{r}_{i} ,         
\end{equation} 
where the correlation $R_{31} =L_{\left(3,2\right)} $ in Fig. \ref{fig:2}. From Eq. \eqref{ZEqnNum952051}, 
\begin{equation}
\frac{\partial Q_{kki} }{\partial r_{i} } =\frac{1}{r^{2} } \left(r^{2} R_{31} \right)_{,r} \quad \textrm{and} \quad \Gamma \left(r\right)=\frac{1}{2ar^{2} } \left(r^{2} R_{31} \right)_{,r}. 
\label{ZEqnNum671139}
\end{equation}

Substitution of $\Gamma \left(r\right)$ into Eq. \eqref{ZEqnNum923677}, an important dynamic relation between second order correlation $R_{2}^{} \left(r\right)$ and third order correlation $R_{31}(r)$ is obtained. To derive an explicit expression for $R_{31} \left(r\right)$, we need the model of $R_{2}(r)$ (see Eq. \eqref{ZEqnNum344034}),
\begin{equation}
\label{ZEqnNum564068} 
R_{2} \left(r\right)=\left\langle \boldsymbol{\mathrm{u}}\cdot \boldsymbol{\mathrm{u}}^{'} \right\rangle =2R\left(r\right)=a_{0} u^{2} \exp \left(-\frac{r}{r_{2} } \right)\left(3-\frac{r}{r_{2} } \right).      
\end{equation} 
The comoving length scale $r_{2}$=23.13Mpc/h is a constant that can be related to the size of the horizon at the matter-radiation equality \citep{Xu:2023-On-the-statistical-theory-of-self-gravitating}. For a matter-dominant model with $a_{0} u^{2} \propto a$ (Eq. \eqref{ZEqnNum971850}), we should have the time variation of $R_2(r)$,
\begin{equation}
\label{ZEqnNum181529} 
\frac{\partial R_{2}^{} }{\partial t} =\frac{\partial R_{2}^{} }{\partial a} Ha=2cR_{2}^{} =HR_{2} .         
\end{equation} 
Substitution of Eq. \eqref{ZEqnNum181529} back to Eq. \eqref{ZEqnNum923677} leads to the relation between $R_{2}^{} \left(r\right)$ and $R_{31} \left(r\right)$,
\begin{equation} 
\label{eq:153} 
\frac{1}{r^{2} } \left(r^{2} R_{31} \right)_{,r} +2a\nu \left(\frac{1}{r^{2} } \frac{\partial }{\partial r} \left(r^{2} \frac{\partial R_{2}^{} }{\partial r} \right)\right)=0,        
\end{equation} 
such that
\begin{equation}
\label{ZEqnNum715845} 
L_{\left(3,2\right)} \left(r\right)=R_{31} \left(r\right)=-2a\nu \frac{\partial R_{2}^{} }{\partial r} .        
\end{equation} 

The density correlation of dark matter $\xi \left(r\right)$ can be related to $R_{2}^{} \left(r\right)$ as \citep[see][Eq. (120)]{Xu:2023-On-the-statistical-theory-of-self-gravitating},
\begin{equation}
\label{ZEqnNum179171} 
\begin{split}
\xi \left(r\right)&=-\frac{1}{\left(aHf\left(\Omega _{m} \right)\right)^{2} } \left[\frac{1}{r^{2} } \frac{\partial }{\partial r} \left(r^{2} \frac{\partial R_{2} }{\partial r} \right)\right]\\
&=\frac{{a_{0}u^{2} }/{(rr_{2}) }}{\left(aHf\left(\Omega _{m} \right)\right)^{2} } \cdot  \exp \left(-\frac{r}{r_{2} } \right)\left[\left(\frac{r}{r_{2} } \right)^{2} -7\left(\frac{r}{r_{2} } \right)+8\right]. \end{split}
\end{equation} 
The energy transfer function $\Gamma \left(r\right)$ can be related to density correlation $\xi(r)$ and reads (using Eq. \eqref{ZEqnNum179171} and \eqref{ZEqnNum671139})
\begin{equation} 
\label{eq:156} 
\begin{split}
\Gamma \left(r\right)&=\nu \left(aHf\left(\Omega _{m} \right)\right)^{2} \xi \left(r\right)\\
&=\frac{\nu a_{0} u^{2} }{rr_{2} } \exp \left(-\frac{r}{r_{2} } \right)\left[\left(\frac{r}{r_{2} } \right)^{2} -7\left(\frac{r}{r_{2} } \right)+8\right]. 
\end{split}
\end{equation} 
where the function $f(\Omega_m)\approx\Omega_m^{0.6}$ depends on the content of matter $\Omega_m$. Here, $\nu<0$ reflects the inverse energy transfer from small to large scales in Eq. \eqref{ZEqnNum133069}. The third order velocity correlation is found to be proportional to the density correlation, i.e. $\Gamma(r)\propto \xi(r)$. The third order total correlation $R_{31} \left(r\right)$ can be related to the density correlation $\xi \left(r\right)$
\begin{equation}
\label{ZEqnNum846484} 
\frac{1}{r^{2} } \left(r^{2} R_{31} \right)_{,r} =2a\nu \xi \left(r\right)\left(aHf\left(\Omega _{m} \right)\right)^{2} .        
\end{equation} 

On large scales, from pair conservation Equation \citep[see][Eq. (47)]{Xu:2022-Two-thirds-law-for-pairwise-ve}, mean pairwise velocity is related to the density correlation $\xi(r)$
\begin{equation}
\label{ZEqnNum990753} 
\left\langle \Delta u_{L} \right\rangle=\left\langle u_{L}^{'}-u_L \right\rangle \approx -\frac{2Ha}{r^{2} } \int _{0}^{r}\xi \left(y\right)y^{2}dy. 
\end{equation} 
Therefore, the mean pairwise velocity $\langle \Delta u_{L} \rangle $ can be modeled as (using Eqs. \eqref{ZEqnNum564068} and \eqref{ZEqnNum179171}),
\begin{equation}
\label{ZEqnNum283672} 
\begin{split}
\left\langle \Delta u_{L} \right\rangle &=2L_{(1,0)}=\frac{2}{aHf\left(\Omega _{m} \right)^{2} } \frac{\partial R_{2} }{\partial r} \\
&=\frac{2a_{0} u^{2} }{aHr_{2} f\left(\Omega _{m} \right)^{2} } \exp \left(-\frac{r}{r_{2} } \right)\left(\frac{r}{r_{2} } -4\right). 
\end{split}
\end{equation} 
Using Eqs. \eqref{ZEqnNum846484} and \eqref{ZEqnNum990753}, we found that $R_{31}(r)\propto \langle\Delta u_{L}\rangle $ as
\begin{equation} 
\label{ZEqnNum216206} 
\begin{split}
L_{(3,2)} = R_{31} =\left\langle u^{2} u_{L}^{'} \right\rangle &=-\nu Ha^{2} f\left(\Omega _{m} \right)^{2} \left\langle \Delta u_{L} \right\rangle\\
&=-\frac{2a_{0} u^{2} a\nu }{r_{2} } \exp \left(-\frac{r}{r_{2} } \right)\left(\frac{r}{r_{2} } -4\right),
\end{split}
\end{equation} 
which provides a simple model for third order correlation $R_{31}$. 

\begin{figure}
\includegraphics*[width=\columnwidth]{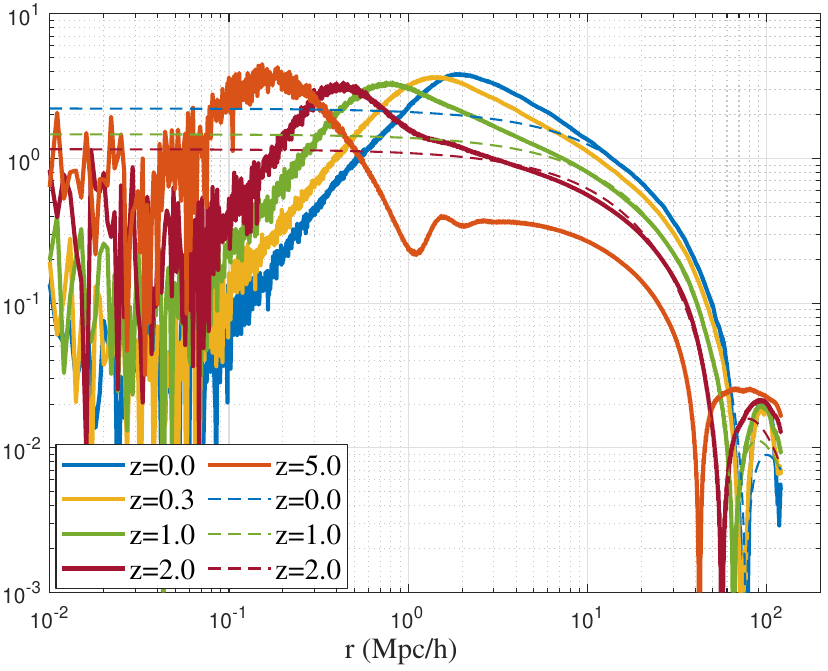}
\caption{The variation of two-point third order velocity correlation function $R_{31} =L_{\left(3,2\right)} $ with scale \textit{r} at different redshifts \textit{z} (normalized by $u^{3}$). The model in Eq. \eqref{ZEqnNum136752} is also plotted in the same plot as the dashed lines.}
\label{fig:8}
\end{figure}

Figure \ref{fig:8} plots the variation of $|R_{31}|$ with scale \textit{r} at several different redshifts \textit{z}. Without loss of generality, the correlation function $R_{31}=L_{(3,2)} $ can be conveniently modeled as (based on Eq. \eqref{ZEqnNum216206})
\begin{equation} 
\label{ZEqnNum136752} 
L_{\left(3,2\right)} =R_{31} =\left\langle u^{2} u_{L}^{'} \right\rangle =a_{3} u^{3} \exp \left(-\frac{r}{r_{2} } \right)\left(\frac{r}{r_{2} } -b_{3} \right),      
\end{equation} 
where coefficients $a_{3} $ and $b_{3} $ are obtained by fitting to the curves in Fig. \ref{fig:8}. The viscosity $\nu(a)$ can be related to $a_{3}$ and $a_{0}$ (Eq. \eqref{ZEqnNum216206}),
\begin{equation}
a_{3} =-\frac{2a_{0} a\nu }{ur_{2} } \quad \textrm{or} \quad \alpha _{\nu } =-\frac{\nu }{u_{0} r_{2} } =\frac{a_{3} u}{2a_{0} au_{0} },     
\label{ZEqnNum529435}
\end{equation}
i.e. the artificial viscosity in the adhesion model is determined by the typical velocity $u$ and the length scale $r_{2}$. Here, $\nu(a)\propto ur_2$ mimics the viscosity of the ideal gas. 

For comparison, in the standard K-epsilon turbulence model, the eddy viscosity $\nu$ depends on the (specific) kinetic energy $k$ ($m^2/s^2$) and the rate of dissipation $\varepsilon$ ($m^2/s^3$), that is, $\nu=\beta_vk^2/\varepsilon$, with $\beta_{\nu} \approx 0.1$ in the K-epsilon model for incompressible flow. Equivalently, we can express the rate of energy dissipation as $\varepsilon \propto {k}/(\nu/k)$, that is, the kinetic energy $k$ is dissipated in a typical time of $\nu/k$.     

In SG-CFD for dark matter flow, the effective viscosity $\nu \left(a\right)$ is required to dissipate the kinetic energy cascade from all scales below $r_t$ \citep{Xu:2023-Dark-matter-halo-mass-functions-and}. This will allow us to write: 
\begin{equation} 
\label{ZEqnNum133069} 
\begin{split}
&\nu \left(a\right)=\frac{-1}{Ha^{2} f\left(\Omega _{m} \right)^{2} } \frac{R_{31} }{\left\langle \Delta u_{L} \right\rangle } \quad \textrm{(from Eq. \eqref{ZEqnNum216206})}\\
&\nu \left(a\right)=\frac{a_{3} r_{2} }{3a_{0} uat} \frac{\left({3u^{2}/2} \right)^{2} }{\varepsilon _{u} } \approx \beta_{\nu}a^{1/2} \frac{\left({3u_0^{2}/2}\right)^{2} }{\varepsilon _{u}} \propto a^{1/2}.
\end{split}
\end{equation} 
Here, ${3u^{2}/2}\equiv k$ is the specific kinetic energy of the entire system. The rate of the energy cascade $\varepsilon_u$ below the scale $r_t$ is \citep{Xu:2023-Universal-scaling-laws-and-density-slope}
\begin{equation} 
\label{ZEqnNum1330699} 
\varepsilon_{u} =-\frac{3}{2}\frac{du^{2}}{dt} \approx -\frac{3}{2}\frac{u_0^{2}}{t_0}=-4.6\times 10^{-7}\frac{m^2}{s^3},
\end{equation}
where $u_0\approx 354$km/s is the typical velocity at $z=0$ and $t_0$ is the age of universe. Negative $\varepsilon_u$ reflects the inverse cascade from small to large scales and leads to a negative effective viscosity $\nu$. The constant $\beta_{\nu} \approx 4.9$ can be obtained from Fig. \ref{fig:12}. The effective viscosity is on the order of $\nu(z=0)\approx -10^4$Mpc km/s (also see \citep{Xu:2022-The-origin-of-MOND-acceleratio} for estimate of viscosity). Its values in terms of the coefficient $\alpha _{\nu}$ in Eq. \eqref{ZEqnNum529435} is also presented in Fig. \ref{fig:12}.

The same model for $L_{(3,2)}$ in Eq. \eqref{ZEqnNum136752} can be generalized to high order correlation functions. Here, we just present the models for the other two correlation functions along the diagonal in Fig. \ref{fig:2},
\begin{equation} 
\label{ZEqnNum475527} 
R_{\left(4,3\right)} =\left\langle u^{2} \boldsymbol{\mathrm{u}}\cdot \boldsymbol{\mathrm{u}}^{'} \right\rangle =a_{4} u^{4} \exp \left(-\frac{r}{r_{2} } \right)\left(b_{4} -\frac{r}{r_{2} } \right),       
\end{equation} 
\begin{equation}
\label{ZEqnNum137224} 
L_{\left(5,4\right)} =\left\langle u^{4} u_{L}^{'} \right\rangle =a_{5} u^{5} \exp \left(-\frac{r}{r_{2} } \right)\left(\frac{r}{r_{2} } -b_{5} \right).       
\end{equation} 

\begin{figure}
\includegraphics*[width=\columnwidth]{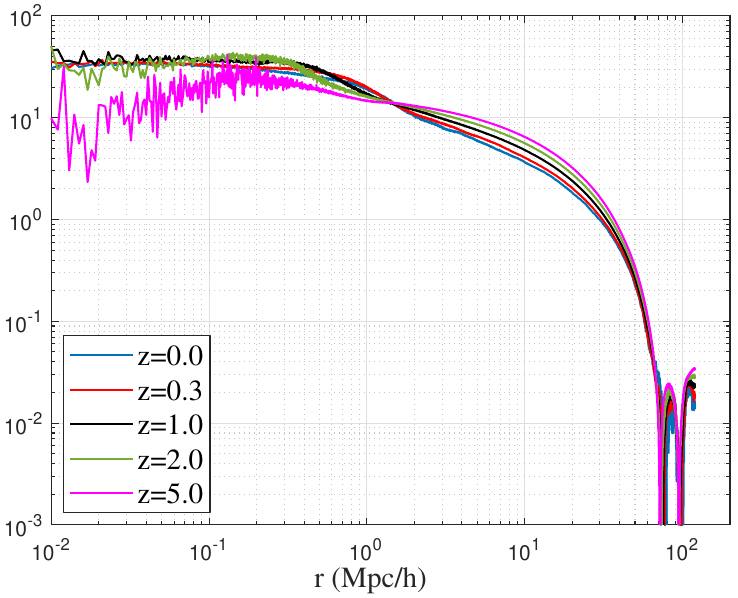}
\caption{The variation of two-point fourth order velocity correlation function $R_{(4,3)}$ with scale \textit{r} at different redshifts \textit{z} (normalized by $u^{4}$). The model for $R_{(4,3)}$ is presented in Eq. \eqref{ZEqnNum475527} and plotted in Fig. \ref{fig:11} for comparison.}
\label{fig:9}
\end{figure}

\begin{figure}
\includegraphics*[width=\columnwidth]{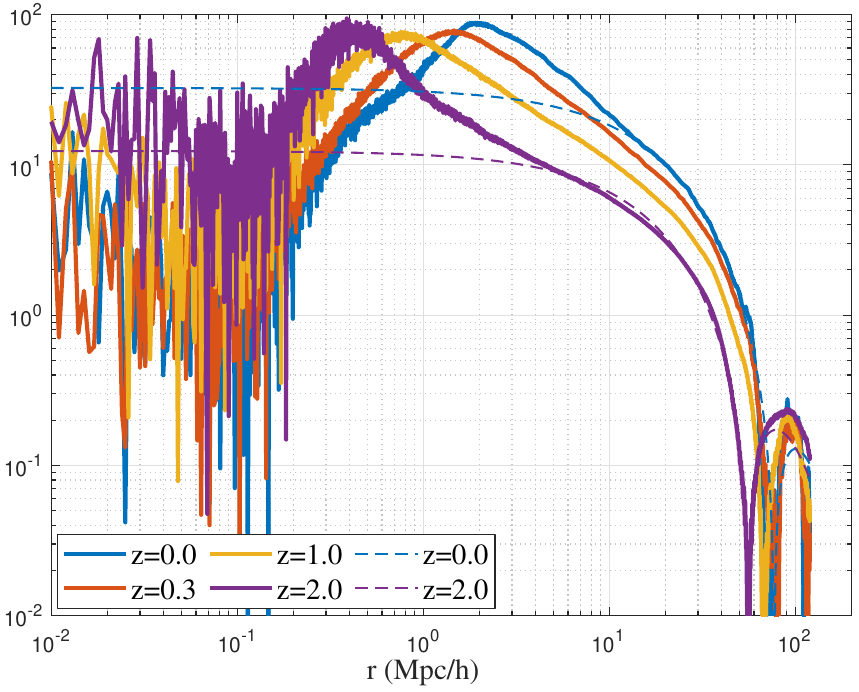}
\caption{The variation of two-point fifth order velocity correlation function $L_{(5,4)} $ with scale \textit{r} at different redshifts \textit{z} (normalized by $u^{5}$). The model in Eq. \eqref{ZEqnNum137224} is also plotted on the same plot as the dashed lines.}
\label{fig:10}
\end{figure}

\begin{figure}
\includegraphics*[width=\columnwidth]{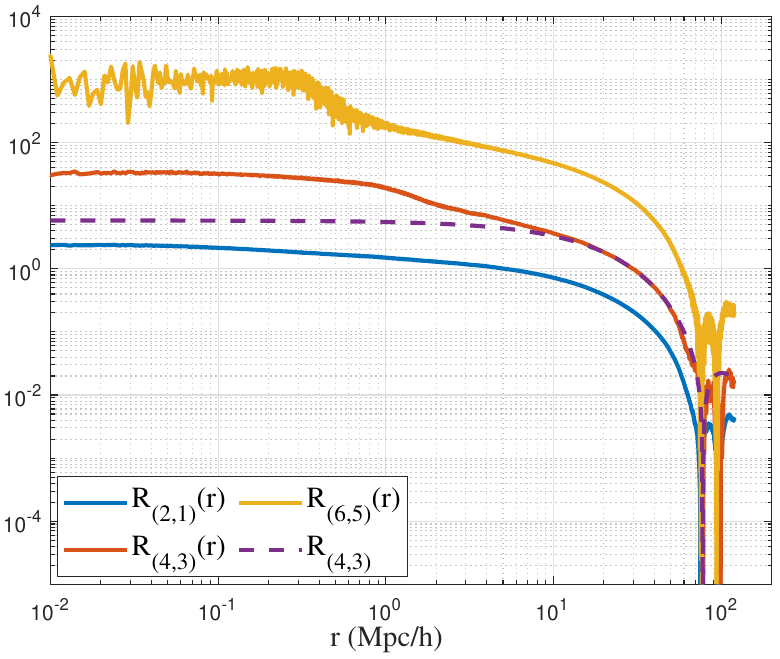}
\caption{The variation of correlation functions $R_{(2,1)} $, $R_{(4,3)}$ and $R_{(6,5)}$ with scale $r$ at \textit{z}=0 (normalized by $u^{2}$, $u^{4}$, and $u^{6}$, respectively). The dashed line shows the model of Eq. \eqref{ZEqnNum475527} for comparison.}
\label{fig:11}
\end{figure}

\begin{figure}
\includegraphics*[width=\columnwidth]{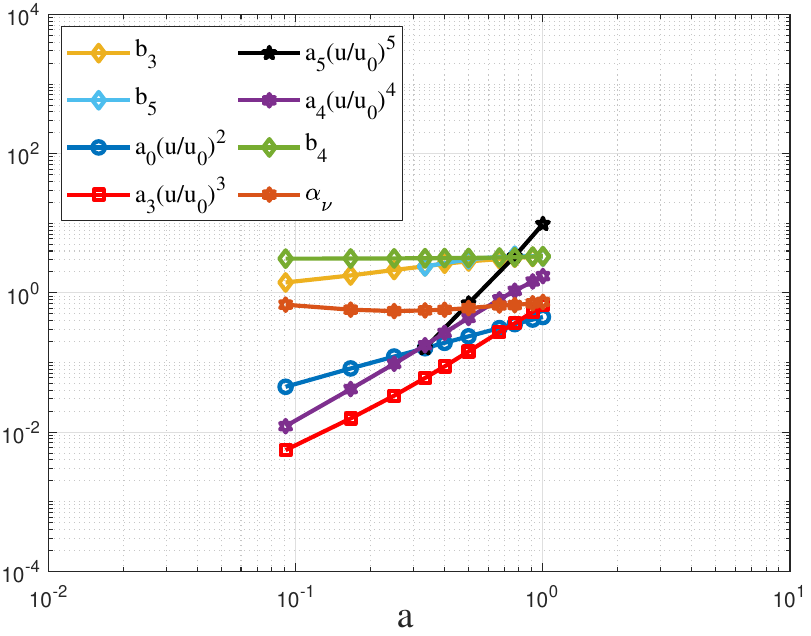}
\caption{The variation of coefficients $a_{m} $ and $b_{m} $ (\textit{m}=3, 4, 5) for velocity correlation functions $L_{(3,2)} $ (Eq. \eqref{ZEqnNum136752}), $R_{(4,3)} $ (Eq. \eqref{ZEqnNum475527}) and $L_{(5,4)} $ (Eq. \eqref{ZEqnNum137224}) with scale factor \textit{a}. it can be confirmed that $a_{0} u^{2} \propto a$, $a_{3} u^{3} \propto a^{{5/2} } $, $a_{4} u^{4} \propto a^{2} $ and $a_{5} u^{5} \propto a^{{7/2} } $. The viscosity coefficient $\nu \left(a\right)=-\alpha _{\nu } u_{0} r_{2} \propto a^{{1/2} } $, where $u_{0} =354.61{km/s} $ and $r_{2} \approx 23{Mpc/h} $. The negative viscosity reflects the inverse energy cascade from smaller to larger scales.}
\label{fig:12}
\end{figure}

Figures \ref{fig:8}, \ref{fig:9}, and \ref{fig:10} plot the variation of the correlation functions $L_{(3,2)} $, $R_{(4,3)} $, and $L_{(5,4)} $ with scale \textit{r} at different redshifts \textit{z}. Models for these correlation functions are presented in Eqs. \eqref{ZEqnNum136752}, \eqref{ZEqnNum475527} and \eqref{ZEqnNum137224} are also plotted in the same plots as the dashed lines (model for $R_{(4,3)}$ from Eq. \eqref{ZEqnNum475527} is presented in Fig. \ref{fig:11}).
Figure \ref{fig:11} plots the variation of correlation functions $R_{(2,1)} $, $R_{(4,3)}$, and $R_{(6,5)} $ with scale \textit{r} at the same redshift $z=0$. The dashed line plots the model of $R_{(4,3)} $ from Eq. \eqref{ZEqnNum475527}. 
Figure \ref{fig:12} plots the variation of the fitted coefficients $a_{m} $ and $b_{m} $ (\textit{m}=3, 4, 5) for the correlations $L_{(3,2)} $ (Eq. \eqref{ZEqnNum136752}), $R_{(4,3)} $ (Eq. \eqref{ZEqnNum475527}) and $L_{(5,4)} $ (Eq. \eqref{ZEqnNum137224}) with the scale factor \textit{a}. It can be easily confirmed that $L_{(3,2)} \propto a^{{5/2} } $, $L_{(5,4)} \propto a^{{7/2} } $, and $R_{(4,3)} \propto a^{2} $. Effective viscosity $\nu(a)=-\alpha _{\nu } u_{0} r_{2} \propto a^{{1/2} } $, where $u_{0} =354.61{km/s} $ and $r_{2} \approx 23.13{Mpc/h} $.

In principle, the same model can be generalized to any order (similar to the dynamic relation in Eq. \eqref{ZEqnNum216206}), 
\begin{equation} 
\label{ZEqnNum995688} 
\begin{split}
L_{\left(q+1,q\right)} &=\left\langle u^{q} u_{L}^{'} \right\rangle \propto u^{q} \left\langle u_{L}^{'} \right\rangle \\
&\propto \left(\nu Ha^{2} \right)^{{q/2} } L_{\left(1,0\right)} \propto a^{{\left(q+3\right)/2}}.    \end{split}
\end{equation} 
The generalized dynamic relations (similar to Eq. \eqref{ZEqnNum715845}) between correlations of different order reads
\begin{equation} 
\label{eq:167} 
L_{\left(q+1,q\right)} \propto -2a\nu \frac{\partial R_{\left(q,q-1\right)} }{\partial r} ,         
\end{equation} 
and
\begin{equation} 
\label{ZEqnNum252855}
\begin{split}
R_{\left(q,q-1\right)} &=\left\langle u^{q-2} \boldsymbol{\mathrm{u}}\cdot \boldsymbol{\mathrm{u}}_{}^{'} \right\rangle \propto u^{q-2} \left\langle \boldsymbol{\mathrm{u}}\cdot \boldsymbol{\mathrm{u}}_{}^{'} \right\rangle \\&\propto \left(\nu Ha^{2} \right)^{{\left(q-2\right)/2} } R_{\left(2,1\right)}\propto a^{{q/2}}.
\end{split}
\end{equation} 

In this section, we focus on the dynamic relations between the velocity correlations of different orders along the diagonal in Fig. \ref{fig:2}. From the dynamic equation on large scales (Eq. \eqref{ZEqnNum709847}), we derive the general dynamic relations in Eqs. \eqref{ZEqnNum715845}, \eqref{ZEqnNum283672}, and \eqref{eq:167}. The time evolution of $p$th-order velocity correlations follows $\propto a^{(p+2)/2}$ for odd order $p$ and $\propto a^{p/2}$ for even order $p$. Effective viscosity $\nu(a)$ on large scales $r>r_t$ in Eq.  \eqref{ZEqnNum709847} can be related to the inverse energy cascade on small scales $r<r_t$ and $\nu(a)\propto a^{1/2}$ (Eq. \eqref{ZEqnNum133069}). 

\subsection{Dynamic relations between velocity and density fields}
\label{sec:6.2}
Similarly to deriving the dynamic relations between third and second order correlations, the relation between second and first order statistics can be obtained by multiplying both sides of dynamic Eqs. \eqref{ZEqnNum373540} and \eqref{ZEqnNum198491} with unit vectors $\hat{r}_{j} ={r_{j}/r} $ and $\hat{r}_{i} ={r_{i}/r}$,
\begin{equation}
\label{eq:169} 
\hat{r}_{j} \frac{\partial v_{j} }{\partial t} -\frac{1}{2a} \frac{\partial \left(v_{i} v_{i} \right)}{\partial r} =cu_{L} -\nu \frac{\partial }{\partial r} \left(\frac{\partial v_{i} }{\partial x_{i} } \right),       
\end{equation} 
and
\begin{equation} 
\label{ZEqnNum409936} 
\hat{r}_{i} \frac{\partial v_{i}^{'} }{\partial t} +\frac{1}{2a} \frac{\partial \left(v_{j}^{'} v_{j}^{'} \right)}{\partial r} =cu_{L}^{'} +\nu \frac{\partial }{\partial r} \left(\frac{\partial v_{j}^{'} }{\partial x_{j}^{'} } \right).       
\end{equation} 
Subtracting two equations and taking the average lead to
\begin{equation}
\label{ZEqnNum870821} 
\left\langle \hat{r}_{i} \frac{\partial v_{i}^{'} }{\partial t} -\hat{r}_{j} \frac{\partial v_{j} }{\partial t} \right\rangle +\frac{1}{a} \frac{\partial \left\langle u^{2} \right\rangle }{\partial r} =c\left\langle \Delta u_{L} \right\rangle +2\nu \frac{\partial \left\langle \theta \right\rangle }{\partial r} ,      
\end{equation} 
where the pairwise velocity $\langle \Delta u_{L} \rangle =\langle u_{L}^{'} -u_{L} \rangle =\langle v_{i}^{'} \hat{r}_{i} -v_{j} \hat{r}_{j} \rangle $. The velocity dispersion $\langle u^{2} \rangle $ and the divergence $\langle \theta \rangle $ on a given scale \textit{r} are defined as
\begin{equation}
\left\langle u^{2} \right\rangle =\frac{1}{2} \left\langle \left|\boldsymbol{\mathrm{u}}\right|^{2} +\left|\boldsymbol{\mathrm{u}}^{'} \right|^{2} \right\rangle \quad \textrm{and} \quad \left\langle \theta \right\rangle =\left\langle \nabla \cdot \boldsymbol{\mathrm{u}}\right\rangle =\frac{1}{2} \left\langle \theta +\theta ^{'} \right\rangle.  
\label{eq:172}
\end{equation}

Second, on large scales, the overdensity $\delta$ can be related to the divergence as \citep [see] [Eq. (119)]{Xu:2023-On-the-statistical-theory-of-self-gravitating},
\begin{equation} 
\label{ZEqnNum749059} 
\delta \approx \eta =-\frac{\nabla \cdot \boldsymbol{\mathrm{u}}}{aHf\left(\Omega _{m} \right)}=-\frac{\theta}{aHf\left(\Omega _{m} \right)} ,          
\end{equation} 
where $\eta \left(\boldsymbol{\mathrm{x}}\right)=\log \left(1+\delta \right)\approx \delta $ is the log-density field. The function $f\left(\Omega _{m} \right)\approx \Omega _{m}^{0.6}$ with $\Omega_m$ being the fraction of matter. The overdensity on the scale $r$ should be $\langle \delta \rangle =\langle \delta +\delta ^{'} \rangle/2$, the average overdensity at two locations $\boldsymbol{\mathrm{x}}$ and $\boldsymbol{\mathrm{x}}^{'}$.

Third, from the identity in the appendix (Eq. \eqref{ZEqnNum517859}) with $p=1$ and $q=0$ such that $\langle u_{s}^{'} \rangle =\langle u_{L}^{'} \hat{r}_{s} \rangle$. Taking the divergence on both sides leads to the relation between the divergence $\langle \theta \rangle $ and the pairwise velocity $\langle \Delta u_{L} \rangle$ on any scale \textit{r} (using identity in Appendix Eq. \eqref{ZEqnNum558872} and product rule for differentiation):
\begin{equation}
\label{ZEqnNum732704} 
\left\langle \theta \right\rangle =\left\langle \nabla \cdot \boldsymbol{\mathrm{u}}\right\rangle =\frac{1}{2r^{2} } \left(r^{2} \left\langle \Delta u_{L} \right\rangle \right)_{,r}.        
\end{equation} 
This is an important kinematic relation good for the entire range of scales. With Eq. \eqref{ZEqnNum990753} for $\langle \Delta u_{L} \rangle $ and Eq. \eqref{ZEqnNum749059} for $\delta $, the divergence on large scale \textit{r} can be finally written as
\begin{equation}
\label{ZEqnNum344666} 
\left\langle \theta \right\rangle =\left\langle \nabla \cdot \boldsymbol{\mathrm{u}}\right\rangle =-aHf\left(\Omega _{m} \right)\left\langle \delta \right\rangle =-Ha\xi \left(r\right).    
\end{equation} 
The divergence is constant on small scales, that is, $\left\langle \theta \right\rangle=-3Ha/2$.

Interestingly, from Eq. \eqref{ZEqnNum344666}, the mean overdensity $\langle\delta\rangle$ at two locations separated by a comoving scale \textit{r} can be related to the density correlation $\xi \left(r\right)$ on the same scale via a dynamic relation, i.e. 
\begin{equation}
\label{ZEqnNum711863} 
f\left(\Omega _{m} \right)\left\langle \delta \right\rangle =f\left(\Omega _{m} \right){\left\langle \delta +\delta ^{'} \right\rangle/2} =\xi \left(r\right)=\left\langle \delta \delta ^{'} \right\rangle ,      
\end{equation} 
where the first and second order statistics are connected via dynamics on large scales (Eq. \eqref{ZEqnNum749059}), the kinematic relation from Eq. \eqref{ZEqnNum732704}, and the pair conservation equation \eqref{ZEqnNum990753}. 

\begin{figure}
\includegraphics*[width=\columnwidth]{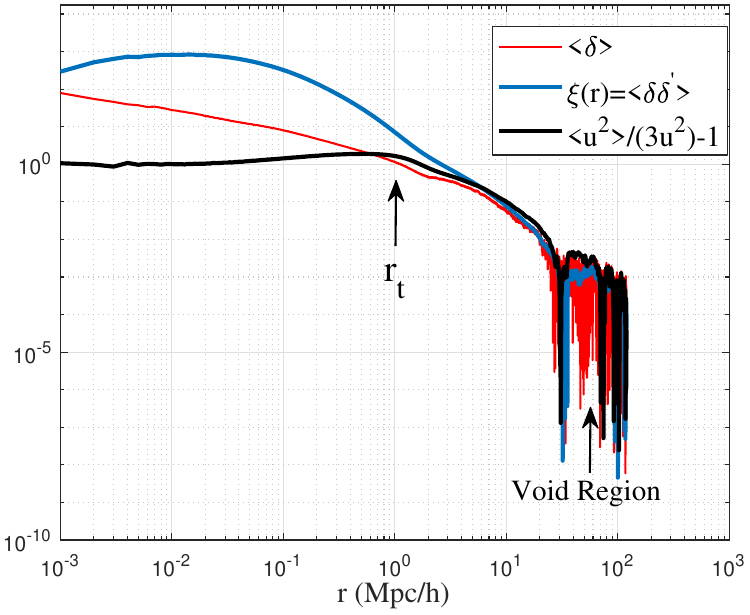}
\caption{The variation of density correlation $\xi (r)=\langle \delta \delta ^{'} \rangle $ and average over-density $\langle \delta \rangle ={\langle \delta +\delta ^{'} \rangle/2} $ with scale \textit{r} from N-body simulation at $z=0$. A relation between first and second order statistics $\langle \delta \rangle \approx \langle \delta \delta ^{'} \rangle $ on large scales (Eq. \eqref{ZEqnNum711863}) can be identified. The plot also shows the relation between an excess velocity dispersion and the density correlation on large scales, that is, $K_{ex}={\langle u^{2} \rangle/(3u^{2} )} -1\propto \xi (r)\propto \langle \delta \rangle $ from Eq. \eqref{ZEqnNum864804}. The lower density $\langle \delta \rangle $ leads to smaller excess velocity dispersion $\langle u^{2} \rangle-3u^2$. On the critical scale of $r_t$, the overdensity is $\xi(r)\approx 1.5$ and $\langle u^{2} \rangle\approx 7.5u^2$ (see Fig. \ref{fig:20}). The region of voids with $\langle \delta \rangle<0$ should have a dispersion $\langle u^{2} \rangle$ less than the asymptotic dispersion $3u^2$.}
\label{fig:13}
\end{figure}

Figure \ref{fig:13} plots the variation of $\langle \delta \rangle$ and $\langle \delta \delta ^{'} \rangle $ with scale \textit{r} in a matter-dominant N-body simulation (Section \ref{sec:2}), where $f(\Omega _{m})=1$ and $\langle \delta \rangle \approx \langle \delta \delta ^{'} \rangle =\xi (r)$ on large scales. With the model of $\xi (r)$ proposed in Eq. \eqref{ZEqnNum179171} \citep[see][Eq. (121)]{Xu:2023-On-the-statistical-theory-of-self-gravitating}, the variation of $\langle \delta \rangle $ on large scales should be the same as $\xi(r)$. Since there exists a negative correlation $\xi(r)<0$ on large scales $r>30{Mpc/h}$, the low-density void region with $\langle \delta \rangle <0$ should be separated by an average distance of around 30 Mpc/h (see Fig. \ref{fig:13}). 

\begin{figure}
\includegraphics*[width=\columnwidth]{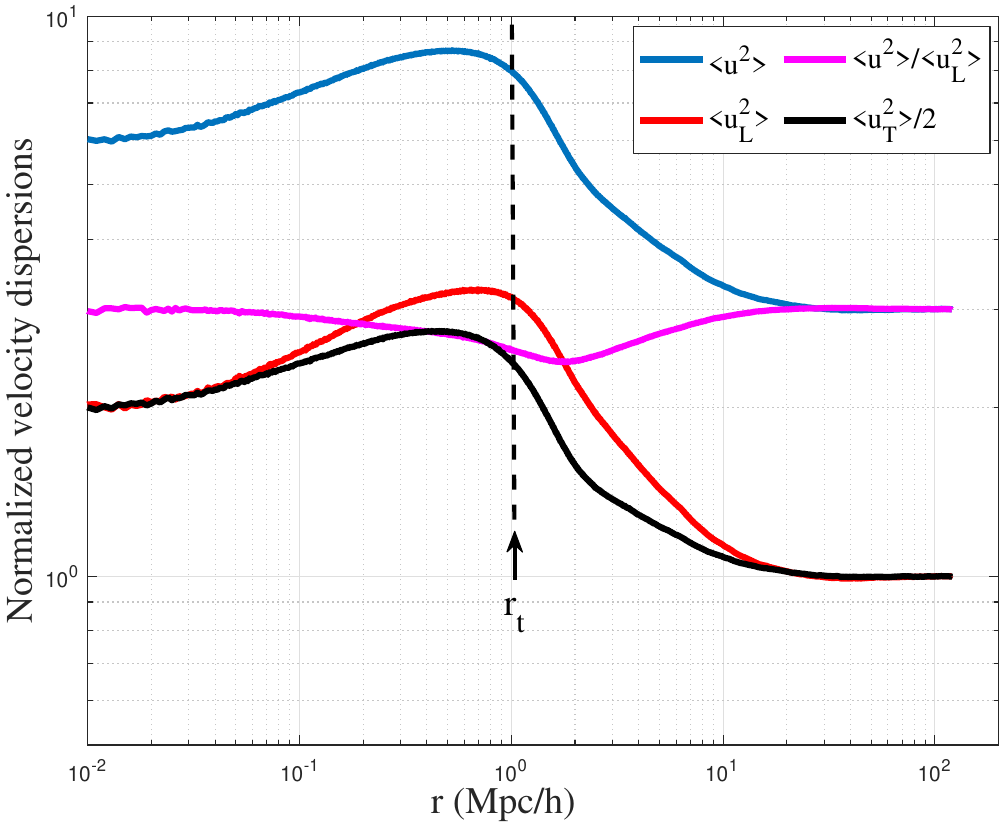}
\caption{Variation of the dispersions of total velocity $\langle u^{2} \rangle =\langle \boldsymbol{\mathrm{u}}\cdot \boldsymbol{\mathrm{u}}\rangle $, longitudinal velocity $\langle u_{L}^{2} \rangle $, and transverse velocity $\langle u_{T}^{2} \rangle =\langle \boldsymbol{\mathrm{u}}_{T} \cdot \boldsymbol{\mathrm{u}}_{T} \rangle $ with scale $r$ at $z=0$ (normalized by $u_{0}^{2}\equiv u^2(z=0)$). The initial increase in all dispersions with \textit{r} for $r<r_{t} $ is mainly due to the increase in velocity dispersion with the size of the haloes. On larger scales $r>r_{t}$, two particles with a given distance $r$ (particle pair) are more likely to come from different haloes, and the dispersion starts to decrease with $r$. On a sufficiently large scale $r\gg r_t$, with all pairs of particles of different haloes, the velocity dispersion reaches a plateau with $\langle u^{2} \rangle =3\langle u_{L}^{2} \rangle =3u^{2}$. The excess dispersion $\langle u^{2} \rangle-3u^2$ is proportional to the overdensity $\delta$ through dynamic relations on large scales (Eq. \eqref{ZEqnNum864804}). On small scales, velocity dispersion approaches $6u^2$, i.e. $\langle u^{2} \rangle =3\langle u_{L}^{2} \rangle =6u^{2}$ and excess dispersion $K_{ex}\rightarrow 1$ (Fig. \ref{fig:13}). }
\label{fig:20}
\end{figure}

Next, the velocity dispersion on any scale of $r$ can be related to the overdensity $\langle \delta \rangle$ on the same scale. In the linear regime, by neglecting the second order advection and viscous term, Eq. \eqref{ZEqnNum709847} reduces to the Zeldovich approximation ${\partial \boldsymbol{\mathrm{v}}/\partial t} =c\left(a\right)\boldsymbol{\mathrm{v}}$ in Eq. \eqref{ZEqnNum911265}. For the pair of velocity at $\boldsymbol{\mathrm{x}}$ and $\boldsymbol{\mathrm{x}}^{'}$, Eq.  \eqref{ZEqnNum911265} leads to (multiplying $\hat{r}_{i}$)
\begin{equation}
\begin{split}
&\hat{r}_{i} \frac{\partial v_{i} }{\partial t} =c\left(a\right)\hat{r}_{i} v_{i} =c\left(a\right)u_{L} \\ 
&\textrm{and}\\ 
&\hat{r}_{i} \frac{\partial v_{i}^{'} }{\partial t} =c\left(a\right)\hat{r}_{i} v_{i}^{'} =c\left(a\right)u_{L}^{'}. \end{split}
\label{ZEqnNum207490}
\end{equation}
Subtracting these two equations leads to
\begin{equation}
\begin{split}
\left\langle \hat{r}_{i} \frac{\partial v_{i}^{'} }{\partial t} -\hat{r}_{i} \frac{\partial v_{i} }{\partial t} \right\rangle = \frac{1}{3} \frac{\partial \left\langle \Delta u_L\right\rangle}{\partial t}= \frac{H}{2}\left\langle \Delta u_{L}\right\rangle. 
\end{split}
\label{ZEqnNum207490-2}
\end{equation}
On large scales, the pairwise velocity $\left\langle \Delta u_L\right\rangle \propto t$ (Eq. \eqref{ZEqnNum283672}). With Eqs. \eqref{ZEqnNum207490-2} and \eqref{ZEqnNum344666} for $\langle \theta \rangle $, the dynamic Eq. \eqref{ZEqnNum870821} gives
\begin{equation} 
\label{eq:178}
\begin{split}
\frac{\partial \left\langle u^{2} \right\rangle }{\partial r} &=2\nu a\frac{\partial \left\langle \theta \right\rangle }{\partial r} \\
&=-2\nu a^{2} Hf\left(\Omega _{m} \right)\frac{\partial \left\langle \delta \right\rangle }{\partial r} =-2\nu Ha^{2} \frac{\partial \xi \left(r\right)}{\partial r}.
\end{split}
\end{equation} 
The velocity dispersion $\langle u^{2} \rangle $ on scale \textit{r} can be related to the density correlation on the same scale as ($\nu<0$ is the effective viscosity),
\begin{equation} 
\label{ZEqnNum864804} 
\begin{split}
&\left\langle u^{2} \right\rangle =3u^{2} -2\nu Ha^{2} f\left(\Omega _{m} \right)\left\langle \delta \right\rangle =3u^{2} -2\nu Ha^{2} \xi \left(r\right), \\
&\textrm{or equivalently, a normalized excess velocity dispersion $K_{ex}$ is}\\
&K_{ex}=\frac{\left\langle u^{2} \right\rangle }{3u^{2} }-1=-\frac{2\nu Ha^{2} \xi \left(r\right)}{3u^{2} } =-\frac{2\nu Ha^{2} }{3u^{2} } f\left(\Omega _{m} \right)\left\langle \delta \right\rangle,\\
&\textrm{such that $\langle u^2 \rangle$ can be analytically modelled (from Eq. \eqref{ZEqnNum179171}) as} \\ 
&\langle u^{2}\rangle = 3u^{2}-\frac{2\nu}{Hf(\Omega _{m})^2} \frac{a_0u^2}{rr_2}\cdot  \exp \left(-\frac{r}{r_{2} } \right)\left[\left(\frac{r}{r_{2} } \right)^{2} -7\left(\frac{r}{r_{2} } \right)+8\right].
\end{split}
\end{equation} 
Here, $\left\langle u^{2} \right\rangle$ is the velocity dispersion on the scale $r$. Figure \ref{fig:20} presents the variation of $\left\langle u^{2} \right\rangle$ with the scale $r$. On a very large scale $r\to \infty$, $\left\langle u^{2} \right\rangle$ approaches the asymptotic dispersion $3u^2$, where $u^2$ is the one-dimensional velocity dispersion of the entire system. Equation \eqref{ZEqnNum864804} describes that the excess velocity dispersion $\left\langle u^{2} \right\rangle-3u^2$ is caused by and proportional to the non-zero overdensity $\langle \delta \rangle$. For the region of the void with $\langle \delta \rangle<0$, the dispersion is less than the asymptotic dispersion or $\left\langle u^{2} \right\rangle<3u^2$. For $r \to \infty$, $\langle u^{2} \rangle=3u^2$ and $\langle \delta \rangle=0$. Figure \ref{fig:13} also presents the variation of the normalized excess velocity dispersion $K_{ex}$ with scale $r$. On large scales, $K_{ex}\propto \xi \left(r\right)$ or $K_{ex}\propto \left\langle \delta \right\rangle $ from Eq. \eqref{ZEqnNum864804}. 

In this section, we focus on the dynamic relations between the first and second order statistics and the relations between velocity and density fields. Based on dynamics on large scales (Eq. \eqref{ZEqnNum749059}), the kinematic relation (Eq. \eqref{ZEqnNum732704}), and the pair conservation equation \eqref{ZEqnNum990753}, we found that overdensity on large scales (first order) is proportional to density correlation (second order), that is, $f(\Omega_m)\langle \delta \rangle \approx \langle \delta \delta ^{'} \rangle =\xi (r)$ (Eq. \eqref{ZEqnNum711863}). Low-density void region with $\langle \delta \rangle<0$ can be identified on scales $r>30Mpc/h$ with a negative density correlation $\xi(r)$. The divergence on the scale $r$ is proportional to the overdensity, that is, $\langle \theta \rangle = -Ha f(\Omega_m)\langle \delta \rangle =-Ha \xi(r)$ (Eq. \eqref{ZEqnNum344666}). Low-density void region with $\langle \delta \rangle<0$ has a positive divergence with matter flowing out of that region. The excess velocity dispersion $\langle u^{2} \rangle-3u^2$ on the scale $r$ is proportional to the overdensity $\langle \delta \rangle$ on the same scale (Eq. \eqref{ZEqnNum864804}). The velocity dispersion $\langle u^{2} \rangle$ in the low-density void region can be smaller than the asymptotic dispersion $3u^2$.

\subsection{Time averaged dynamic equations on large scales}
\label{sec:7.1.2}
The large scale dynamics in Eq. \eqref{ZEqnNum709847} includes an effective viscosity $\nu(a)$ %(<0)
due to the inverse energy cascade on small scales $r<r_t$ (Figs. \ref{fig:19} and \ref{fig:20}). This effective viscosity is required to transfer the kinetic energy cascaded \citep{Xu:2023-Dark-matter-halo-mass-functions-and,Xu:2022-Postulating-dark-matter-partic} from scales below $r_t$ to scales above $r_t$ and therefore is related to the rate of the energy cascade in Eq. \eqref{ZEqnNum133069}. The time-averaged large-scale dynamics will provide more information on the effective viscosity on scales $r>r_t$, similar to the formulation of Reynolds-averaged Navier–Stokes equations (RANS). 

By decomposing the total velocity into the mean and (temporal) fluctuation $\boldsymbol{\mathrm{v}}=\bar{\boldsymbol{\mathrm{v}}}+\boldsymbol{\mathrm{v}}^{'}$, plugging that into Eq. \eqref{ZEqnNum709847} and taking the time average through filtering on a given time scale $\tau$, we obtain a time-averaged dynamic equation for the mean velocity $\bar{\boldsymbol{\mathrm{v}}}$,
\begin{comment}
\begin{equation} 
\label{ZEqnNum959459} 
\begin{split}
\frac{\partial \bar{\boldsymbol{\mathrm{v}}}}{\partial t}&+\frac{1}{a} \left(1-\gamma \right)\bar{\boldsymbol{\mathrm{v}}}\cdot \nabla \bar{\boldsymbol{\mathrm{v}}}+\frac{\gamma }{2a} \nabla \left(\bar{\boldsymbol{\mathrm{v}}}\cdot \bar{\boldsymbol{\mathrm{v}}}\right)+H\bar{\boldsymbol{\mathrm{v}}}\\
&=-\frac{1}{a} \nabla \overline{\phi ^{*} }-\left(\frac{1-\gamma }{a} \underbrace{\overline{\boldsymbol{\mathrm{v}}^{'} \cdot \nabla \boldsymbol{\mathrm{v}}^{'} }}_{1}+\frac{\gamma }{2a} \underbrace{\overline{\nabla \left(\boldsymbol{\mathrm{v}}^{'} \cdot \boldsymbol{\mathrm{v}}^{'} \right)}}_{2}\right).
\end{split}
\end{equation} 
\end{comment}
\begin{equation} 
\label{ZEqnNum959459} 
\begin{split}
\frac{\partial \bar{\boldsymbol{\mathrm{v}}}}{\partial t}+\frac{1}{2a} \nabla \left(\bar{\boldsymbol{\mathrm{v}}}\cdot \bar{\boldsymbol{\mathrm{v}}}\right)
=\frac{H}{2}\bar{\boldsymbol{\mathrm{v}}}+\nu(a)\nabla ^{2} \bar{\boldsymbol{\mathrm{v}}} -\frac{1}{2a} \underbrace{\nabla \overline{\left(\boldsymbol{\mathrm{v}}^{'} \cdot \boldsymbol{\mathrm{v}}^{'} \right)}}_{1}.
\end{split}
\end{equation} 
The eddy viscosity $\nu_{edd}$ in turbulence literature originates from velocity fluctuations on scales below the filtering scale $\tau$ 
\begin{equation}
\nu_{edd} \nabla ^{2} \bar{\boldsymbol{\mathrm{v}}}=\overline{\boldsymbol{\mathrm{v}}^{'} \cdot \nabla \boldsymbol{\mathrm{v}}^{'} }=\nabla \cdot \overline{\boldsymbol{\mathrm{v}}^{'} \otimes \boldsymbol{\mathrm{v}}^{'}}
\end{equation}
for incompressible flow (see Reynolds-averaged Navier–Stokes equations). The effect of velocity fluctuations $\boldsymbol{\mathrm{v}}^{'}$ below the filtering scale $\tau$ on the mean velocity field is equivalent to an "effective" viscosity $\nu_{edd}$ applied to the mean velocity field $\bar{\boldsymbol{\mathrm{v}}}$. The same idea also applies to the irrotational flow on large scales in SG-CFD. Similarly to the "Reynolds stress" in turbulence \citep{Xu:2022-The-mean-flow--velocity-disper}, the velocity fluctuation (${\boldsymbol{\mathrm{v}^{'}}}$) below the filtering scale $\tau$ leads to a "artificial stress" term (LHS of Eq. \eqref{eq:223}) applied onto the mean velocity field ($\bar{\boldsymbol{\mathrm{v}}}$) in the averaged Eq. \eqref{ZEqnNum959459}. Therefore, we can write 
\begin{equation} 
\label{eq:223} 
(\Delta \nu) \nabla ^{2} \bar{\boldsymbol{\mathrm{v}}}=-\frac{1}{2a} \nabla \left(\overline{\boldsymbol{\mathrm{v}}^{'} \cdot \boldsymbol{\mathrm{v}}^{'} }\right),
\end{equation} 
which is the closure relating velocity fluctuation below filtering scale $\tau$ to the mean velocity above scale $\tau$ through an artificial viscosity $\Delta \nu = \nu_\tau-\nu$ that depends on the filtering scale $\tau$. The dynamics for the mean velocity (Eq. \eqref{ZEqnNum959459}) now reads:
\begin{equation} 
\label{ZEqnNum959459-2} 
\begin{split}
\frac{\partial \bar{\boldsymbol{\mathrm{v}}}}{\partial t}+\frac{1}{2a} \nabla \left(\bar{\boldsymbol{\mathrm{v}}}\cdot \bar{\boldsymbol{\mathrm{v}}}\right)
=\frac{H}{2}\bar{\boldsymbol{\mathrm{v}}}+\nu_{\tau}(\tau,a)\nabla ^{2} \bar{\boldsymbol{\mathrm{v}}}.
\end{split}
\end{equation} 
Here $\nu_\tau(\tau,a)$ is the effective viscosity that increases with the filtering scale $\tau$. For filtering scale $\tau\to 0$, $\nu_\tau \to \nu$ such that Eq. \eqref{ZEqnNum959459-2} reduces to the original dynamic Eq. \eqref{ZEqnNum709847}, as expected.

Using Eq. \eqref{ZEqnNum749059} for relation between overdensity and divergence, 
\begin{equation} 
\label{ZEqnNum471934}
\begin{split}
&\nabla \left(\overline{\boldsymbol{\mathrm{v}}^{'} \cdot \boldsymbol{\mathrm{v}}^{'} }\right)=2a^{2} Hf\left(\Omega _{m} \right) (\Delta \nu) \nabla \bar{\delta },\\ 
&\textrm{or equivalently (after integration)}\\  
&\overline{\boldsymbol{\mathrm{v}}^{'} {}^{2} }=F\left(t\right)+2(\Delta\nu) a^{2} Hf\left(\Omega _{m} \right)\bar{\delta }, \end{split}
\end{equation} 
where $F\left(t\right)$ is a function emerging after integration and $\bar{\delta}$ is the (time) average density above the filtering scale $\tau$. Equation \eqref{ZEqnNum471934} is essentially a subgrid closure for large-scale dynamics, where the unresolved velocity fluctuations $\boldsymbol{\mathrm{v}}^{'}{}^{2}$ below the filtering scale $\tau$ can be related to the resolved overdensity $\bar{\delta}$ on scale $\tau$ through an artificial viscosity $\Delta \nu = \nu_\tau-\nu$. 

The total kinetic energy can be decomposed into the kinetic energy in mean flow (resolved) and in fluctuations (unresolved), 
\begin{equation} 
\label{eq:225} 
\begin{split}
\overline{\boldsymbol{\mathrm{v}}^{2} } =\bar{\boldsymbol{\mathrm{v}}}^{2} +\overline{\boldsymbol{\mathrm{v}}^{'} {}^{2} }=\bar{\boldsymbol{\mathrm{v}}}^{2} +F\left(t\right)+2(\Delta\nu) a^{2} Hf\left(\Omega _{m} \right)\bar{\delta },
\end{split}
\end{equation} 
such that the kinetic energy in mean flow reads (same as Eq. \eqref{ZEqnNum864804}), 
\begin{equation} 
\label{eq:226} 
\bar{\boldsymbol{\mathrm{v}}}^{2} =\overline{\boldsymbol{\mathrm{v}}^{2} }-F\left(t\right)-2(\Delta\nu) Ha^{2} f\left(\Omega _{m} \right)\bar{\delta }.        
\end{equation} 
On large scales, gravitational collapse leads to a decreasing dispersion in mean flow $\bar{\boldsymbol{\mathrm{v}}}^{2}$ due to increasing $\bar\delta$ over time. There is continuous energy transfer from mean flow ($\bar{\boldsymbol{\mathrm{v}}}^{2}$) to fluctuation (random motion in $\overline{\boldsymbol{\mathrm{v}}^{'} {}^{2}}$) on large scales. The energy transfer from mean flow to fluctuation on both large scales and halo scale %\citep{Xu:2022-The-mean-flow--velocity-disper} 
continuously increases system entropy in non-equilibrium dark matter flow \citep{Xu:2023-Maximum-entropy-distributions-of-dark-matter}. 

\section{Exponential correlations on large scales}
\label{sec:6.3}
The exponential function was proposed for the second order transverse velocity correlation $T_{(2,0)}\propto ae^{-r/r_2}$ on large scales (Eq. \eqref{ZEqnNum971850}), while the power-law function was proposed for velocity correlations on small scales (Eq. \eqref{ZEqnNum955991}). These are not coincidences and must be deeply rooted in the dynamics and kinematics on large and small scales, which is demonstrated in this and the next sections. 

We first look at the distribution of kinetic energy on different scales via the one-dimensional variance of the smoothed velocity (the bulk flow) using a spherical filter of radius \textit{r}, i.e. velocity dispersion functions $\sigma _{u}^{2} \left(r\right)$ and $\sigma _{d}^{2} \left(r\right)$ for the kinetic energy contained in all scales above or below \textit{r} \citep[see][Eqs. (27) and (29)]{Xu:2023-On-the-statistical-theory-of-self-gravitating},
\begin{equation}
\begin{split}
&\sigma _{u}^{2} \left(r\right)+\sigma _{d}^{2} \left(r\right)=u^2,\\
&\sigma _{u}^{2} \left(r\right)=\frac{1}{3} \int _{-\infty }^{\infty }E_{u} \left(k\right)W\left(kr\right)^{2} dk \\
&\textrm{and}\\ 
&\sigma _{d}^{2} \left(r\right)=\frac{1}{3} \int _{-\infty }^{\infty }E_{u} \left(k\right)\left[1-W\left(kr\right)^{2} \right]dk,
\end{split}
\label{ZEqnNum451933}
\end{equation}

\noindent where $W(x\equiv kr)$ is a window function. For a tophat spherical filter, the window function $W(x)$ reads
\begin{equation}
\label{ZEqnNum605014} 
W\left(x\right)=\frac{3}{x^{3} } \left[\sin \left(x\right)-x\cos \left(x\right)\right]=3\frac{j_{1} \left(x\right)}{x} ,       
\end{equation} 
where $j_{1}(x)$ is the \textit{first} order spherical Bessel function of \textit{first} kind. 

\begin{figure}
\includegraphics*[width=\columnwidth]{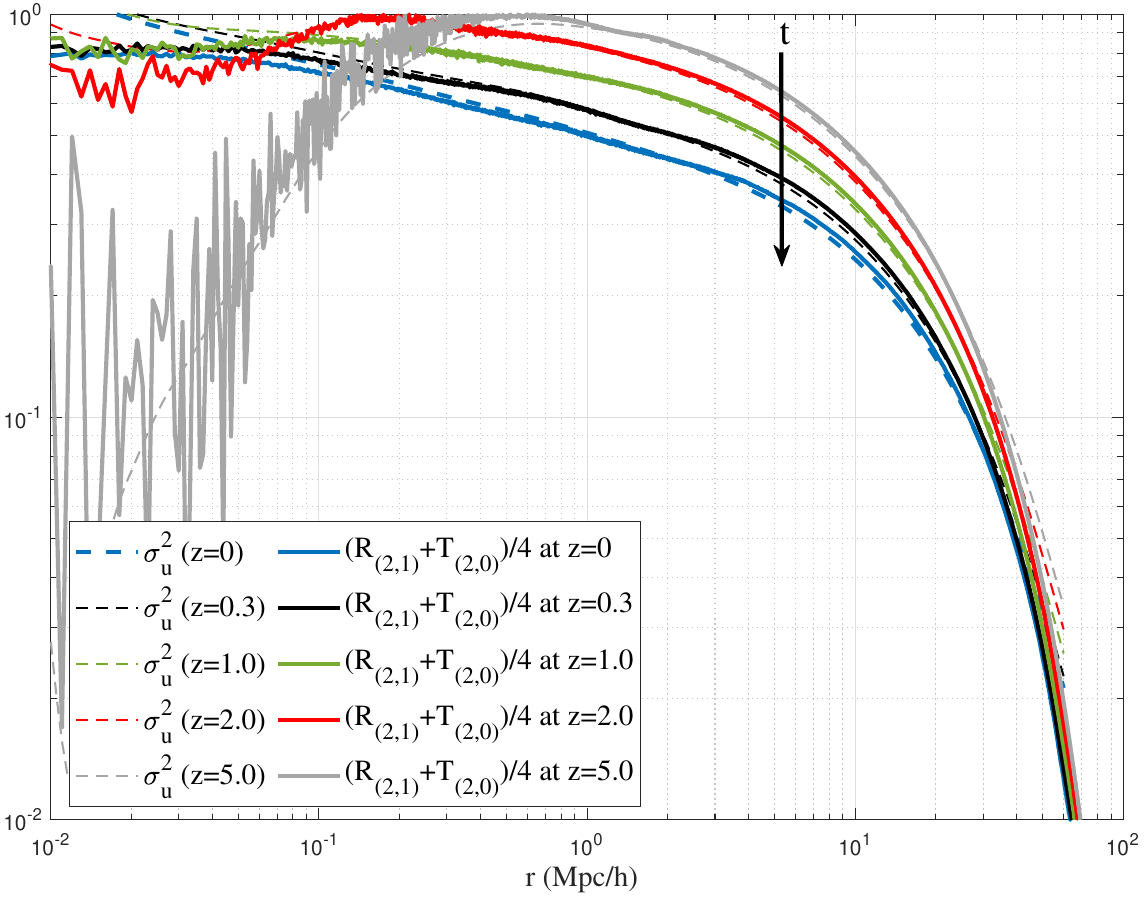}
\caption{The variation of velocity dispersion $\sigma _{u}^{2} \left(r\right)$ normalized by velocity dispersion $u^2$ (kinetic energy contained in scales above $r$) with scale \textit{r} at different redshifts $z$. The function $\sigma _{u}^{2} \left(r\right)$ is obtained from N-body simulation using Eq. \eqref{ZEqnNum451933}. Approximation of $\sigma _{u}^{2} \left(r\right)$ by correlations (Eq. \eqref{ZEqnNum369683}) is also presented as solid lines for comparison.}
\label{fig:14}
\end{figure}

On large scales, the dispersion function $\sigma _{u}^{2}(r)$ is well approximated by $R_{(2,1)}(r)$ and $T_{(2,0)}(r)$ as
\begin{equation} 
\label{ZEqnNum369683} 
\bar\sigma _{u}^{2} \left(r\right)\approx \frac{1}{4} \left[R_{\left(2,1\right)}^{} \left(r\right)+T_{\left(2,0\right)}^{} \left(r\right)\right],      
\end{equation} 
which is a manifestation of energy equipartition including three translational degrees of freedom in total correlation $R_{(2,1)}$ and one rotational degree of freedom in transverse correlation $T_{(2,0)}$. Correlations on large scales characterize the motion of haloes. For a pair of haloes, the transverse motion is dominant, because of weak gravity on large scales. The contribution of longitudinal correlation is relatively small on large scales r>$r_t$ (Fig. 5 in \citep{Xu:2023-On-the-statistical-theory-of-self-gravitating}). 

Figure \ref{fig:14} plots the variation of velocity dispersion $\sigma _{u}^{2} \left(r\right)$ at different redshifts \textit{z}. The function $\sigma _{u}^{2} \left(r\right)$ is obtained directly from N-body simulations \citep[also see][Fig. 10]{Xu:2023-On-the-statistical-theory-of-self-gravitating}. The approximation $\bar\sigma _{u}^{2} \left(r\right)$ by Eq. \eqref{ZEqnNum369683} is also presented as solid lines, in good agreement with $\sigma _{u}^{2} \left(r\right)$ from the simulation.

\begin{figure}
\includegraphics*[width=\columnwidth]{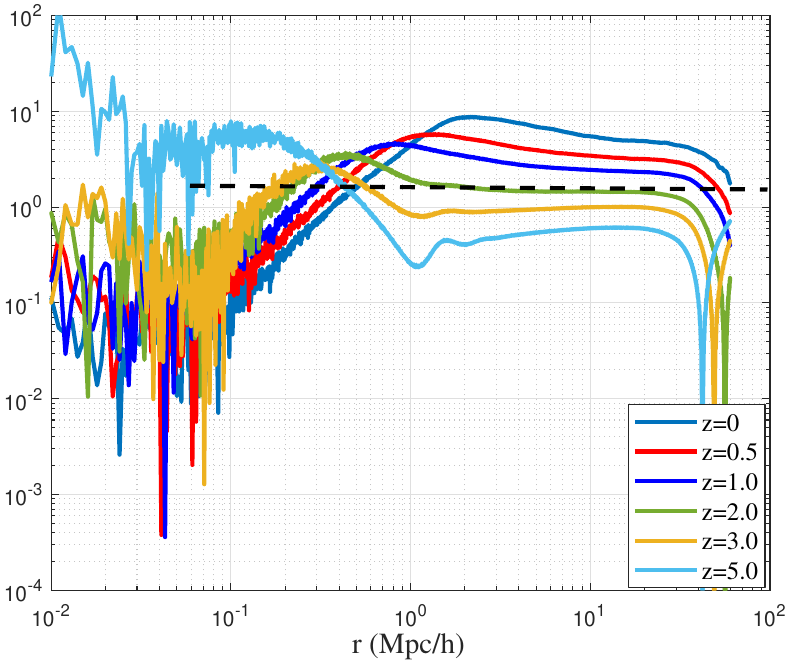}
\caption{The variation of the ratio $L_{(3,2)}/(u\sigma _{u}^{2})$ with scale $r$ at different redshifts $z$ obtained from N-body simulations. That ratio is relatively independent of scale $r$ on large scales as predicted in Eq. \eqref{ZEqnNum445071}.}
\label{fig:22}
\end{figure}

The large scale dynamics involves a characteristic scale $r_2$. The rate of energy change $\varepsilon$ on large scales can be expressed in terms of either the correlation $L_{(3,2)}(r)$ or the dispersion function $\sigma_u^2(r)$
\begin{equation} 
\label{ZEqnNum4450711}
\begin{split}
&\varepsilon \propto -\frac{L_{\left(3,2\right)} \left(r\right)}{r_2}\propto \nu(a) \frac{\sigma_u^2(r)}{r_2^2} ,     
\end{split}
\end{equation} 
where $\sigma_u/r_2$ represents the strain rate. Equation \eqref{ZEqnNum4450711} describes the rate of kinetic energy dissipated by viscosity $\nu$ on the scale $r$ equals the rate of energy transferred to scales greater than $r$ by $L_{(3,2)}$. Using Eq. \eqref{ZEqnNum4450711}, the effective viscosity in Eq. \eqref{ZEqnNum529435}, and the approximation in Eq. \eqref{ZEqnNum369683}), we can write
\begin{equation} 
\label{ZEqnNum445071} 
-L_{\left(3,2\right)} \left(r\right) \approx \alpha _{r} u\sigma _{u}^{2} \left(r\right) = \frac{\alpha _{r}}{4} \left[R_{\left(2,1\right)}^{} \left(r\right)+T_{\left(2,0\right)}^{} \left(r\right)\right]u,     
\end{equation} 
where $\alpha _{r}$ is a proportional constant and $L_{(3,2)}(r)<0$. Figure \ref{fig:22} plots the ratio between $L_{(3,2)}$ and $\sigma_u^2$ from N-body simulations at different redshifts, which is scale independent on large scales. With dynamic relation (Eq. \eqref{ZEqnNum715845}), Eq. \eqref{ZEqnNum445071} finally becomes

\begin{equation}
\label{ZEqnNum731622} 
\frac{8\nu a}{\alpha _{r} u} \frac{\partial R_{\left(2,1\right)}^{} }{\partial r} =\left[R_{\left(2,1\right)}^{} \left(r\right)+T_{\left(2,0\right)}^{} \left(r\right)\right].        
\end{equation} 
Using the kinematic relationship between $R_{(2,1)}$ and $T_{(2,0)}$ in Eq. \eqref{ZEqnNum320035} for second order correlations, the exponential transverse correlation function on large scales is recovered:
\begin{equation}
T_{\left(2,0\right)}^{} =Const\cdot \exp \left(-\frac{r}{r_{2} } \right) \quad \textrm{with} \quad r_{2} =-\frac{8\nu a}{\alpha _{r} u}.    
\label{eq:196}
\end{equation}
\noindent Comparing with Eq. \eqref{ZEqnNum529435}, we found the proportional constant $\alpha _{r} \equiv {4a_{3}/a_{0} }\propto a^{3/4}$.

\begin{figure}
\includegraphics*[width=\columnwidth]{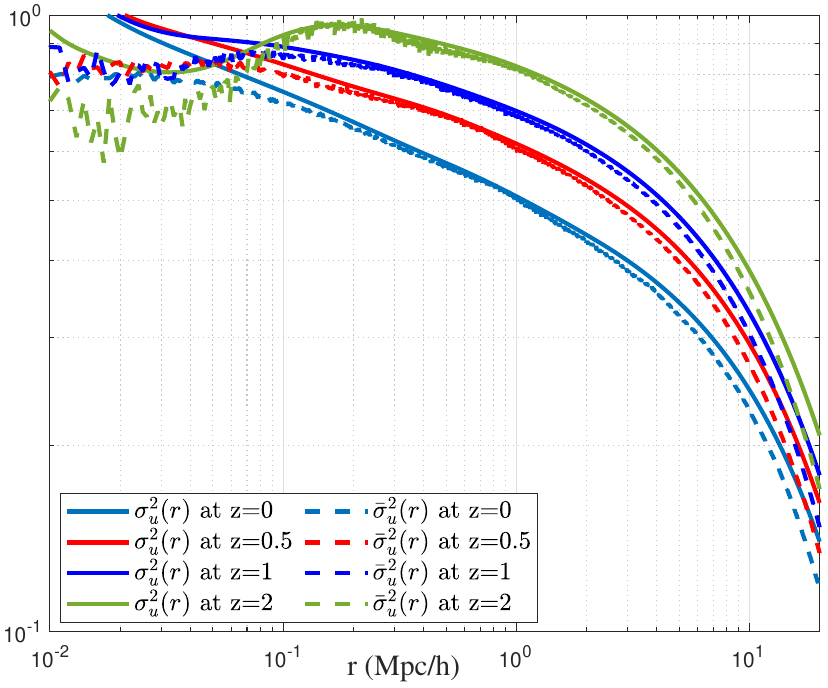}
\caption{The variation of velocity dispersion $\sigma _{u}^{2} \left(r\right)$ normalized by velocity dispersion $u^2$ (kinetic energy contained in scales above $r$) with scale $r$ at different redshifts $z$. The approximation $\bar\sigma _{u}^{2} \left(r\right)$ are also obtained from N-body simulation using Eq. \eqref{eq:188} and presented as dashed lines for comparison. Good agreement can be found on small scales.}
\label{fig:21}
\end{figure}

\section{Power-law correlations on small scales}
\label{sec:6.3-2}
So far, we have discussed the dynamic origin of exponential correlations on large scales. This is based on Eq. \eqref{ZEqnNum369683} for dispersion function $\sigma _{u}^{2}(r)$, Eq. \eqref{ZEqnNum445071} for relation between $L_{(3,2)}$ and $\sigma _{u}^{2}$, and kinematic relations (Eqs. \eqref{ZEqnNum715845} and \eqref{ZEqnNum320035}). Next, we focus on the dynamic origin of power law correlations on small scales. 

First, correlations on small scales reflect the motion of particles in the same halo, so that longitudinal and transverse correlations are comparable \citep[see][Figs. 3 and 4]{Xu:2023-On-the-statistical-theory-of-self-gravitating}. Therefore, the velocity dispersion function $\sigma _{u}^{2} (r)$ on small scales can be approximated by
\begin{equation} 
\label{eq:188} 
\bar\sigma _{u}^{2} \left(r\right)\approx \frac{1}{5} \left[R_{\left(2,1\right)}^{} \left(r\right)+T_{\left(2,0\right)}^{} \left(r\right)+L_{\left(2,0\right)}^{} \left(r\right)\right].      
\end{equation} 
with an additional degree of freedom from the longitudinal correlation (different from Eq. \eqref{ZEqnNum369683}). Figure \ref{fig:21} presents both the velocity dispersion $\sigma_u^2$ and the approximation $\bar\sigma_u^2$ from the N-body simulations. Good agreement can be found on small scales.

The exact relation between the velocity correlation $R_{2}(r)$ and $\sigma _{u}^{2} \left(r\right)$ was obtained previously as (\citep[see][Eq. (30)]{Xu:2023-On-the-statistical-theory-of-self-gravitating})
\begin{equation} 
\label{ZEqnNum910224} 
R_{2}^{} \left(2r\right)=\frac{1}{24r^{2} } \frac{\partial }{\partial r} \left(\frac{1}{r^{2} } \frac{\partial }{\partial r} \left(r^{3} \frac{\partial }{\partial r} \left(\sigma _{u}^{2} \left(r\right)r^{4} \right)\right)\right).      
\end{equation} 
On small scales, for velocity correlation $R_2(r)$ from Eq. \eqref{ZEqnNum955991}, the power law velocity dispersion function ($\sigma_d^2(r) \propto r^n$) should read (\citep[also see][Eq. (141)]{Xu:2023-On-the-statistical-theory-of-self-gravitating})
\begin{equation} 
\label{eq:141-2} 
\sigma _{d}^{2} \left(r\right)=u^2-\sigma _{u}^{2} \left(r\right)=\frac{24\cdot 2^{n} }{\left(4+n\right)\left(6+n\right)} u^{2} \left(\frac{r}{r_{1} } \right)^{n}.  
\end{equation} 
When combined with kinematic relations for constant divergence flow on small scales (Eqs. \eqref{ZEqnNum178092} and \eqref{ZEqnNum955991}), Equation \eqref{eq:188} can be used to derive the equation for the power law exponent $n$
\begin{equation}
\label{eq:189}    
\begin{split}
(10+3n)(4+n)(6+n) = 15\cdot 2^{n+4},
\end{split}    
\end{equation}
which gives $n \approx 1/4$. This provides the dynamic and kinematic origin of the "one-fourth" scaling law ($\sigma_d^2(r) \propto r^{1/4}$) for the constant divergence flow on small scales \citep[see][Section 5.2]{Xu:2023-On-the-statistical-theory-of-self-gravitating}.

\section{Dynamic relations on small scales}
\label{sec:7}
\subsection{Dynamic equations for velocity on small scales}
\label{sec:7.1}
To the author's knowledge, self-closed equations for velocity evolution on small scales do not exist. However, self-closed velocity evolution (like Eq. \eqref{ZEqnNum709847} on large scales) is required to derive the dynamic relations. In this section, we will first formulate the self-close equations for velocity. These equations are then applied to derive the dynamic relations on small scales. 

\subsubsection{Self-closed dynamic equations for velocity}
\label{sec:7.1.1}
on small scales, the flow is of constant divergence, i.e. $\nabla \cdot \boldsymbol{\mathrm{v}}=\theta $ . We focus on the momentum equation (Jeans' equation) for peculiar velocity in comoving coordinate,
\begin{equation}
\label{ZEqnNum987321} 
\frac{\partial \boldsymbol{\mathrm{v}}}{\partial t} +\frac{1}{a} \boldsymbol{\mathrm{v}}\cdot \nabla \boldsymbol{\mathrm{v}}+H\boldsymbol{\mathrm{v}}=-\frac{1}{a} \frac{\nabla \cdot \boldsymbol{\mathrm{p}}}{\rho } -\frac{1}{a} \nabla \phi  ,       
\end{equation} 
where $\rho$ is density, $\boldsymbol{\mathrm{p}}=\rho \boldsymbol{\mathrm{\sigma }}^{\boldsymbol{\mathrm{2}}} $ is the pressure tensor, $\boldsymbol{\mathrm{\sigma }}^{\boldsymbol{\mathrm{2}}} $ is the velocity dispersion tensor, and $\phi $ is the gravitational potential. 

It is well known that this equation is not closed, and closure must be developed. Starting from the halo-based description of the entire system, the peculiar velocity can be decomposed into velocity due to the motion of haloes ($\boldsymbol{\mathrm{v}}_{h} $) and velocity due to the motion in haloes, i.e., intrahalo motion $\boldsymbol{\mathrm{v}}_{v} $ \citep [also see] [Eq. (1)]{Xu:2023-Maximum-entropy-distributions-of-dark-matter}, 
\begin{equation} 
\label{ZEqnNum643096} 
\boldsymbol{\mathrm{v}}\left(\boldsymbol{\mathrm{x}},t\right)=\boldsymbol{\mathrm{v}}_{h} \left(\boldsymbol{\mathrm{x}}_{h} ,t\right)+\boldsymbol{\mathrm{v}}_{v} \left(\boldsymbol{\mathrm{r}},t\right),         
\end{equation} 
where $\boldsymbol{\mathrm{x}}_{h} $ is the center of mass of a given halo and $\boldsymbol{\mathrm{r}}$ is the vector relative to the halo center such that particle position $\boldsymbol{\mathrm{x}}=\boldsymbol{\mathrm{x}}_{h} +\boldsymbol{\mathrm{r}}$. All particles in the same halo should have the same halo velocity $\boldsymbol{\mathrm{v}}_{h} $. The spatial variation of $\boldsymbol{\mathrm{v}}_{h} $ is on a much larger length scale, compared to the variation of $\boldsymbol{\mathrm{v}}_{v} $. Therefore, $\boldsymbol{\mathrm{v}}_{h}$ is relatively constant on the halo scale. In spherical coordinates, motion in the halo (velocity $\boldsymbol{\mathrm{v}}_{v}$) can be further decomposed into the radial flow and azimuthal flow,% where the polar flow can be neglected \citep[see][Fig. 2]{Xu:2022-The-mean-flow--velocity-disper}, 
\begin{equation}
\label{ZEqnNum743696} 
\boldsymbol{\mathrm{v}}_{v} =\boldsymbol{\mathrm{v}}_{r} +\boldsymbol{\mathrm{v}}_{\varphi }.           
\end{equation} 
where the polar flow (meridional flow) is neglected \citep{Xu:2022-The-mean-flow--velocity-disper}. In spherical coordinate, the mean (peculiar) radial flow and azimuthal flow in virialized haloes (i.e. small scales) read %(see solutions for small haloes with low peak height $\nu$) \citep[see][Section 3.4]{Xu:2022-The-mean-flow--velocity-disper},
\begin{equation}
\boldsymbol{\mathrm{v}}_{r} =-Ha\boldsymbol{\mathrm{r}}, \quad \boldsymbol{\mathrm{v}}_{\varphi } =\boldsymbol{\mathrm{\omega }}_{h} \times \boldsymbol{\mathrm{r}},\quad \textrm{and} \quad \boldsymbol{\mathrm{\omega }}_{h} ={\left(\nabla \times \boldsymbol{\mathrm{v}}_{\varphi } \right)/2}.     
\label{ZEqnNum110870}
\end{equation}

\noindent where $\boldsymbol{\mathrm{\omega }}_{h}$ is the angular velocity of that halo and should be the same for all particles in the same halo. The evolution of the velocity field on small scales can be formulated based on Eq. \eqref{ZEqnNum110870}. Obviously, the radial and azimuthal flow have the following properties: 
\begin{equation}
\begin{split}
&\nabla \boldsymbol{\mathrm{v}}_{r} =-Ha\nabla \boldsymbol{\mathrm{r}}=-Ha\boldsymbol{\mathrm{I}}, \quad \nabla \boldsymbol{\mathrm{v}}_{\varphi } +\left(\nabla \boldsymbol{\mathrm{v}}_{\varphi } \right)^{T} =0,\\ 
&\nabla \times \boldsymbol{\mathrm{v}}_{r} =0, \quad \textrm{and} \quad \boldsymbol{\mathrm{v}}_{\varphi } =\boldsymbol{\mathrm{r}}\cdot \nabla \boldsymbol{\mathrm{v}}_{\varphi } ,
\end{split}
\label{ZEqnNum946038}
\end{equation}
\noindent where $\boldsymbol{\mathrm{I}}$ is an identity matrix and $\nabla \boldsymbol{\mathrm{v}}_{\varphi } $ is antisymmetric. It can be easily confirmed that the radial flow satisfies (from Eq. \eqref{ZEqnNum946038})
\begin{equation} 
\label{ZEqnNum309912} 
\frac{\partial \boldsymbol{\mathrm{v}}_{r} }{\partial t} +\frac{1}{a} \boldsymbol{\mathrm{v}}_{r} \cdot \nabla \boldsymbol{\mathrm{v}}_{r} +H\boldsymbol{\mathrm{v}}_{r} =\frac{\partial \boldsymbol{\mathrm{v}}_{r} }{\partial t} ,        
\end{equation} 
and the azimuthal flow satisfies (from Eq. \eqref{ZEqnNum110870})
\begin{equation} 
\label{eq:203} 
H\boldsymbol{\mathrm{v}}_{\varphi } +\frac{1}{a} \boldsymbol{\mathrm{\omega }}_{h} \times \boldsymbol{\mathrm{v}}_{r} =0.          
\end{equation} 

Radial and azimuthal flow in Eq. \eqref{ZEqnNum110870} also satisfy
\begin{equation}
\frac{1}{a} \boldsymbol{\mathrm{v}}_{\varphi } \cdot \nabla \boldsymbol{\mathrm{v}}_{r} +H\boldsymbol{\mathrm{v}}_{\varphi } =0 \quad \textrm{(from Eq. \eqref{ZEqnNum946038})},   
\label{ZEqnNum755386}
\end{equation}
\begin{equation}
\frac{1}{a} \boldsymbol{\mathrm{v}}_{r} \cdot \nabla \boldsymbol{\mathrm{v}}_{\varphi } =\frac{1}{a} \boldsymbol{\mathrm{\omega }}_{h} \times \boldsymbol{\mathrm{v}}_{r} \quad \textrm{(from Eq. \eqref{ZEqnNum946038})},      
\label{ZEqnNum321682}
\end{equation}
\begin{equation}
\frac{\partial \boldsymbol{\mathrm{v}}_{\varphi } }{\partial t} +\frac{1}{a} \boldsymbol{\mathrm{v}}_{\varphi } \cdot \nabla \boldsymbol{\mathrm{v}}_{\varphi } =\frac{1}{a} \boldsymbol{\mathrm{\omega }}_{h} \times \boldsymbol{\mathrm{v}}_{\varphi } 
\quad \textrm{(Newton's second law)}.    
\label{ZEqnNum157890}
\end{equation}
\noindent By adding Eqs. \eqref{ZEqnNum309912}, \eqref{ZEqnNum755386}, \eqref{ZEqnNum321682} and \eqref{ZEqnNum157890} together, the equation for intra-halo motion $\boldsymbol{\mathrm{v}}_{v} $ is
\begin{equation} 
\label{ZEqnNum611390}
\begin{split}
\frac{\partial \boldsymbol{\mathrm{v}}_{v} }{\partial t} +\frac{1}{a} \boldsymbol{\mathrm{v}}_{v} \cdot \nabla \boldsymbol{\mathrm{v}}_{v} +H\boldsymbol{\mathrm{v}}_{v} &=\frac{\partial \boldsymbol{\mathrm{v}}_{r} }{\partial t} +\frac{1}{a} \boldsymbol{\mathrm{\omega }}_{h} \times \boldsymbol{\mathrm{v}}_{v}\\ &=\frac{\partial \boldsymbol{\mathrm{v}}_{r} }{\partial t} +\frac{1}{2a} \left(\nabla \times \boldsymbol{\mathrm{v}}_{v} \right)\times \boldsymbol{\mathrm{v}}_{v}.
\end{split}
\end{equation} 

The motion of halo $\boldsymbol{\mathrm{v}}_{h} $ is spatially varying on a much larger scale than the size of the halo, that is, the motion of halo $\boldsymbol{\mathrm{v}}_{h} $ can be treated as a constant on the halo scale such that
\begin{equation}
\frac{1}{a} \boldsymbol{\mathrm{v}}_{v} \cdot \nabla \boldsymbol{\mathrm{v}}_{h} \approx 0 \quad \textrm{and} \quad \frac{1}{a} \boldsymbol{\mathrm{v}}_{h} \cdot \nabla \boldsymbol{\mathrm{v}}_{h} \approx 0.      
\label{ZEqnNum478697}
\end{equation}
\noindent In addition, using the identity for two vectors \textbf{A} and \textbf{B}, 
\begin{equation} 
\label{eq:209} 
\boldsymbol{\mathrm{A}}\times \left(\nabla \times \boldsymbol{\mathrm{B}}\right)=\boldsymbol{\mathrm{A}}\cdot \left[\left(\nabla \boldsymbol{\mathrm{B}}\right)^{T} -\nabla \boldsymbol{\mathrm{B}}\right],        
\end{equation} 
halo velocity $\boldsymbol{\mathrm{v}}_{h} $ satisfies ($\nabla \boldsymbol{\mathrm{v}}_{\varphi }$ is antisymmetric and Eq. \eqref{ZEqnNum946038})
\begin{equation}
\boldsymbol{\mathrm{v}}_{h} \cdot \nabla \boldsymbol{\mathrm{v}}_{\varphi } =\boldsymbol{\mathrm{\omega }}_{h} \times \boldsymbol{\mathrm{v}}_{h} \quad \textrm{and} \quad\frac{1}{a} \boldsymbol{\mathrm{v}}_{h} \cdot \nabla \boldsymbol{\mathrm{v}}_{r} +H\boldsymbol{\mathrm{v}}_{h} =0.  
\label{ZEqnNum863234}
\end{equation}
Equation for $\boldsymbol{\mathrm{v}}_{h}$ is finally written as (using Eqs. \eqref{ZEqnNum110870} and \eqref{ZEqnNum863234}),
\begin{equation} 
\label{ZEqnNum701799} 
\begin{split}
\frac{\partial \boldsymbol{\mathrm{v}}_{h} }{\partial t} +\frac{1}{a} \boldsymbol{\mathrm{v}}_{h} \cdot \nabla \boldsymbol{\mathrm{v}}_{v} +H\boldsymbol{\mathrm{v}}_{h} &=\frac{\partial \boldsymbol{\mathrm{v}}_{h} }{\partial t} +\frac{1}{a} \boldsymbol{\mathrm{v}}_{h} \cdot \nabla \boldsymbol{\mathrm{v}}_{\varphi }\\
&=\frac{\partial \boldsymbol{\mathrm{v}}_{h} }{\partial t} +\frac{1}{2a} \left(\nabla \times \boldsymbol{\mathrm{v}}_{\varphi } \right)\times \boldsymbol{\mathrm{v}}_{h}.
\end{split}
\end{equation} 

Adding Eq. \eqref{ZEqnNum611390} for intra-halo motion $\boldsymbol{\mathrm{v}}_{v} $ and Eqs. \eqref{ZEqnNum478697} and \eqref{ZEqnNum701799} for the motion of the halo $\boldsymbol{\mathrm{v}}_{h} $ together with the relation $\nabla \times \boldsymbol{\mathrm{v}}_{\varphi } =\nabla \times \boldsymbol{\mathrm{v}}_{v} =\nabla \times \boldsymbol{\mathrm{v}}$, the dynamic equation for the total velocity $\boldsymbol{\mathrm{v}}$ reads
\begin{equation} 
\label{eq:212} 
\frac{\partial \boldsymbol{\mathrm{v}}}{\partial t} +\frac{1}{a} \boldsymbol{\mathrm{v}}\cdot \nabla \boldsymbol{\mathrm{v}}+H\boldsymbol{\mathrm{v}}=\frac{\partial \boldsymbol{\mathrm{v}}_{r} }{\partial t} +\frac{\partial \boldsymbol{\mathrm{v}}_{h} }{\partial t} +\frac{1}{2a} \left(\nabla \times \boldsymbol{\mathrm{v}}\right)\times \boldsymbol{\mathrm{v}}.  
\end{equation} 
With $\nabla \times \boldsymbol{\mathrm{v}}_{h} =0$ and $\nabla \times \boldsymbol{\mathrm{v}}_{r} =0$, both $\boldsymbol{\mathrm{v}}_{r} $ and $\boldsymbol{\mathrm{v}}_{h} $ are of irrotational nature that can be expressed as gradient of a scalar field (velocity potential) such that 
\begin{equation} 
\label{eq:213} 
\begin{split}
\frac{\partial \boldsymbol{\mathrm{v}}_{r} }{\partial t} +\frac{\partial \boldsymbol{\mathrm{v}}_{h} }{\partial t}=-\frac{1}{a} \nabla \frac{\partial }{\partial t} \left(\phi _{r} +\phi _{h} \right)=-\frac{1}{a} \nabla \phi ^{*},
\end{split}
\end{equation} 
where $\phi _{r}$ and $\phi _{h}$ are the velocity potential for intra-halo radial flow $\boldsymbol{\mathrm{v}}_{r}$ and halo motion $\boldsymbol{\mathrm{v}}_{h}$ and total potential $\phi^{*}={\partial (\phi _{r} +\phi _{h} )}/{\partial t}$.

The final complete set of self-closed dynamic equations for velocity field on small scales reads,
\begin{equation} 
\label{ZEqnNum598457} 
\begin{split}
&\nabla \cdot \boldsymbol{\mathrm{v}}=\theta \left(t\right)\equiv-3Ha\\ 
&\textrm{and}\\
&\frac{\partial \boldsymbol{\mathrm{v}}}{\partial t} +\frac{1}{a} \boldsymbol{\mathrm{v}}\cdot \nabla \boldsymbol{\mathrm{v}}+H\boldsymbol{\mathrm{v}}=-\frac{1}{a} \nabla \phi ^{*} +\gamma \frac{1}{a} \underbrace{\left(\nabla \times \boldsymbol{\mathrm{v}}\right)\times \boldsymbol{\mathrm{v}}}_{1},
\end{split}
\end{equation} 
where we have four equations for velocity $\boldsymbol{\mathrm{v}}$ and potential $\phi^{*}$ with a constant (in space) divergence $\theta={-3Ha}$. A trivial solution that satisfies Eq. \eqref{ZEqnNum598457} can be obtained for virialized spherical haloes:
\begin{equation} 
\label{ZEqnNum598457-2} 
\begin{split}
&v_x=-Hax-\omega y, \quad v_y=-Hay+\omega x, \quad v_z = -Haz,\\
&\phi^* = F(t)-H^2a^2(x^2+y^2+z^2)/4,
\end{split}
\end{equation} 
where $\omega$ is an arbitrary angular velocity of haloes around $z$ axis, $F(t)$ is an integration constant. With appropriate boundary and initial conditions, Eq. \eqref{ZEqnNum598457} can be solved numerically for the velocity field $\boldsymbol{\mathrm{v}}$ and the potential field $\phi ^{*} $. Since dark matter is assumed to be collisionless, there is no viscosity in Eq. \eqref{ZEqnNum598457}. Dynamic relations for correlations of different orders can be obtained from Eq. \eqref{ZEqnNum598457}.

\begin{figure}
\includegraphics*[width=\columnwidth]{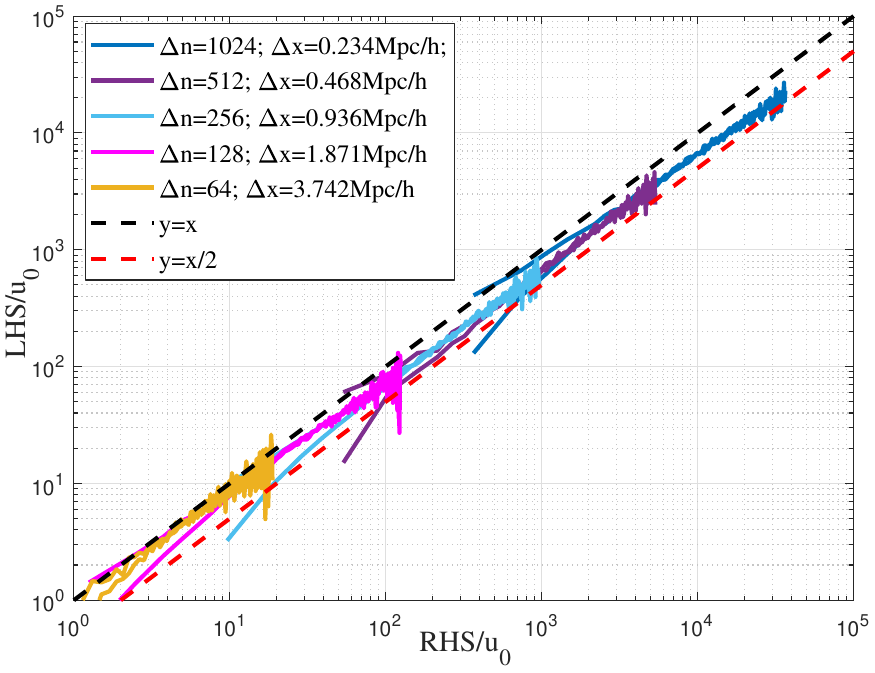}
\caption{The value of parameter $\gamma$ on different scales determined by Eq. \eqref{ZEqnNum312598} from N-body simulations at $z=0$. A small shift from $\gamma =1$ on large scales (yellow line) to $\gamma =0.5$ on small scales (blue line) can be clearly identified. The value of $\gamma =1$ indicates the irrotational flow on large scales. $\gamma =0.5$ corresponds to the constant divergence flow on small scales.}
\label{fig:16}
\end{figure}

Here, the coefficient $\gamma$ depends on the scale. For velocity on small scales $\gamma ={1/2}$ (comparing Eqs. \eqref{ZEqnNum598457} and \eqref{eq:212}). In principle, $\gamma$ can be determined from the simulation (Fig. \ref{fig:16}). Term 1 in Eq. \eqref{ZEqnNum598457} represents the centripetal acceleration from the mean flow $\boldsymbol{\mathrm{v}}$.

Using the identity 
\begin{equation} 
\label{eq:215} 
\nabla \cdot \left(\boldsymbol{\mathrm{v}}\otimes \boldsymbol{\mathrm{v}}\right)=\boldsymbol{\mathrm{v}}\left(\nabla \cdot \boldsymbol{\mathrm{v}}\right)+\boldsymbol{\mathrm{v}}\cdot \nabla \boldsymbol{\mathrm{v}},         
\end{equation} 
and the identity in Eq. \eqref{ZEqnNum683197}, Eq. \eqref{ZEqnNum598457} can be equivalently transformed into other forms for convenience of numerical solutions,
\begin{equation}
\label{ZEqnNum657666} 
\frac{\partial \boldsymbol{\mathrm{v}}}{\partial t} +\frac{1}{a} \left(1-\gamma \right)\boldsymbol{\mathrm{v}}\cdot \nabla \boldsymbol{\mathrm{v}}+\frac{\gamma }{2a} \nabla \left(\boldsymbol{\mathrm{v}}\cdot \boldsymbol{\mathrm{v}}\right)+H\boldsymbol{\mathrm{v}}=-\frac{1}{a} \nabla \phi ^{*} ,      
\end{equation} 
and
\begin{equation} 
\label{ZEqnNum793273} 
\begin{split}
\frac{\partial \boldsymbol{\mathrm{v}}}{\partial t}&+\frac{1}{a} \left(1-\gamma \right)\nabla \cdot \left(\boldsymbol{\mathrm{v}}\otimes \boldsymbol{\mathrm{v}}\right)\\
&+\frac{\gamma }{2a} \nabla \left(\boldsymbol{\mathrm{v}}\cdot \boldsymbol{\mathrm{v}}\right)+\left[H-\frac{1}{a} \left(1-\gamma \right)\theta \right]\boldsymbol{\mathrm{v}}=-\frac{1}{a} \nabla \phi ^{*}.
\end{split}
\end{equation} 
By setting $\gamma =1$ and 
\begin{equation} 
\label{eq:218} 
\nabla \phi ^{*} =-\frac{3}{2} Ha\boldsymbol{\mathrm{v}}-a\nu \nabla ^{2} \boldsymbol{\mathrm{v}}=-\frac{3}{2} Ha\boldsymbol{\mathrm{v}}-a\nu \nabla \left(\nabla \cdot \boldsymbol{\mathrm{v}}\right),       
\end{equation} 
Eqs. \eqref{ZEqnNum657666} and \eqref{ZEqnNum793273} reduce to the dynamic equation on large scales (Eq. \eqref{ZEqnNum284543}). Setting the parameter $\gamma =0$, Eq. \eqref{ZEqnNum657666} reduces to the inviscid Euler equation. 

The equation for the vorticity field $\boldsymbol{\mathrm{\omega }}=\nabla \times \boldsymbol{\mathrm{v}}$ can be obtained by taking the curl on both sides of Eqs. \eqref{ZEqnNum598457} and \eqref{ZEqnNum657666}, 
\begin{equation}
\label{ZEqnNum312598} 
\underbrace{\frac{\partial \boldsymbol{\mathrm{\omega }}}{\partial t} +\frac{1}{a} \nabla \times \left(\boldsymbol{\mathrm{v}}\cdot \nabla \boldsymbol{\mathrm{v}}\right)+H\boldsymbol{\mathrm{\omega }}}_{LHS}=\gamma \underbrace{\frac{1}{a} \nabla \times \left[\left(\nabla \times \boldsymbol{\mathrm{v}}\right)\times \boldsymbol{\mathrm{v}}\right]}_{RHS},      
\end{equation} 
and
\begin{equation} 
\label{ZEqnNum550342} 
\frac{\partial \boldsymbol{\mathrm{\omega }}}{\partial t} +H\boldsymbol{\mathrm{\omega }}=\frac{1}{a} \left(\gamma -1\right)\nabla \times \left(\boldsymbol{\mathrm{v}}\cdot \nabla \boldsymbol{\mathrm{v}}\right).        
\end{equation} 
For $\gamma =1$ on large scales in Eq. \eqref{ZEqnNum550342}, we found that vorticity $\omega\propto a^{-1}$ decays over time and there is no source to generate vorticity on large scales. However, on small scales with $\gamma=1/2$, vorticity is continuously generated (see Eq. \eqref{ZEqnNum968516}). Both equations can be used to determine the value of $\gamma $ from the N-body simulations. 

Figure \ref{fig:16} plots the variation of $\gamma $ with scale (or the size of the grid $\Delta x$). The particle velocity field was projected onto grids of size $\Delta x$ using Cloud-in-Cell (CIC) \citep{Hockney:1988-Computer-Simulation-Using-Part}, or equivalently $\Delta x\cdot \Delta n=L$, where $\Delta n$ is the number of grids and $L=239.5{Mpc/h} $ is the size of the simulation domain. The LHS and RHS terms in Eq. \eqref{ZEqnNum312598} were computed using the projected velocity field and plotted in Fig. \ref{fig:16}. The slope gives the value of $\gamma $ from the simulation. The two dashed lines for $\gamma =1$ and $\gamma ={1/2} $ are also plotted for comparison. For large grid size $\Delta x$ (large scales), $\gamma $ approaches 1. However, for a small grid size $\Delta x$ (small scales), $\gamma $ approaches 1/2. For the entire range of scales,  ${1/2} \le \gamma \le 1$ is expected with a smooth transition from $\gamma=1/2$ on small scales to $\gamma=1$ on large scales.

\subsubsection{Equations for the evolution of vorticity and enstrophy}
\label{sec:7.1.3}
The vorticity field is important on small scales, as every halo has finite spin. With identity in Eq. \eqref{ZEqnNum683197} and the identity for two arbitrary vectors \textbf{A} and \textbf{B,}
\begin{equation} 
\label{eq:227} 
\nabla \times \left(\boldsymbol{\mathrm{{\rm A} }}\times \boldsymbol{\mathrm{B}}\right)=\boldsymbol{\mathrm{A}}\nabla \cdot \boldsymbol{\mathrm{B}}-\boldsymbol{\mathrm{B}}\nabla \cdot \boldsymbol{\mathrm{A}}+\boldsymbol{\mathrm{B}}\cdot \nabla \boldsymbol{\mathrm{A}}-\boldsymbol{\mathrm{A}}\cdot \nabla \boldsymbol{\mathrm{B}},      
\end{equation} 
Eq. \eqref{ZEqnNum550342} can be transformed to ($\nabla \cdot \boldsymbol{\mathrm{\omega }}=0$):  
\begin{equation}
\label{ZEqnNum968516} 
\frac{\partial \boldsymbol{\mathrm{\omega }}}{\partial t} +\frac{1-\gamma }{a} \underbrace{\boldsymbol{\mathrm{v}}\cdot \nabla \boldsymbol{\mathrm{\omega }}}_{1}+\left[1+\frac{\left(1-\gamma \right)}{Ha}\theta  \right]\underbrace{H\boldsymbol{\mathrm{\omega }}}_{2}=\frac{1-\gamma }{a} \underbrace{\boldsymbol{\mathrm{\omega }}\cdot \nabla \boldsymbol{\mathrm{v}}}_{3}.     
\end{equation} 
From the identity \eqref{ZEqnNum248930}, the velocity field can be expressed in terms of the vorticity field for the constant divergence flow, 
\begin{equation} 
\label{eq:229} 
\nabla ^{2} \boldsymbol{\mathrm{v}}=-\nabla \times \boldsymbol{\mathrm{\omega }}.          
\end{equation} 
For an infinite domain, this can be solved by Green's function,
\begin{equation} 
\label{eq:230} 
\boldsymbol{\mathrm{v}}\left(\boldsymbol{\mathrm{x}}\right)=-\frac{1}{4\pi } \int \frac{\left[\nabla \times \boldsymbol{\mathrm{\omega }}\right]^{'} }{\left|\boldsymbol{\mathrm{x}}-\boldsymbol{\mathrm{x}}^{'} \right|}  d\boldsymbol{\mathrm{x}}^{'} ,         
\end{equation} 
where vorticity is a local quantity, but velocity field can be non-local. The velocity field is affected by the vorticity from other locations. 

Three terms in Eq. \eqref{ZEqnNum968516} represent the transport of vorticity (term 1), the decaying of vorticity due to expanding background (term 2), and vorticity generation (term 3) similar to vortex stretching in hydrodynamic turbulence \citep[see][Eq. (1)]{Xu:2021-Inverse-and-direct-cascade-of-}. On large scales, $\gamma \approx 1$ and the dominant mode is the decay of vorticity, whereas vorticity is strongly generated with $\gamma \to {1/2} $ on small scales. For comparison, the vorticity equation for an incompressible flow is
\begin{equation} 
\label{eq:231} 
\frac{\partial \boldsymbol{\mathrm{\omega }}}{\partial t} +\boldsymbol{\mathrm{v}}\cdot \nabla \boldsymbol{\mathrm{\omega }}=\boldsymbol{\mathrm{\omega }}\cdot \nabla \boldsymbol{\mathrm{v}}+\nu \nabla ^{2} \boldsymbol{\mathrm{\omega }},         
\end{equation} 
where $\nu $ is the viscosity leading to the destruction of vorticity. 

Finally, the evolution of enstrophy can be obtained by the scalar product of Eq. \eqref{ZEqnNum968516} with vorticity vector $\boldsymbol{\mathrm{\omega }}$,
\begin{equation} 
\label{ZEqnNum517069} 
\begin{split}
\frac{\partial {\boldsymbol{\mathrm{\omega }}^{2}/2} }{\partial t} +\frac{1-\gamma }{a} \underbrace{\boldsymbol{\mathrm{v}}\cdot \nabla \frac{\boldsymbol{\mathrm{\omega }}^{2} }{2} }_{1}&+\left[1+\left(1-\gamma \right)\frac{\theta }{Ha} \right]\underbrace{H\boldsymbol{\mathrm{\omega }}^{2} }_{2}\\
&=\frac{1-\gamma }{a} \underbrace{\boldsymbol{\mathrm{\omega }}\cdot \left(\boldsymbol{\mathrm{\omega }}\cdot \nabla \boldsymbol{\mathrm{v}}\right)}_{3},
\end{split}
\end{equation} 
where the enstrophy is generated on the small scale (term 3), transported (term 1) and decaying on large scales (term 2).

\subsection{Dynamic relations from dynamics on small scales}
\label{sec:7.2}
In this section, we are ready to develop dynamic relations on small scales. Starting from the index notation of dynamic Eq. \eqref{ZEqnNum793273} at two different locations $\boldsymbol{\mathrm{x}}$ and $\boldsymbol{\mathrm{x}}^{'} $,
\begin{equation} 
\label{eq:244} 
\begin{split}
\frac{\partial v_{i} }{\partial t} +\frac{1-\gamma }{a} \frac{\partial \left(v_{i} v_{k} \right)}{\partial x_{k} } &+\frac{\gamma }{2a} \frac{\partial \left(v_{k} v_{k} \right)}{\partial x_{i} }\\ 
&+\left[1-\frac{\left(1-\gamma \right)}{aH} \theta \right]Hv_{i} =-\frac{1}{a} \frac{\partial \phi ^{*} }{\partial x_{i} },
\end{split}
\end{equation} 
\begin{equation} 
\label{eq:245} 
\begin{split}
\frac{\partial v_{j}^{'} }{\partial t} +\frac{1-\gamma }{a} \frac{\partial \left(v_{j}^{'} v_{k}^{'} \right)}{\partial x_{k}^{'} } 
&+\frac{\gamma }{2a} \frac{\partial \left(v_{k}^{'} v_{k}^{'} \right)}{\partial x_{j}^{'} }\\
&+\left[1-\frac{\left(1-\gamma \right)}{aH} \theta \right]Hv_{j}^{'} =-\frac{1}{a} \frac{\partial \phi ^{*'} }{\partial x_{j}^{'} },     
\end{split}
\end{equation} 
multiplying $v_{j}^{'} $ and $v_{i} $ to both sides of two equations, adding them together, and taking average at a given scale \textit{r} leads to
\begin{equation} 
\label{eq:246} 
\begin{split}
&\frac{\partial \left\langle v_{i} v_{j}^{'} \right\rangle }{\partial t} +\frac{1-\gamma }{a} \left(\frac{\partial \left\langle v_{i} v_{k} v_{j}^{'} \right\rangle }{\partial x_{k} } +\frac{\partial \left\langle v_{j}^{'} v_{k}^{'} v_{i} \right\rangle }{\partial x_{k}^{'} } \right)\\
&+\frac{\gamma }{2a} \left(\frac{\partial \left\langle v_{k} v_{k} v_{j}^{'} \right\rangle }{\partial x_{i} }+\frac{\partial \left\langle v_{k}^{'} v_{k}^{'} v_{i} \right\rangle }{\partial x_{j}^{'} } \right)\\
&+2\left[1-\frac{\left(1-\gamma \right)}{aH} \theta \right]H\left\langle v_{i} v_{j}^{'} \right\rangle=-\frac{1}{a} \left(\frac{\partial \left\langle \phi ^{*} v_{j}^{'} \right\rangle }{\partial x_{i} } +\frac{\partial \left\langle \phi _{}^{*'} v_{i} \right\rangle }{\partial x_{j}^{'} } \right), 
\end{split}
\end{equation} 
or in terms of the second and third order velocity correlation tensors (see definition in Eq. \eqref{ZEqnNum883006}),
\begin{equation} 
\label{ZEqnNum539624} 
\begin{split}
&\frac{\partial Q_{ij} }{\partial t}+2\left[1-\frac{\left(1-\gamma \right)}{aH} \theta \right]HQ_{ij}\\
&=\frac{2-2\gamma }{a} \frac{\partial Q_{ikj} }{\partial r_{k} } +\frac{\gamma }{a} \frac{\partial Q_{kkj} }{\partial r_{i} }-\frac{1}{a} \left[\frac{\partial \left\langle \phi ^{*} v_{j}^{'} \right\rangle }{\partial x_{i} } +\frac{\partial \left\langle \phi ^{*'} v_{i} \right\rangle }{\partial x_{j}^{'} } \right].
\end{split}
\end{equation} 

Again, second order correlation tensor $Q_{ij}$ is related to the third order correlation tensor $Q_{ijk}$. Multiplying both sides by $\delta _{ij} $, using Eqs. \eqref{ZEqnNum952051} and \eqref{ZEqnNum671139} for third-order correlations with $R_{3} \equiv R_{(3,1)} $ and $R_{31} \equiv L_{(3,2)} $, and using the fact that the first-order correlation tensor for constant divergence flow \citep[see][Eq. (9)]{Xu:2023-On-the-statistical-theory-of-self-gravitating} satisfying,
\begin{equation} 
\label{eq:248} 
\frac{\partial \left\langle \phi ^{*} v_{j}^{'} \right\rangle }{\partial x_{j} } =-\frac{\partial \left\langle \phi ^{*} v_{j}^{'} \right\rangle }{\partial r_{j} } =-\theta \left\langle \phi ^{*} \right\rangle,     
\end{equation} 
the evolution of second order correlation on small scales reads
\begin{equation} 
\label{ZEqnNum168148} 
\begin{split}
&\frac{\partial R_{\left(2,1\right)} }{\partial t} +2\left[1-\frac{\left(1-\gamma \right)}{aH} \theta \right]HR_{\left(2,1\right)}\\
&=\frac{1}{ar^{2} } \left[\frac{\partial }{\partial r} \left(r^{2} \left[\left(2-2\gamma \right)R_{\left(3,1\right)} +\gamma L_{\left(3,2\right)} \right]\right)\right]+\frac{2}{a} \theta \left\langle \phi ^{*} \right\rangle,
\end{split}
\end{equation} 
where third order correlation functions are (Eqs. \eqref{ZEqnNum952051} and \eqref{ZEqnNum671139})
\begin{equation}
\frac{\partial Q_{iki} }{\partial r_{k} } =\frac{1}{r^{2} } \frac{\partial }{\partial r} \left(r^{2} R_{\left(3,1\right)} \right), \quad \frac{\partial Q_{kkj} }{\partial r_{j} } =\frac{1}{r^{2} } \frac{\partial }{\partial r} \left(r^{2} L_{\left(3,2\right)} \right). 
\label{eq:250}
\end{equation}
\noindent Equation \eqref{ZEqnNum168148} provides a dynamic relation between second and third order correlation functions (similar to Eq. \eqref{ZEqnNum923677} on large scales), which can be integrated and equivalently transformed to
\begin{equation} 
\label{ZEqnNum330309} 
\begin{split}
&\frac{-\left\langle \phi ^{*} \right\rangle }{u^{2} } = \frac{3}{2\theta ru^{2} } \left[\left(2-2\gamma \right)R_{\left(3,1\right)} +\gamma L_{\left(3,2\right)} \right]\\
&+\left[\left(1-\gamma \right)\frac{2\theta }{aH} -2-\frac{\partial \ln R_{\left(2,1\right)} }{\partial \ln a} \right]\frac{3Ha}{2u^{2} \theta r^{3} } \int _{0}^{r}R_{\left(2,1\right)}  \left(y\right)y^{2}dy. 
\end{split}
\end{equation} 
On small scales, the function $R_{\left(2,1\right)} $ can be modeled as (Eq. \eqref{ZEqnNum955991}),
\begin{equation}
R_{\left(2,1\right)} =u^{2} \left[3-\left(3+n\right)\left(\frac{r}{r_{1} } \right)^{n} \right] \quad \textrm{and} \quad \frac{\partial \ln R_{\left(2,1\right)} }{\partial \ln a} =\frac{3}{2},   
\label{eq:252}
\end{equation}
\noindent with $u^{2} \sim t$, where $u^{2} $ is the one-dimensional velocity dispersion of entire system and $n\approx {1/4}$ (Eq. \eqref{eq:189}). The correlation function $R_{(2,1)} \propto a$ on large scales and $R_{(2,1)} \propto a^{{3/2}}$ on small scales. 

On small scales, we expect $\gamma ={1/2} $ and $\theta ={-3Ha/2} $, and relation $\langle u^{2} \rangle +\beta ^{*} \langle \phi ^{*} \rangle =0$, Eq. \eqref{ZEqnNum330309} reduces to
\begin{equation} 
\label{ZEqnNum443061} 
\begin{split}
&\frac{-\left\langle \phi ^{*} \right\rangle }{u^{2} } =\frac{\left\langle u^{2} \right\rangle }{\beta ^{*} u^{2} }\\
&=\underbrace{\frac{5}{u^{2} r^{3}} \int _{0}^{r}R_{\left(2,1\right)}  \left(y\right)y^{2} dy}_{1}\underbrace{-\frac{1}{Haru^{2} } \left(R_{\left(3,1\right)} +\frac{1}{2} L_{\left(3,2\right)} \right)}_{2},
\end{split}
\end{equation} 
where $\beta ^{*} $ is a virial ratio on scale \textit{r}. 

\begin{figure}
\includegraphics*[width=\columnwidth]{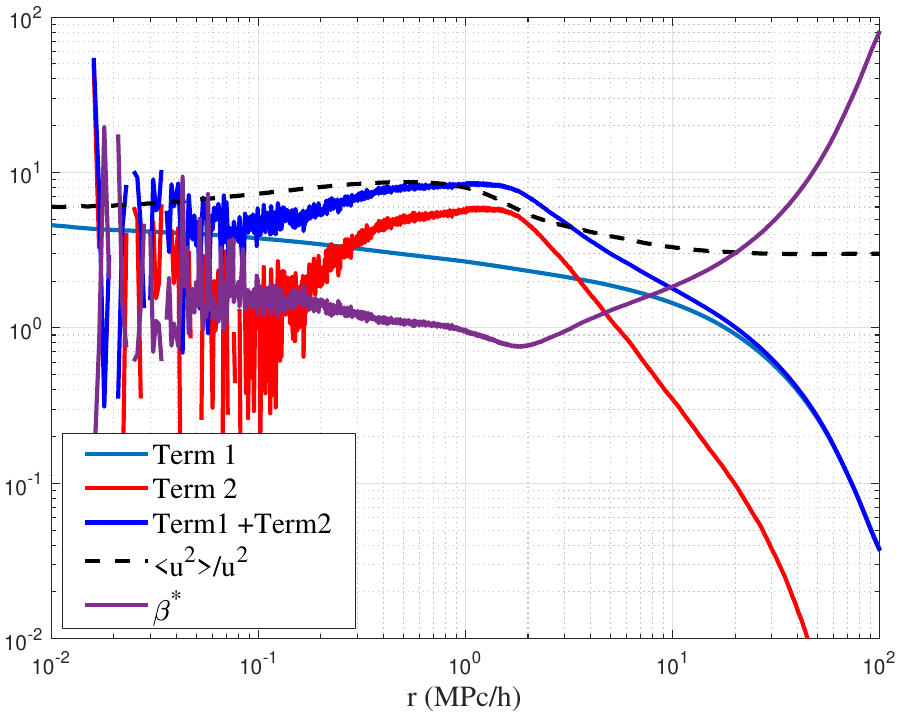}
\caption{The variation of different terms in Eq. \eqref{ZEqnNum443061} with scale \textit{r} at \textit{z}=0. On small scales, term 1 approaches 5 and term2 approaches 1. The sum of term1 and term2 approaches the velocity dispersion ${\left\langle u^{2} \right\rangle /u^{2} } $. The variation of the virial coefficient $\beta ^{*} $ with scale \textit{r} (slightly decreasing with \textit{r}) is also presented with ${\left\langle u^{2} \right\rangle/u^{2}} $ directly obtained from simulation and Eq. \eqref{ZEqnNum443061}.}
\label{fig:17}
\end{figure}

To validate Eq. \eqref{ZEqnNum443061}, different terms were obtained directly from N-body simulations and plotted in Fig. \ref{fig:17}. The sum of terms 1 and 2 in the RHS of Eq. \eqref{ZEqnNum443061} approaches LHS on small scales. The variation of $\beta ^{*} $ with scale \textit{r} (slightly decreasing with \textit{r}) is also presented with ${\left\langle u^{2} \right\rangle/u^{2} } $ directly obtained from simulation. For $r\to 0$, we have ${\mathop{\lim }\limits_{r\to 0}} R_{(2,1)} ={\mathop{\lim }\limits_{r\to 0}} \langle \boldsymbol{\mathrm{u}}\cdot \boldsymbol{\mathrm{u}}^{'} \rangle =3u^{2} $ and ${\mathop{\lim }\limits_{r\to 0}} \langle u^{2} \rangle ={\mathop{\lim }\limits_{r\to 0}} \langle \boldsymbol{\mathrm{u}}\cdot \boldsymbol{\mathrm{u}}\rangle =6u^{2}$, where $\langle u^{2} \rangle \approx 6u^{2} $ on small scales (Fig. \ref{fig:20}) and $\beta ^{*} \approx 1$, term 1 in the RHS of Eq. \eqref{ZEqnNum443061} approaches 5 such that term 2 becomes
\begin{equation}
\label{ZEqnNum558518} 
\left(R_{\left(3,1\right)} +\frac{1}{2} L_{\left(3,2\right)} \right)=-Hau^{2} r=\left\langle \Delta u_{L} \right\rangle u^{2} =\frac{4}{9} \varepsilon _{u} ar,      
\end{equation} 
where the rate of kinetic energy cascade $\varepsilon _{u}$ is negative (inverse energy cascade) and relatively a constant of time \citep{Xu:2022-Postulating-dark-matter-partic}, 
\begin{equation} 
\label{eq:255} 
\varepsilon _{u} =-\frac{3}{2} \frac{\partial u^{2} }{\partial t} \approx -\frac{3}{2} \frac{u^{2} }{t} =-\frac{3}{2} \frac{u_{0}^{2} }{t_{0} } \approx -4.6\times 10^{-7} {m^{2}/s^{3} } .      
\end{equation} 

\begin{figure}
\includegraphics*[width=\columnwidth]{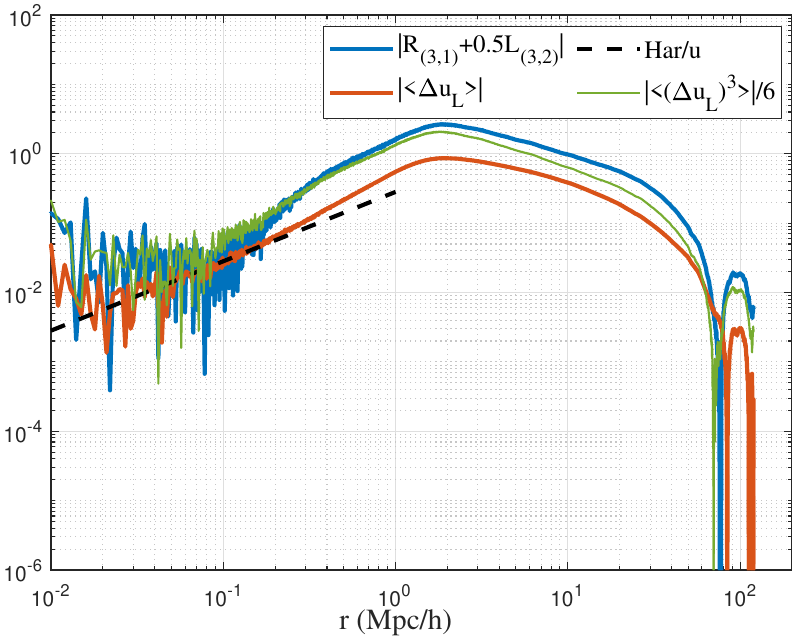}
\caption{The variation of correlation function $R_{(3,1)} +{L_{(3,2)}/2}$ at $z=0$ (normalized by $u_0^3$), the pairwise velocity (first order longitudinal structure function) $\langle \Delta u_{L} \rangle$ (normalized by $u_0$), and the third order longitudinal structure function $\langle (\Delta u_{L} )^{3} \rangle $ (normalized by $u_0^3$) with scale r (from Eqs. \eqref{ZEqnNum558518} and \eqref{ZEqnNum260646}). All functions are normalized, and $\langle (\Delta u_{L} )^{3} \rangle ={8\varepsilon _{u} ar/3}$.}
\label{fig:18}
\end{figure}

The velocity dispersion in the present epoch is about $u_{0} \approx 354.61{km/s} $ in the current N-body simulation. The third moment of pairwise velocity (third order structure function) %in Eq. \eqref{ZEqnNum772554}) 
satisfies from the generalized stable clustering hypothesis (GSCH \citep{Xu:2021-A-non-radial-two-body-collapse}),
\begin{equation} 
\label{eq:256} 
\left\langle \left(\Delta u_{L} \right)^{3} \right\rangle =3\left\langle \left(\Delta u_{L} \right)^{2} \right\rangle \left\langle \Delta u_{L} \right\rangle ,         
\end{equation} 
and ${\mathop{\lim }\limits_{r\to 0}} \left\langle \left(\Delta u_{L} \right)^{2} \right\rangle \approx 2u^{2} $, %\citep[see][Fig. 21]{Xu:2022-Two-thirds-law-for-pairwise-ve}, 
we may write  
\begin{equation} 
\label{ZEqnNum260646} 
\left(R_{\left(3,1\right)} +\frac{1}{2} L_{\left(3,2\right)} \right)=\frac{1}{6} \left\langle \left(\Delta u_{L} \right)^{3} \right\rangle =\left\langle \Delta u_{L} \right\rangle u^{2} =\frac{4}{9} \varepsilon _{u} ar,      
\end{equation} 
where the third moment of pairwise velocity determines the rate of energy production on the smallest scale, i.e. 
\begin{equation}
\left\langle \left(\Delta u_{L} \right)^{3} \right\rangle =\frac{8}{3} \varepsilon _{u} ar \quad \textrm{or} \quad \varepsilon _{u} =\frac{3}{8} \frac{\left\langle \left(\Delta u_{L} \right)^{3} \right\rangle }{ar}. 
\label{eq:258}
\end{equation}
Figure \ref{fig:18} presents the relevant correlation and structure functions in Eqs. \eqref{ZEqnNum558518} and \eqref{ZEqnNum260646}. On small scales, all these functions can be directly related to the rate of the energy cascade $\varepsilon _{u}$. 

\begin{figure}
\includegraphics*[width=\columnwidth]{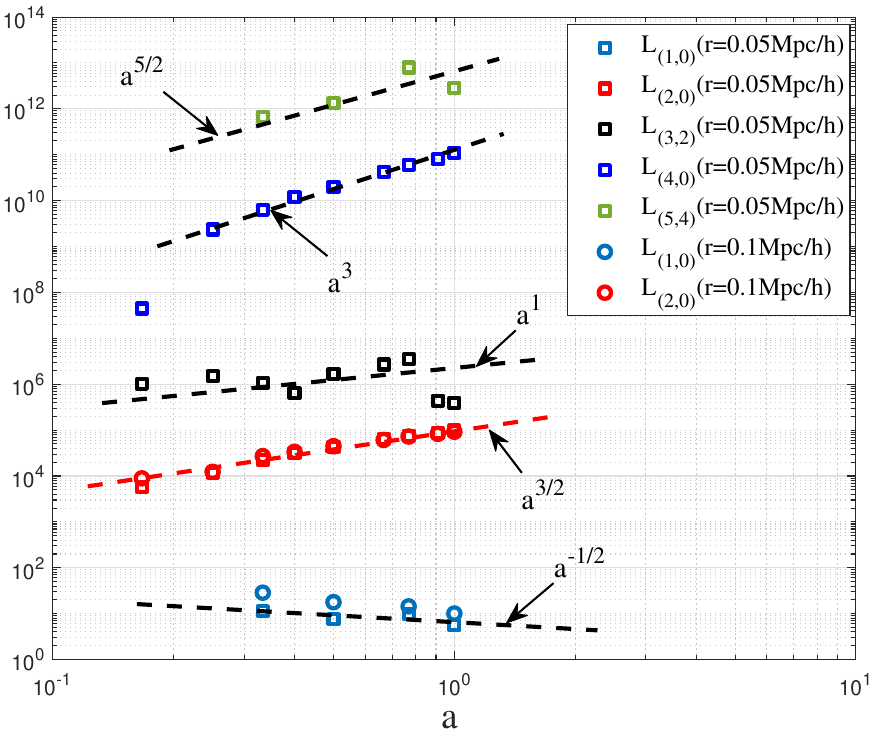}
\caption{The variation of the velocity correlation functions $L_{(p,q)}(r)$ (normalized by (km/s)$^p$) of different order $p$ with scale factor $a$ at a fixed small scale $r=0.05Mpc/h$. These correlation functions follow $\propto a^{(3p-5)/4}$ for odd order $p$ and $\propto a^{p/2}$ for even order $p$. To ensure that results are not affected by numerical softening effect, the correlation functions ($L_{(1,0)}$ and $L_{(2,0)}$) at a different scale $r=0.1Mpc/h$ are also presented as circles.}
\label{fig:23}
\end{figure}

Finally, the time evolution of the velocity correlation functions $L_{(p,q)}$ on small scales can be obtained from the dynamic relation in Eq. \eqref{ZEqnNum168148}, which follows $\propto a^{(3p-5)/4}$ for odd order $p$ and $\propto a^{p/2}$ for even order $p$. Figure \ref{fig:23} presents the variation of the correlation function $L_{(p,q)}(r)$ of different order $p$ at a fixed $r=0.05Mpc/h$. All correlation functions are normalized by (km/s)$^p$. To ensure that results are not affected by the numerical softening effect, the correlation functions ($L_{(1,0)}$ and $L_{(2,0)}$) on a different scale $r=0.1Mpc/h$ are also presented as circles. The results of N-body simulations confirm the predicted time evolution of the velocity correlation functions on small scales. %We note that this is also consistent with the generalized stable clustering hypothesis \citep{Xu:2021-A-non-radial-two-body-collapse}
%\begin{equation}
%\left\langle \left(\Delta u_{L} \right)^{2m+1} \right\rangle=(2m+1)\left\langle \left(\Delta u_{L} \right)^{2m} \right\rangle \left\langle \left(\Delta u_{L} %\right)\right\rangle 
%\label{eq:258-2}
%\end{equation}

\section{Conclusions}
\label{sec:8}
A statistical theory is presented for the self-gravitating collisionless dark matter flow (SG-CFD), which is constant divergence on small scales and is irrotational on large scales. The size of the largest halo leads to a critical transition scale $r_t$ between two flow regimes (Fig. \ref{fig:19}). To understand the nature of flow across the entire range of scales, kinematic and dynamic relations for statistical measures of different orders need to be developed for different types of flow. 

By extending the two-point second order statistics in our previous work \citep{Xu:2023-On-the-statistical-theory-of-self-gravitating} to high order, we present the third order statistical measures and associated kinematic relations for different types of flow. In principle, the incompressible and constant divergence flows share the same kinematic relations for even-order correlations, while they can be different for odd-order correlations (Eq. \eqref{M_ZEqnNum365670}). For the third order velocity correlation tensor $Q_{ijk}(r)$, four scalar correlation functions (total correlations $R_{3} $ and $R_{31} $, longitudinal correlation $L_{3} $, and transverse correlation $T_{3}$) can be obtained by contraction of the indices from $Q_{ijk}(r)$ (Eqs. \eqref{ZEqnNum550536}-\eqref{ZEqnNum796638}). Kinematic relations are developed for incompressible flow (Eqs. \eqref{ZEqnNum374800} and \eqref{ZEqnNum218978}), constant divergence flow (Eqs. \eqref{ZEqnNum647632} and \eqref{ZEqnNum487892}), and irrotational flow (Eq. \eqref{ZEqnNum337667}). The correlation functions from the N-body simulation are presented in Figs. \ref{fig:3} and \ref{fig:4} for the third and fourth order, respectively. 

To formulate kinematic relations of any order, a compact derivation involving high-order tensor and vector calculus is presented. This is a challenging task (see Appendix \ref{appendix:A}). Starting from reformulating the kinematic relations of second and third order, general kinematic relations of any order are developed, i.e. Eqs. \eqref{M_ZEqnNum234089}-\eqref{M_ZEqnNum533624} for incompressible flow, Eq. \eqref{M_ZEqnNum591192} for constant divergence on small scales, and Eqs. \eqref{M_ZEqnNum900611}-\eqref{M_ZEqnNum433592} for irrotational flow on large scales. To validate these relations by N-body simulations, the original differential equations are transformed to integral form (Eqs. \eqref{ZEqnNum167557} to \eqref{ZEqnNum467352}). The comparison is presented in Figs. \ref{fig:6} and \ref{fig:7} with good agreement.

The dynamic relations between correlation functions of different orders can only be determined from the dynamic equation of velocity evolution on relevant scales. On large scales, the Zeldovich and adhesion approximations govern the dynamics of velocity (Eq. \eqref{ZEqnNum709847}). Third-order correlations are related to second-order velocity/density correlations and mean pairwise velocity ($L_{(3,2)} \propto -\nu \langle \Delta u_{L} \rangle$) through an effective viscosity $\nu(a)$ (Eqs. \eqref{ZEqnNum715845}, \eqref{ZEqnNum846484}, and \eqref{ZEqnNum216206}). The negative viscosity $\nu (a)\propto -ur_{2} \propto a^{{1/2}}$ originates from the inverse cascade of kinetic energy on scales below the transition scale $r_t$ (Eq. \eqref{ZEqnNum133069} and Fig. \ref{fig:12}). 

On large scales, the redshift dependence of the correlation functions on any order is obtained as $L_{(p,q)} \propto a^{{(p+2)/2}}$ for odd order $p$ and $L_{(p,q)} \propto a^{{p/2}}$ for even order $p$ (Eqs. \eqref{ZEqnNum995688}-\eqref{ZEqnNum252855} and Figs. \ref{fig:8}-\ref{fig:12}). The velocity divergence can be determined by the mean pairwise velocity (Eq. \eqref{ZEqnNum732704}). The mean overdensity on a given scale \textit{r} is proportional to the density correlation $f(\Omega _{m})\langle \delta \rangle \approx \langle \delta \delta ^{'} \rangle $ (Eq. \eqref{ZEqnNum711863} and Fig. \ref{fig:13})). An excess velocity dispersion is also proportional to the density correlation, that is, ${\langle u^{2} \rangle/(3u^{2})} -1\propto \langle \delta \delta ^{'} \rangle $ (Eq. \eqref{ZEqnNum864804} and Fig. \ref{fig:13})). The low density void region with $\langle \delta \rangle<0$ can be related to the negative density correlation $\langle\delta\delta{'}\rangle<0$ on scales around 30Mpc/h. The velocity dispersion in the void region $\langle u^{2} \rangle$ is less than the asymptotic dispersion $3u^2$. Finally, both exponential velocity correlations on large scales and the "one-fourth" scaling law for correlations on small scales are direct results of combined dynamics and kinematic relations (Eqs. \eqref{ZEqnNum445071}, \eqref{ZEqnNum731622}, and \eqref{eq:189}). 

On small scales, a self-closed equation for velocity evolution is developed to derive dynamic relations. Based on solutions for virialized haloes, the self-closed dynamic equation includes an additional term to reflect the effect of local vorticity (term 1 in Eq. \eqref{ZEqnNum598457}) that is controlled by a parameter $\gamma $. From large to small scales, $\gamma $ gradually changes from 1 to 1/2 (Fig. \ref{fig:16}). Vorticity and enstrophy evolution are all derived from the self-closed dynamic equation (Eqs. \eqref{ZEqnNum968516}, \eqref{ZEqnNum517069}).%, \eqref{ZEqnNum921459}, and \eqref{ZEqnNum183585}
The dynamic relations are derived to relate second and third order correlations in Eq. \eqref{ZEqnNum168148}. The same relation in integral form (Eqs. \eqref{ZEqnNum330309} and \eqref{ZEqnNum443061}) can be directly validated by N-body simulations (Fig. \ref{fig:17}). Third-order velocity correlations are related to the energy production rate $\varepsilon_{u} $ (Eq. \eqref{ZEqnNum260646} and Fig. \ref{fig:18}) and the third moment of pairwise velocity $\langle (\Delta u_{L} )^{3} \rangle ={8\varepsilon _{u} ar/3} $ (Eq. \eqref{eq:258}). Finally, the redshift dependence of velocity correlations on small scales satisfies $L_{(p,q)} \propto a^{{(3p-5)/4}}$ for odd order $p$ and $L_{(p,q)} \propto a^{{3p/4}}$ for even order $p$ (Fig. \ref{fig:23}).

%\section*{Acknowledgements}
\section*{AUTHOR DECLARATIONS}
The authors have no conflicts to disclose.

\section*{Acknowledgements}
This research was supported by the Laboratory Directed Research and Development at Pacific Northwest National Laboratory (PNNL). PNNL is a multiprogram national laboratory operated for the U.S. Department of Energy (DOE) by Battelle Memorial Institute under contract no. DE-AC05-76RL01830.

\section*{Data Availability}
Two datasets underlying this article, that is, halo-based and correlation-based statistics of dark matter flow, are available on Zenodo \citep{Xu:2022-Dark_matter-flow-dataset-part1,Xu:2022-Dark_matter-flow-dataset-part2}, along with the accompanying presentation slides 'A comparative study of dark matter flow \& hydrodynamic turbulence and its applications' \citep{Xu:2022-Dark_matter-flow-and-hydrodynamic-turbulence-presentation}. All data are also available on GitHub \citep{Xu:Dark_matter_flow_dataset_2022_all_files}.

\bibliographystyle{Papers}
\bibliography{Papers}

\appendix
\addtocontents{toc}{\protect\setcounter{tocdepth}{1}}% This turns off subsections
\section{Kinematic relations for correlations of arbitrary order}
\label{appendix:A}
In this Appendix, a compact derivation for kinematic relations of arbitrary order is presented for incompressible, constant divergence, and irrotational flow.
\subsection{Introduction to some general identities}
\label{appendix:A1}
Let's first introduce some identities, which are the generalization of identities in Eq. \eqref{ZEqnNum679713},
\begin{equation} 
\label{ZEqnNum829225} 
\left(\hat{r}_{i} \right)_{,j} =\frac{\partial \hat{r}_{i} }{\partial r_{j} } =\frac{1}{r} \left(\delta _{ij} -\hat{r}_{i} \hat{r}_{j} \right)=\left(\hat{r}_{j} \right)_{,i} ,         
\end{equation} 
\begin{equation} 
\label{ZEqnNum105713} 
\left(\hat{r}_{i} \hat{r}_{j} \right)_{,k} =\frac{1}{r} \left(\hat{r}_{i} \delta _{jk} +\hat{r}_{j} \delta _{ik} -2\hat{r}_{i} \hat{r}_{j} \hat{r}_{k} \right),         
\end{equation} 
\begin{equation} 
\label{eq:46} 
\left(\hat{r}_{i} \hat{r}_{j} \hat{r}_{k} \right)_{,l} =\frac{1}{r} \left(\hat{r}_{i} \hat{r}_{j} \delta _{kl} +\hat{r}_{i} \hat{r}_{k} \delta _{jl} +\hat{r}_{j} \hat{r}_{k} \delta _{il} -3\hat{r}_{i} \hat{r}_{j} \hat{r}_{k} \hat{r}_{l} \right),       
\end{equation} 
\begin{equation} 
\label{ZEqnNum424097} 
\begin{split}
\left(\hat{r}_{i} \hat{r}_{j} \hat{r}_{k} \hat{r}_{l} \right)_{,m} &=\frac{1}{r} \left(\hat{r}_{i} \hat{r}_{j} \hat{r}_{l} \delta _{km} +\hat{r}_{i} \hat{r}_{k} \hat{r}_{l} \delta _{jm}\right.\\
&\left.+\hat{r}_{i} \hat{r}_{j} \hat{r}_{k} \delta _{lm} +\hat{r}_{j} \hat{r}_{k} \hat{r}_{l} \delta _{im} -4\hat{r}_{i} \hat{r}_{j} \hat{r}_{k} \hat{r}_{l} \hat{r}_{m} \right),
\end{split}
\end{equation} 
\begin{equation} 
\label{ZEqnNum669596} 
\left(\hat{r}_{i} \right)_{,j} \hat{r}_{j} =0, \left(\hat{r}_{i} \hat{r}_{j} \right)_{,k} \hat{r}_{k} =0,  \left(\hat{r}_{i} \hat{r}_{j} \hat{r}_{k} \right)_{,l} \hat{r}_{l} =0,       
\end{equation} 
\begin{equation} 
\label{ZEqnNum558872} 
\left(\hat{r}_{i} \right)_{,i} =\frac{2}{r} , \left(\hat{r}_{i} \hat{r}_{j} \right)_{,j} =\frac{2}{r} \hat{r}_{i} , \left(\hat{r}_{i} \hat{r}_{j} \hat{r}_{k} \right)_{,k} =\frac{2}{r} \hat{r}_{i} \hat{r}_{j} ,      
\end{equation} 
\begin{equation} 
\label{eq:50} 
\left(r_{i} \right)_{,j} =\delta _{ij} ,  \left(\hat{r}_{i} r_{j} \right)_{,j} =3\hat{r}_{i} ,  \left(\hat{r}_{i} \hat{r}_{j} r_{k} \right)_{,k} =3\hat{r}_{i} \hat{r}_{j} .      
\end{equation} 
In the following, we first reformulate kinematic relations in a more general but compact way for second and third correlation functions to get familiar with the general idea and tensor notations, followed by the generalization to arbitrary orders.  

\subsection{Reformulating second order kinematic relations}
\label{appendix:A2}
The starting point of new formulation is definitions of correlation tensor $Q_{ij} $ and associated correlation functions $R_{2} $, $L_{2} $, and $T_{2} $,
\begin{equation}
\begin{split}
&Q_{ij} \left(r\right)=\left\langle u_{i} u_{j}^{'} \right\rangle, \quad R_{2} =\left\langle u_{i} u_{i}^{'} \right\rangle, \\
&L_{2} =\left\langle u_{L}^{} u_{L}^{'} \right\rangle, \quad \textrm{and} \quad R_{2} =L_{2} +2T_{2}.
\end{split}   
\label{eq:51}
\end{equation}
\noindent With isotropic homogeneous correlation tensor defined as Eq. \eqref{ZEqnNum195930}, 
\begin{equation} 
\label{eq:52} 
Q_{ij} \left(\boldsymbol{\mathrm{r}}\right)=Q_{ij} \left(r\right)=A_{2} \left(r\right)r_{i} r_{j} +B_{2} \left(r\right)\delta _{ij} ,       
\end{equation} 
we should have an identity valid for any type of flow,
\begin{equation} 
\label{ZEqnNum900457} 
\begin{split}
Q_{ij} \hat{r}_{i} \left(\delta _{jk} -\hat{r}_{j} \hat{r}_{k} \right)&=\left\langle u_{L} u_{j}^{'} \right\rangle \left(\delta _{jk} -\hat{r}_{j} \hat{r}_{k} \right)\\
&=\left\langle u_{L} u_{k}^{'} \right\rangle -\left\langle u_{L} u_{L}^{'} \hat{r}_{k} \right\rangle =0.
\end{split}
\end{equation} 
This is an important identity that we will use repeatedly. 

\subsubsection{Kinematic relations for incompressible flow}
\label{appendix:A2.1}
Using identity in Eq. \eqref{ZEqnNum829225}, the relation $u_{L} =u_{i} \hat{r}_{i} $, and the product rule of differentiation,
\begin{equation} 
\label{ZEqnNum535326} 
\begin{split}
Q_{ij} {}_{,j} \hat{r}_{i} =\left\langle u_{i} u_{j}^{'} \right\rangle _{,j} \hat{r}_{i} &=\left\langle u_{i} \hat{r}_{i} u_{j}^{'} \right\rangle _{,j} -\left\langle u_{i} u_{j}^{'} \right\rangle \left(\hat{r}_{i} \right)_{j} \\
&=\left\langle u_{L} u_{j}^{'} \right\rangle _{,j} -\frac{1}{r} \left\langle u_{i} u_{j}^{'} \right\rangle \left(\delta _{ij} -\hat{r}_{i} \hat{r}_{j} \right).
\end{split}
\end{equation} 
Using identity in Eq. \eqref{ZEqnNum900457}, Eq. \eqref{ZEqnNum535326} can be further written as 
\begin{equation} 
\label{ZEqnNum723761} 
\begin{split}
&Q_{ij} {}_{,j} \hat{r}_{i} =\left\langle u_{L} u_{L}^{'} \hat{r}_{j} \right\rangle _{,j} -\frac{1}{r} \left\langle u_{i} u_{j}^{'} \right\rangle \left(\delta _{ij} -\hat{r}_{i} \hat{r}_{j} \right)\\&=\left\langle u_{L} u_{L}^{'} \right\rangle _{,j} \hat{r}_{j} +\left\langle u_{L} u_{L}^{'} \right\rangle \left(\hat{r}_{j} \right)_{,j} -\frac{1}{r} \left\langle u_{i} u_{j}^{'} \right\rangle \left(\delta _{ij} -\hat{r}_{i} \hat{r}_{j} \right).
\end{split}
\end{equation} 
For incompressible flow, $Q_{ij} {}_{,j} =0$. From Eq. \eqref{ZEqnNum723761} and identity \eqref{ZEqnNum558872}, kinematic relations between second order correlation functions (defined in Eqs. \eqref{ZEqnNum935612}, \eqref{ZEqnNum950112} and \eqref{ZEqnNum991811}) can be obtained in a very simple and compact way,
\begin{equation} 
\label{ZEqnNum386533} 
\begin{split}
&2T_{2} =\frac{1}{r} \left(r^{2} L_{2} \right)_{,r}, R_{2} =\frac{1}{r^{2} } \left(r^{3} L_{2} \right)_{,r},\\
&\textrm{and}\\
&\left(r^{2} L_{2}^{} \right)_{,r} =r\left(R_{2} -L_{2} \right),
\end{split}
\end{equation} 
which are exact the same as our previous result (see Eq. \eqref{ZEqnNum314105}) in \citep{Xu:2023-On-the-statistical-theory-of-self-gravitating}. Here the relation
\begin{equation} 
\label{eq:57} 
\left\langle u_{L} u_{L}^{'} \right\rangle _{,j} \hat{r}_{j} =\left\langle u_{L} u_{L}^{'} \right\rangle _{,r} \hat{r}_{j} \hat{r}_{j} =\left\langle u_{L} u_{L}^{'} \right\rangle _{,r}  
\end{equation} 
is applied to Eq. \eqref{ZEqnNum723761} for results in Eq. \eqref{ZEqnNum386533}. 

\subsubsection{Kinematics relations for constant divergence flow}
\label{appendix:A2.2}
On small scales, the SG-CFD flow is of constant divergence. For even order correlation functions, kinematic relations for constant divergence flow should be the same as those for incompressible flow, i.e. Eq. \eqref{ZEqnNum386533} is still valid.

\subsubsection{Kinematic relations for irrotational flow}
\label{appendix:A2.3}
On large scales, SG-CFD is of irrotational nature. The starting point to formulate kinematic relations for irrotational flow is the identity,
\begin{equation} 
\label{ZEqnNum564359} 
\begin{split}
Q_{ij} {}_{,k} \varepsilon _{mik} \varepsilon _{mjn} \hat{r}_{n} &=Q_{ij,k} \left(\delta _{ij} \delta _{kn} -\delta _{in} \delta _{jk} \right)\hat{r}_{n} \\
&=Q_{ij,k} \left(\hat{r}_{k} \delta _{ij} -\hat{r}_{i} \delta _{jk} \right),
\end{split}
\end{equation} 
where $\varepsilon _{ijk}$ is the Levi-Civita symbol and $\hat{r}_{n}=\textbf{r}/|\textbf{r}|$ is a unit vector. With curl-free condition $Q_{ij} {}_{,k} \varepsilon _{mik} =0$ and the product rule of differentiation, Eq. \eqref{ZEqnNum564359} becomes
\begin{equation} 
\label{eq:59} 
\begin{split}
Q_{ij,k} &\left(\hat{r}_{k} \delta _{ij} - \hat{r}_{i} \delta _{jk} \right)=\left\langle u_{i} u_{i}^{'} \right\rangle _{,k} \hat{r}_{k} -\left\langle u_{i} u_{j}^{'} \right\rangle _{,j} \hat{r}_{i} \\
&=\left\langle u_{i} u_{i}^{'} \right\rangle _{,k} \hat{r}_{k} -\left\langle u_{i} u_{j}^{'} \hat{r}_{i} \right\rangle _{,j} +\left\langle u_{i} u_{j}^{'} \right\rangle \left(\hat{r}_{i} \right)_{,j} =0.
\end{split}
\end{equation} 
With identity in Eq. \eqref{ZEqnNum829225}, $u_{L}=u_{i} \hat{r}_{i}$, and the help of Eq. \eqref{ZEqnNum900457}, 
\begin{equation} 
\label{ZEqnNum474481} 
\begin{split}
&\left\langle u_{i} u_{i}^{'} \right\rangle _{,r} -\left\langle u_{L} u_{j}^{'} \right\rangle _{,j} +\frac{1}{r} \left\langle u_{i} u_{j}^{'} \right\rangle \left(\delta _{ij} -\hat{r}_{i} \hat{r}_{j} \right)\\
&=\left\langle u_{i} u_{i}^{'} \right\rangle _{,r} -\left\langle u_{L} u_{L}^{'} \hat{r}_{j} \right\rangle _{,j} +\frac{1}{r} \left\langle u_{i} u_{j}^{'} \right\rangle \left(\delta _{ij} -\hat{r}_{i} \hat{r}_{j} \right)=0.
\end{split}
\end{equation} 
Using the product rule of differentiation, Eq. \eqref{ZEqnNum474481} leads to
\begin{equation} 
\label{ZEqnNum908347} 
\begin{split}
\left\langle u_{i} u_{i}^{'} \right\rangle _{,r} -\left\langle u_{L} u_{L}^{'} \right\rangle _{,j} \hat{r}_{j} &-\left\langle u_{L} u_{L}^{'} \right\rangle \left(\hat{r}_{j} \right)_{,j} \\
&+\frac{1}{r} \left\langle u_{i} u_{j}^{'} \right\rangle \left(\delta _{ij} -\hat{r}_{i} \hat{r}_{j} \right)=0.
\end{split}
\end{equation} 
Finally, for the irrotational flow, with $R_{2} =L_{2} +2T_{2} $ and identity \eqref{ZEqnNum558872}, Eq. \eqref{ZEqnNum908347} leads to the relations
\begin{equation} 
\label{ZEqnNum382437} 
\left(R_{2}^{} r\right)_{,r} =\frac{1}{r^{2} } \left(r^{3} L_{2}^{} \right)_{,r} , R_{2} =\frac{1}{r^{2} } \left(r^{3} T_{2} \right)_{,r} , L_{2} =\left(rT_{2} \right)_{,r} ,       
\end{equation} 
which are the same as previous result (see Eq. \eqref{ZEqnNum320035}) in \citep{Xu:2023-On-the-statistical-theory-of-self-gravitating}, but it is obtained in a much simpler and more compact way.

\subsection{Reformulating third order kinematic relations}
\label{appendix:A3}
Kinematic relations for third-order correlations can be derived in a similar way. Starting from the definitions of third order correlation tensor and correlation functions
\begin{equation}
\begin{split}
&Q_{ijk} \left(r\right)=\left\langle u_{i} u_{j} u_{k}^{'} \right\rangle, \quad R_{3} =\left\langle u_{L}^{} u_{i} u_{i}^{'} \right\rangle,\\
&R_{31} =\left\langle u_{i}^{} u_{i} u_{L}^{'} \right\rangle, \quad L_{3} =\left\langle u_{L}^{2} u_{L}^{'} \right\rangle, \quad R_{3} =L_{3} +2T_{3},
\end{split}
\label{eq:63}
\end{equation}
\noindent we can easily verify identities (similar to Eq. \eqref{ZEqnNum900457})
\begin{equation} 
\label{ZEqnNum213105} 
Q_{ijk} \hat{r}_{i} \hat{r}_{j} \left(\delta _{kl} -\hat{r}_{k} \hat{r}_{l} \right)=\left\langle u_{L}^{2} u_{l}^{'} \right\rangle -\left\langle u_{L}^{2} u_{L}^{'} \hat{r}_{l} \right\rangle =0,       
\end{equation} 
\begin{equation} 
\label{ZEqnNum933594} 
Q_{ijk} \hat{r}_{j} \hat{r}_{k} \left(\delta _{il} -\hat{r}_{i} \hat{r}_{l} \right)=\left\langle u_{l} u_{L} u_{L}^{'} \right\rangle -\left\langle u_{L}^{2} u_{L}^{'} \hat{r}_{l} \right\rangle =0,       
\end{equation} 
\begin{equation} 
\label{ZEqnNum805450} 
Q_{ijk} \delta _{ij} \left(\delta _{kl} -\hat{r}_{k} \hat{r}_{l} \right)=\left\langle u_{i} u_{i} u_{l}^{'} \right\rangle -\left\langle u_{i} u_{i} u_{L}^{'} \hat{r}_{l} \right\rangle =0,       
\end{equation} 
where the identity \eqref{ZEqnNum805450} only exists for odd order correlation tensors.

\subsubsection{Kinematic relations for incompressible flow}
\label{appendix:A3.1}
Using the identity in Eqs. \eqref{ZEqnNum105713} and \eqref{ZEqnNum213105}, $u_{L} =u_{i} \hat{r}_{i}$, and the product rule of differentiation,
\begin{equation} 
\label{eq:67} 
\begin{split}
&Q_{ij} {}_{k,k} \hat{r}_{i} \hat{r}_{j} =\left\langle u_{i} u_{j} u_{k}^{'} \hat{r}_{i} \hat{r}_{j} \right\rangle _{,k} -\left\langle u_{i} u_{j} u_{k}^{'} \right\rangle \left(\hat{r}_{i} \hat{r}_{j} \right)_{,k}\\
&=\left\langle u_{L}^{2} u_{k}^{'} \right\rangle _{,k} -\frac{1}{r} \left\langle u_{i} u_{j} u_{k}^{'} \right\rangle \left(\hat{r}_{j} \delta _{ik} +\hat{r}_{i} \delta _{jk} -2\hat{r}_{i} \hat{r}_{j} \hat{r}_{k} \right).
\end{split}
\end{equation} 
Using Eqs. \eqref{ZEqnNum558872} and \eqref{ZEqnNum213105}, and the product rule of differentiation, 
\begin{equation}
\label{ZEqnNum572382} 
\begin{split}
&Q_{ij} {}_{k,k} \hat{r}_{i} \hat{r}_{j} =\left\langle u_{L}^{2} u_{L}^{'} \hat{r}_{k} \right\rangle _{,k} -\frac{1}{r} \left(2\left\langle u_{L} u_{i} u_{i}^{'} \right\rangle -2\left\langle u_{L}^{2} u_{L}^{'} \right\rangle \right)\\ 
&=\left\langle u_{L}^{2} u_{L}^{'} \right\rangle _{,k} \hat{r}_{k} +\frac{2}{r} \left\langle u_{L}^{2} u_{L}^{'} \right\rangle -\frac{1}{r} \left(2\left\langle u_{L} u_{i} u_{i}^{'} \right\rangle -2\left\langle u_{L}^{2} u_{L}^{'} \right\rangle \right).
\end{split}
\end{equation} 
For incompressible flow, $Q_{ij} {}_{k,k} =0$. Kinematic relations for third order correlation functions can be easily derived from Eq. \eqref{ZEqnNum572382}
\begin{equation}
\begin{split}
&2R_{3} =\frac{1}{r^{3} } \left(r^{4} L_{3} \right)_{,r}, 4T_{3} =\frac{1}{r} \left(r^{2} L_{3} \right)_{,r},\\
&\textrm{and}\\ 
&\left(r^{2} L_{3}^{} \right)_{,r} =r\left(2R_{3} -2L_{3} \right),
\end{split}
\label{ZEqnNum772849}
\end{equation}

\noindent which are the same as our original results in Eq. \eqref{ZEqnNum374800}.

\subsubsection{Kinematic relations for constant divergence flow}
\label{appendix:A3.2}
Equation \eqref{ZEqnNum572382} is still valid for constant divergence flow such that
\begin{equation} 
\label{eq:70}
\begin{split}
Q_{ij} {}_{k,k} \hat{r}_{i} \hat{r}_{j} =\theta \left\langle u_{L}^{2} \right\rangle =& \left\langle u_{L}^{2} u_{L}^{'} \right\rangle _{,k} \hat{r}_{k}+\frac{2}{r} \left\langle u_{L}^{2} u_{L}^{'} \right\rangle \\
&-\frac{1}{r} \left(2\left\langle u_{L} u_{i} u_{i}^{'} \right\rangle -2\left\langle u_{L}^{2} u_{L}^{'} \right\rangle \right),
\end{split}
\end{equation} 
from which an exact kinematic relation reads (same as Eq. \eqref{ZEqnNum647632})
\begin{equation} 
\label{eq:71} 
R_{3} +\frac{1}{2} \left\langle u_{L}^{2} \right\rangle \theta r=\frac{1}{2r^{3} } \left(r^{4} L_{3} \right)_{,r} .         
\end{equation} 
On the other hand, from identity \eqref{ZEqnNum805450}, we have
\begin{equation} 
\label{eq:72}
\begin{split}
Q_{ij} {}_{k,k} \delta _{ij} =\theta \left\langle u^{2} \right\rangle &=\left\langle u_{i} u_{i} u_{k}^{'} \right\rangle _{,k} =\left\langle u_{i} u_{i} u_{L}^{'} \hat{r}_{k} \right\rangle _{,k} \\
&=\left\langle u^{2} u_{L}^{'} \right\rangle _{,r} +\frac{2}{r} \left\langle u^{2} u_{L}^{'} \right\rangle.
\end{split}
\end{equation} 
With $R_{31} =\langle u^{2} u_{L}^{'} \rangle$, another exact relation (same as in Eq. \eqref{ZEqnNum487892}) is
\begin{equation} 
\label{eq:73} 
\left\langle u^{2} \right\rangle \theta =\frac{1}{r^{2} } \left(r^{2} R_{31} \right)_{,r} .          
\end{equation} 
With $\langle u^{2} \rangle \approx 3 \langle u_{L}^{2} \rangle $, the kinematic relation reads (same as Eq. \eqref{ZEqnNum552000})
\begin{equation} 
\label{eq:74} 
R_{3} +\frac{1}{6r} \left(r^{2} R_{31} \right)_{,r} =\frac{1}{2r^{3} } \left(r^{4} L_{3} \right)_{,r} .        
\end{equation} 
\textbf{}

\noindent 
\subsubsection{Kinematic relations for irrotational flow}
\label{appendix:A3.3}
Like the derivation for second order correlations in Eq. \eqref{ZEqnNum564359}, the starting point is the expression,
\begin{equation} 
\label{eq:75} 
\begin{split}
Q_{ij} {}_{k,l} \varepsilon _{mkl} \varepsilon _{min} \hat{r}_{j} \hat{r}_{n} &=Q_{ijk,l} \left(\delta _{ik} \delta _{nl} -\delta _{nk} \delta _{il} \right)\hat{r}_{j} \hat{r}_{n} \\
&=Q_{ijk,l} \left(\hat{r}_{l} \delta _{ik} -\hat{r}_{k} \delta _{il} \right)\hat{r}_{j}.
\end{split}
\end{equation} 
where an identity for Levi-Civita symbol is used
\begin{equation} 
\label{eq:76} 
\varepsilon _{mkl} \varepsilon _{min} =\delta _{ik} \delta _{nl} -\delta _{nk} \delta _{il} .         
\end{equation} 
The curl free condition leads to $Q_{ij} {}_{k,l} \varepsilon _{mkl} =0$. Using the product rule of differentiation,
\begin{equation}
\label{ZEqnNum155733} 
\begin{split}
Q_{ijk,l} &\left(\hat{r}_{l} \delta _{ik} -\hat{r}_{k} \delta _{il} \right)\hat{r}_{j} =\left\langle u_{i} u_{j} u_{i}^{'} \right\rangle _{,l} \hat{r}_{j} \hat{r}_{l} -\left\langle u_{i} u_{j} u_{k}^{'} \right\rangle _{,i} \hat{r}_{j} \hat{r}_{k}\\
& =\left\langle u_{i} u_{j} u_{i}^{'} \hat{r}_{j} \right\rangle _{,l} \hat{r}_{l} -\left\langle u_{i} u_{j} u_{i}^{'} \right\rangle \left(\hat{r}_{j} \right)_{,l} \hat{r}_{l}\\
& -\left\langle u_{i} u_{j} u_{k}^{'} \hat{r}_{j} \hat{r}_{k} \right\rangle _{,i}+\left\langle u_{i} u_{j} u_{k}^{'} \right\rangle \left(\hat{r}_{j} \hat{r}_{k} \right)_{,i} =0.
\end{split}
\end{equation} 
Because $\langle u_{i} u_{j} u_{i}^{'} \rangle (\hat{r}_{j})_{,l} \hat{r}_{l} =\langle u_{i} u_{j} u_{i}^{'} \rangle (\hat{r}_{j})_{,r} =0$ (from identity \eqref{ZEqnNum669596}), Eq. \eqref{ZEqnNum155733} becomes
\begin{equation} 
\label{eq:78} 
\begin{split}
\left\langle u_{L} u_{i} u_{i}^{'} \right\rangle _{,l} \hat{r}_{l} &-\left\langle u_{i} u_{L} u_{L}^{'} \right\rangle _{,i} \\
&+\frac{1}{r} \left\langle u_{i} u_{j} u_{k}^{'} \right\rangle \left(\hat{r}_{j} \delta _{ik} +\hat{r}_{k} \delta _{ij} -2\hat{r}_{i} \hat{r}_{j} \hat{r}_{k} \right)=0.
\end{split}
\end{equation} 
With identity in Eq. \eqref{ZEqnNum829225}, $u_{L} =u_{i} \hat{r}_{i} $, and the help of Eq. \eqref{ZEqnNum933594}, 
\begin{equation} 
\label{ZEqnNum976643} 
\begin{split}
\left\langle u_{L} u_{i} u_{i}^{'} \right\rangle _{,r} &-\left\langle u_{L}^{2} u_{L}^{'} \hat{r}_{i} \right\rangle _{,i} \\
&+\frac{1}{r} \left(\left\langle u_{L} u_{i} u_{i}^{'} \right\rangle +\left\langle u_{i} u_{i} u_{L}^{'} \right\rangle -2\left\langle u_{L}^{2} u_{L}^{'} \right\rangle \right)=0.
\end{split}
\end{equation} 
With the production rule of differentiation for the second term in Eq. \eqref{ZEqnNum976643}, we should have
\begin{equation} 
\label{eq:80} 
R_{3} {}_{,r} -\frac{2}{r} L_{3} -L_{3,r} +\frac{1}{r} \left(R_{3} +R_{31} -2L_{3} \right)=0.       
\end{equation} 
Finally, with $R_{3} =L_{3} +2T_{3}$, kinematic relations are
\begin{equation}
\begin{split}
&\left(R_{3}^{} r\right)_{,r} +R_{31} =\frac{1}{r^{3} } \left(L_{3}^{} r^{4} \right)_{,r}, 3R_{3} -R_{31} =\frac{2}{r^{3} } \left(r^{4} T_{3} \right)_{,r},\\
&\textrm{and}\\ 
&3L_{3} -R_{31} =2\left(rT_{3} \right)_{,r}, 
\end{split}
\label{eq:81}
\end{equation}
\noindent which are the same as our previous results (Eq. \eqref{ZEqnNum337667}) but in a much more compact way.
 
\subsection{Formulating kinematic relations of arbitrary order}
\label{appendix:A4}
The new compact formulation for second and third order kinematic relations can be generalized to arbitrary order. Some tensor and vector algebra are involved, and readers can jump to the main results. Equations \eqref{ZEqnNum747023}-\eqref{ZEqnNum950434} present the limiting values of correlation functions on small and large scales. General kinematic relations for different types of flow are also presented, i.e. Eqs. \eqref{ZEqnNum733395}-\eqref{ZEqnNum533624} for incompressible flow, Eqs. \eqref{ZEqnNum383689}-\eqref{ZEqnNum591192} for constant divergence flow on small scales, and Eqs. \eqref{ZEqnNum900611}-\eqref{ZEqnNum433592} for irrotational flow on large scales. 

Just like in Sections \ref{appendix:A2} and \ref{appendix:A3}, some general tensors and identities are introduced first\textit{.} The \textit{p}th order $\Omega $ tensor and its derivative reads
\begin{equation} 
\label{eq:82} 
{}_{(p)} \Omega _{ij...kl} =(\underbrace{\hat{r}_{i} \hat{r}_{j} ...\hat{r}_{k} \hat{r}_{l} }_{p}),          
\end{equation} 
\begin{equation} 
\label{ZEqnNum107942}
\begin{split}
\left({}_{(p)} \Omega _{ij...kl}\right)&_{,m} =\frac{\partial \left({}_{\left(p\right)} \Omega _{ij...kl} \right)}{\partial r_{m} } \\
&=\frac{1}{r} \left({}_{\left(p+1\right)} \Pi _{ij...klm} -p\left({}_{\left(p+1\right)} \Omega _{ij...klm} \right)\right),
\end{split}
\end{equation} 
where the $\Pi $ tensor is written as
\begin{equation} 
\label{ZEqnNum991520} 
{}_{\left(p+1\right)} \Pi _{ij...klm} =\sum _{a,b..,d,e\in \left[S_{p} \right]^{p-1} }\hat{r}_{a} \hat{r}_{b...} \hat{r}_{d} \delta _{em}.   
\end{equation} 
Equation \eqref{ZEqnNum107942} is a generalization of identities in Eqs. \eqref{ZEqnNum829225}-\eqref{ZEqnNum424097}. Here $[S_{p}]^{p-1}$ includes all subsets of size (\textit{p}-1) in set $[S_{p}]=\{i,j,...k,l\}$ of size \textit{p}. There is a total of \textit{p} subsets of size (\textit{p}-1) from set $S_{p} $ of size \textit{p}. Indices $a,b,..d,e$ represent all possible combinations of size (\textit{p}-1) from set $S_{p} $ of size \textit{p}. Here, the Kronecker delta $\delta _{\bullet m} $ in $\Pi $ tensor should always have \textit{m} as one of its indices. There will be \textit{p} terms for the tensor ${}_{(p+1)} \Pi _{ij...klm} $ of order (p+1). 

One example of 4$^{th}$ order $\Omega$ tensor is:
\begin{equation} 
\label{eq:85}
\begin{split}
\left({}_{\left(4\right)} \Omega _{ijkl} \right)_{,m} &=\left(\hat{r}_{i} \hat{r}_{j} \hat{r}_{k} \hat{r}_{l} \right)_{,m}\\ &=\frac{1}{r} \left({}_{\left(5\right)} \Pi _{ijklm} -4\left({}_{\left(5\right)} \Omega _{ijklm} \right)\right),
\end{split}
\end{equation} 
where 5${}^{th}$ order tensor $\Pi $ is
\begin{equation} 
\label{eq:86} 
\begin{split}
{}_{\left(5\right)} \Pi _{ijklm} =\hat{r}_{i} \hat{r}_{j} \hat{r}_{l} \delta _{km} &+\hat{r}_{i} \hat{r}_{k} \hat{r}_{l} \delta _{jm}\\ &+\hat{r}_{i} \hat{r}_{j} \hat{r}_{k} \delta _{lm} +\hat{r}_{j} \hat{r}_{k} \hat{r}_{l} \delta _{im}.
\end{split}
\end{equation} 

For odd number \textit{p}, two additional tensors $\Lambda $ and $\Sigma $ of (\textit{p}-1) order can be introduced consisting of Kronecker delta, 
\begin{equation}
\begin{split}
&{}_{\left(p-1\right)} \Lambda _{ijkl...mn} =\delta _{ij} \delta _{kl} ...\delta _{mn}\\
&\textrm{with} \quad \left(p-1\right)/2 \quad \textrm{terms in multiplication},
\end{split}
\label{eq:87}
\end{equation}
\begin{equation}
\begin{split}
&{}_{\left(p-1\right)} \Sigma _{ijkl...mn} =\delta _{ij} \delta _{kl} ...+\delta _{ik} \delta _{jl} ...+.... \\
&\textrm{with} \quad \frac{\left(p-1\right)!}{2^{{\left(p-1\right)/2} } \left({\left(p-1\right)/2} \right)!} \quad \textrm{terms in summation}. 
\end{split}
\label{ZEqnNum230977}
\end{equation}
Examples of 4${}^{th}$ order $\Lambda $ and $\Sigma $ tensors are:
\begin{equation}
\begin{split}
&{}_{\left(4\right)} \Lambda _{ijkl} =\delta _{ij} \delta _{kl} \\
&\textrm{and}\\
&{}_{\left(4\right)} \Sigma _{ijkl} =\delta _{ij} \delta _{kl} +\delta _{ik} \delta _{jl} +\delta _{il} \delta _{jk}.
\end{split}
\label{eq:89}
\end{equation}

With the help of Eq. \eqref{ZEqnNum107942}, two additional identities can be established for tensor $\Omega $,
\begin{equation}
\label{ZEqnNum430198}
\begin{split}
\left({}_{\left(p\right)} \Omega _{ij...kl} \right)_{,r} &=\left({}_{\left(p\right)} \Omega _{ij...kl} \right)_{,m} \hat{r}_{m}\\
&=\left(\hat{r}_{i} \hat{r}_{j} ...\hat{r}_{k} \hat{r}_{l} \right)_{,m} \hat{r}_{m} =0,
\end{split}
\end{equation} 
\begin{equation} 
\label{eq:91} 
\left({}_{\left(p\right)} \Omega _{ij...kl} \right)_{,l} =\left({}_{\left(p\right)} \Omega _{ij...kl} \right)_{,m} \delta _{lm} =\frac{2}{r} \left({}_{\left(p-1\right)} \Omega _{ij...k} \right) .      
\end{equation} 

\subsubsection{Correlation functions and identities for any type of flow}
\label{appendix:A4.1}
The two-point velocity correlation tensor $Q$ of arbitrary order \textit{p} can be defined as,
\begin{equation}
\label{eq:92} 
\left({}_{\left(p\right)} Q_{ij} {}_{k..mn} \right)=\left\langle u_{i} u_{j} u_{k} ...u_{m} u_{n}^{'} \right\rangle .         
\end{equation} 
Two identities for $Q$ tensor of any order \textit{p} are (similar to Eq. \eqref{ZEqnNum213105}),
\begin{equation}
\label{ZEqnNum517859} 
\begin{split}
&\left\langle u_{i} u_{j} u_{k} ...u_{m} ...u_{n}^{'} \right\rangle \underbrace{\delta _{ij} \delta _{..} ..}_{q}\underbrace{\left(\hat{r}_{k} ...\hat{r}_{m} \right)}_{p-q-1}\left(\delta _{ns} -\hat{r}_{n} \hat{r}_{s} \right)\\
&=\left\langle u^{q} u_{L}^{p-q-1} u_{n}^{'} \right\rangle \left(\delta _{ns} -\hat{r}_{n} \hat{r}_{s} \right)\\ 
&=\left\langle u^{q} u_{L}^{p-q-1} u_{s}^{'} \right\rangle -\left\langle u^{q} u_{L}^{p-q-1} u_{L}^{'} \hat{r}_{s} \right\rangle =0, 
\end{split}
\end{equation} 
\noindent and
\begin{equation}
\label{ZEqnNum781857}
\begin{split}
&\left\langle u_{i} u_{j} u_{k} ...u_{m} ...u_{n}^{'} \right\rangle \underbrace{\delta _{jk} \delta _{..} ..}_{q}\underbrace{\left(\hat{r}_{m} ...\hat{r}_{n} \right)}_{p-q-1}\left(\delta _{is} -\hat{r}_{i} \hat{r}_{s} \right)\\
&=\left\langle u^{q} u_{L}^{p-q-2} u_{i} u_{L}^{'} \right\rangle \left(\delta _{is} -\hat{r}_{i} \hat{r}_{s} \right)\\ 
&=\left\langle u^{q} u_{L}^{p-q-2} u_{s} u_{L}^{'} \right\rangle -\left\langle u^{q} u_{L}^{p-q-1} u_{L}^{'} \hat{r}_{s} \right\rangle =0,
\end{split}
\end{equation} 
where $q$ is an even number that stands for \textit{q} indices in term $(\delta _{ij} \delta _{..} ..)$. Scalar correlation functions are defined by the tensor contraction of $Q$. 

For even number \textit{q}, the total correlation functions of order $(p,q+1)$ is defined as
\begin{equation} 
\label{ZEqnNum442788} 
R_{\left(p,q+1\right)} =\left\langle u^{q} u_{L}^{p-q-2} u_{i} u_{i}^{'} \right\rangle =\left\langle u^{q} u_{L}^{p-q-2} \boldsymbol{\mathrm{u}}\cdot \boldsymbol{\mathrm{u}}_{}^{'} \right\rangle .       
\end{equation} 
For even number \textit{q}, the longitudinal and transverse correlation functions of order $(p,q)$ are 
\begin{equation} 
\label{eq:96} 
L_{\left(p,q\right)} =\left\langle u^{q} u_{L}^{p-q-1} u_{L}^{'} \right\rangle  
\end{equation} 
and
\begin{equation} 
\label{ZEqnNum340773} 
T_{\left(p,q\right)} ={\left(R_{\left(p,q+1\right)} -L_{\left(p,q\right)} \right)/2} .         
\end{equation} 
Figure \ref{fig:2} lists the velocity correlation functions up to the sixth order. Just like the second order correlations, all these correlation functions can be similarly computed from N-body simulations.

\subsubsection{Correlation functions in the limit \texorpdfstring{$r\to 0$ and $r\to \infty $}{}}
\label{appendix:A4.2}
In the limit $r\to 0$ on small scales, the \textit{r} dependence is eliminated and isotropic homogeneous velocity correlation tensor $\boldsymbol{\mathrm{Q}}$ of odd order \textit{p} should satisfy (see Eq. \eqref{ZEqnNum405014} as an example): 
\begin{equation} 
\label{ZEqnNum949171} 
{\mathop{\lim }\limits_{r\to 0}} \left({}_{\left(p\right)} Q_{ij} {}_{..klm} \right)_{,m} =C_{p} \cdot {}_{\left(p-1\right)} \sum _{ij..kl} ,        
\end{equation} 
such that with the definition of $\Sigma $ tensor in Eq. \eqref{ZEqnNum230977}, we have
\begin{equation}
\label{eq:99}
\begin{split}
{\mathop{\lim }\limits_{r\to 0}} \left\langle u^{q} u_{L}^{p-q-1} \right\rangle \theta &={\mathop{\lim }\limits_{r\to 0}} \left({}_{\left(p\right)} Q_{ij} {}_{k..lm} \right)_{,m} \underbrace{\delta _{ij} \delta _{..} ..}_{q}\underbrace{\left(\hat{r}_{k} ...\hat{r}_{l} \right)}_{p-q-1}\\
&=C_{p} \frac{p!}{2^{{\left(p-1\right)/2} } \left({\left(p-1\right)/2} \right)!\left(p-q\right)}.
\end{split}
\end{equation} 
where $C_{p} $ is a parameter that is only dependent on the order \textit{p.} Here $\theta =u_{i,i} $ is the divergence on small scales. 

The limiting ratio for odd order \textit{p} reads
\begin{equation}
{\mathop{\lim }\limits_{r\to 0}} \frac{\left\langle u^{q} u_{L}^{p-q-1} \right\rangle }{\left\langle u_{L}^{p-1} \right\rangle } =\frac{p}{p-q} \quad \textrm{with q=0...p-1}.     
\label{ZEqnNum747023}
\end{equation}
\noindent Equation \eqref{ZEqnNum747023} is also valid for $r\to \infty $where velocity distributions are independent of scale \textit{r}, i.e. 
\begin{equation}
{\mathop{\lim }\limits_{r\to \infty }} \frac{\left\langle u^{q} u_{L}^{p-q-1} \right\rangle }{\left\langle u_{L}^{p-1} \right\rangle } =\frac{p}{p-q} \quad \textrm{with q=0...p-1}.     
\label{ZEqnNum300527}
\end{equation}
\noindent Finally, using the definition of correlation functions from index contraction, for correlation functions of odd order \textit{p},
\begin{equation} 
\label{eq:102}
\begin{split}
{\mathop{\lim }\limits_{r\to 0,\infty }} \frac{L_{\left(p,q\right)} }{L_{\left(p,0\right)} }={\mathop{\lim }\limits_{r\to 0,\infty }} \frac{\left\langle u^{q} u_{L}^{p-q-1} \right\rangle }{\left\langle u_{L}^{p-1} \right\rangle } =\frac{p}{p-q}. \end{split} 
\end{equation} 
Similar relations can be obtained for correlation functions of even order \textit{p} (from Eq. \eqref{ZEqnNum300527}),
\begin{equation} 
\label{eq:103} 
{\mathop{\lim }\limits_{r\to 0}} \frac{R_{\left(p,q+1\right)} }{L_{\left(p,0\right)} } ={\mathop{\lim }\limits_{r\to 0}} \frac{\left\langle u^{q} u_{L}^{p-q-2} \boldsymbol{\mathrm{u}}\cdot \boldsymbol{\mathrm{u}}_{}^{'} \right\rangle }{\left\langle u_{L}^{p-1} u_{L}^{'} \right\rangle } =\frac{p+1}{p-q-1} ,       
\end{equation} 
and
\begin{equation} 
\label{ZEqnNum950434} 
{\mathop{\lim }\limits_{r\to 0,\infty }} \frac{L_{\left(p,q\right)} }{L_{\left(p,0\right)} } ={\mathop{\lim }\limits_{r\to 0,\infty }} \frac{\left\langle u^{q} u_{L}^{p-q-1} u_{L}^{'} \right\rangle }{\left\langle u_{L}^{p-1} u_{L}^{'} \right\rangle } =\frac{p+1}{p+1-q} .       
\end{equation} 

\subsubsection{Kinematic relations for incompressible flow}
\label{appendix:A4.3}
The incompressibility condition requires a vanishing divergence of correlation tensor $Q$
\begin{equation} 
\label{ZEqnNum129631} 
\left({}_{\left(p\right)} Q_{ij} {}_{..mn} \right)_{,n} =\left\langle u_{i} u_{j} ...u_{m} u_{n}^{'} \right\rangle _{,n} =0.        
\end{equation} 
Using Eq. \eqref{ZEqnNum129631}, the relation $u_{L} =u_{i} \hat{r}_{i} $, and the product rule of differentiation, 
\begin{equation}
\label{ZEqnNum743715}
\begin{split}
&\left\langle u_{i} u_{j} u_{k} ...u_{m} u_{n}^{'} \right\rangle _{,n} \underbrace{\delta _{ij} \delta _{..} ..}_{q}\underbrace{\left(\hat{r}_{k} ...\hat{r}_{m} \right)}_{p-q-1}\\
&=\left[\left\langle u_{i} u_{j} u_{k} ...u_{m} u_{n}^{'} \right\rangle \delta _{ij} ...\left(\hat{r}_{k} ...\hat{r}_{m} \right)\right]_{,n}\\
&-\left\langle u_{i} u_{j} u_{k} ...u_{m} u_{n}^{'} \right\rangle \delta _{ij} ...\left(\hat{r}_{k} ...\hat{r}_{m} \right)_{,n} =0, 
\end{split}
\end{equation} 
where $q$ is an even number that stands for \textit{q} indices in term $(\delta _{ij} \delta_{..}...)$. Using identities \eqref{ZEqnNum517859} and \eqref{ZEqnNum107942}, Eq. \eqref{ZEqnNum743715} becomes
\begin{equation}
\label{ZEqnNum946495}
\begin{split}
&\left\langle u^{q} u_{L}^{p-q-1} u_{L}^{'} \hat{r}_{n} \right\rangle _{,n} =\left\langle u^{q} u_{k} ...u_{m} u_{n}^{'} \right\rangle \left(\hat{r}_{k} ...\hat{r}_{m} \right)_{,n}\\&=\left\langle u^{q} u_{k} ...u_{m} u_{n}^{'} \right\rangle \frac{1}{r} \left[{}_{\left(p-q\right)} \Pi _{k...mn} -\left(p-q-1\right)\left(\hat{r}_{k} ...\hat{r}_{m} \hat{r}_{n} \right)\right].
\end{split}
\end{equation} 
Applying the product rule of differentiation on the left side of Eq. \eqref{ZEqnNum946495} and the definition of velocity correlation functions in Eqs. \eqref{ZEqnNum442788}, \eqref{ZEqnNum340773} and the tensor $\Pi $ in Eq. \eqref{ZEqnNum991520}, we have
\begin{equation} 
\label{ZEqnNum733395} 
\left(L_{\left(p,q\right)}^{} \right)_{,r} +\frac{2}{r} L_{\left(p,q\right)} =\frac{1}{r} \left(p-q-1\right)\left(R_{\left(p,q+1\right)} -L_{\left(p,q\right)} \right)      
\end{equation} 
Finally, kinematic relations for correlations of order \textit{p} read
\begin{equation}
\left(p-q-1\right)R_{\left(p,q+1\right)} =\frac{1}{r^{p-q} } \left(r^{p-q+1} L_{\left(p,q\right)} \right)_{,r},      \label{ZEqnNum234089}
\end{equation}
\begin{equation} 
\label{eq:110} 
2\left(p-q-1\right)T_{\left(p,q\right)} =\frac{1}{r} \left(r^{2} L_{\left(p,q\right)} \right)_{,r} ,        
\end{equation} 
\begin{equation} 
\label{ZEqnNum533624} 
\left(r^{2} R_{\left(p,q+1\right)} \right)_{,r} =\frac{2}{r^{p-q-1} } \left(r^{p-q+1} T_{\left(p,q\right)} \right)_{,r} .  \end{equation} 

\subsubsection{Kinematic relations for constant divergence flow}
\label{appendix:A4.4}
Just like the incompressible flow, we start from the equation
\begin{equation}
\label{ZEqnNum703630}
\begin{split}
&\left\langle u^{q} u_{L}^{p-q-1} \right\rangle \theta =\left\langle u_{i} u_{j} u_{k} ...u_{m} u_{n}^{'} \right\rangle _{,n} \underbrace{\delta _{ij} \delta _{..} ..}_{q}\underbrace{\left(\hat{r}_{k} ...\hat{r}_{m} \right)}_{p-q-1}\\
&=\left[\left\langle u_{i} u_{j} u_{k} ...u_{m} u_{n}^{'} \right\rangle \delta _{ij} \delta _{..} ..\left(\hat{r}_{k} ...\hat{r}_{m} \right)\right]_{,n} \\
&-\left\langle u_{i} u_{j} u_{k} ...u_{m} u_{n}^{'} \right\rangle \delta _{ij} \delta _{..} ..\left(\hat{r}_{k} ...\hat{r}_{m} \right)_{,n},
\end{split}
\end{equation} 
where $q$ is an even number. Using identities \eqref{ZEqnNum517859} and \eqref{ZEqnNum107942}, Eq. \eqref{ZEqnNum703630} becomes
\begin{equation} 
\label{eq:113} 
\begin{split}
&\left\langle u^{q} u_{L}^{p-q-1} \right\rangle \theta =\left\langle u^{q} u_{L}^{p-q-1} u_{n}^{'} \right\rangle _{,n}\\ &-\left\langle u^{q} u_{k} ...u_{m} u_{n}^{'} \right\rangle \frac{1}{r} \left[{}_{\left(p-q\right)} \Pi _{k...mn} -\left(p-q-1\right)\left(\hat{r}_{k} ...\hat{r}_{m} \hat{r}_{n} \right)\right].
\end{split}
\end{equation} 
Applying the product rule of differentiation to the first item on the RHS and the definition of correlation functions in Eqs. \eqref{ZEqnNum442788} and \eqref{ZEqnNum340773}, a general relation for constant divergence flow is
\begin{equation} 
\label{ZEqnNum365670} 
\begin{split}
\left(p-q-1\right)R_{\left(p,q+1\right)} &+\left\langle u^{q} u_{L}^{p-q-1} \right\rangle \theta r \\ &=\frac{1}{r^{p-q} } \left(r^{p-q+1} L_{\left(p,q\right)} \right)_{,r} .   
\end{split}
\end{equation} 
Equation \eqref{ZEqnNum365670} is a general relation for correlations of any order \textit{p}. For correlations of even order \textit{p} (\textit{q} is always an even number), 
\begin{equation}
\label{ZEqnNum136987} 
{\mathop{\lim }\limits_{r\to 0}} \left\langle u^{q} u_{L}^{p-q-1} \right\rangle =0.          
\end{equation} 
Therefore, the kinematic relations for even order correlations in constant divergence flow should be the same as that of incompressible flow, i.e. using Eqs. \eqref{ZEqnNum365670} and \eqref{ZEqnNum136987}, Eqs. \eqref{ZEqnNum234089}-\eqref{ZEqnNum533624} are still valid for correlation functions of even order \textit{p} in a constant divergence flow. 

For odd order \textit{p}, two special cases are considered with $q=p-1$ and $q=0$ from Eq. \eqref{ZEqnNum365670},
\begin{equation} 
\label{ZEqnNum383689} 
\left\langle u^{p-1} \right\rangle \theta r=\frac{1}{r} \left(r^{2} L_{\left(p,p-1\right)} \right)_{,r}  
\end{equation} 
and
\begin{equation} 
\label{ZEqnNum213847} 
\left(p-1\right)R_{\left(p,1\right)} +\left\langle u_{L}^{p-1} \right\rangle \theta r=\frac{1}{r^{p} } \left(r^{p+1} L_{\left(p,0\right)} \right)_{,r} .       
\end{equation} 
For $r\to 0$, the correlation function $L_{\left(p,p-1\right)} $ can be solved from Eq. \eqref{ZEqnNum383689}  (use Eq. \eqref{ZEqnNum747023})
\begin{equation}
\label{ZEqnNum719510} 
L_{\left(p,p-1\right)} =\frac{p}{3} \theta \left\langle u_{L}^{p-1} \right\rangle r=\frac{1}{3} \theta \left\langle u_{}^{p-1} \right\rangle r.        
\end{equation} 
For $p=1$ and $q=0$ in Eq. \eqref{ZEqnNum213847}, the mean pairwise velocity $S_{1}^{lp} \left(r\right)=\left\langle \Delta u_{L} \right\rangle =\left\langle u_{L}^{'} -u_{L} \right\rangle =2\left\langle u_{L}^{'} \right\rangle $ can be directly related to divergence,
\begin{equation} 
\label{ZEqnNum643394} 
\theta =\frac{1}{2r^{2} } \left(r^{2} \left\langle \Delta u_{L} \right\rangle \right)_{,r} .          
\end{equation} 

With $\langle \Delta u_{L} \rangle =-Har$ from the stable clustering hypothesis, the divergence $\theta ={-3Ha/2} $ on small scales. Equation \eqref{ZEqnNum643394} is derived for a constant divergence flow. With Eq. \eqref{ZEqnNum747023}, Eqs. \eqref{ZEqnNum365670} and \eqref{ZEqnNum213847}, the kinematic relations for odd order \textit{p} should read
\begin{equation} 
\label{ZEqnNum591192} 
\begin{split}
\left(p-q-1\right)R_{\left(p,q+1\right)} &+\frac{1}{p-q} \frac{1}{r} \left(r^{2} L_{\left(p,p-1\right)} \right)_{,r}\\ &=\frac{1}{r^{p-q} } \left(r^{p-q+1} L_{\left(p,q\right)} \right)_{,r} .     
\end{split}
\end{equation} 

\subsubsection{Kinematic relations for irrotational flow}
\label{appendix:A4.5}
The irrotational flow requires a vanishing curl, i.e. $Q_{ij...} {}_{kl,m} \varepsilon _{nlm} =0$, such that \textbf{}
\begin{equation} 
\label{ZEqnNum391500} 
\begin{split}
&\left\langle u_{i} u_{j} ...u_{k} u_{o} u_{l}^{'} \right\rangle _{,m} \varepsilon _{nlm} \varepsilon _{nis} \underbrace{\delta _{jk} \delta _{..} ..}_{q}\underbrace{\left(...\hat{r}_{o} \hat{r}_{s} \right)}_{p-q-1}\\
&=\left\langle u_{i} u_{j} ...u_{k} u_{o} u_{l}^{'} \right\rangle _{,m} \left(\delta _{il} \delta _{ms} -\delta _{im} \delta _{ls} \right)\underbrace{\delta _{jk} \delta _{..} ..}_{q}\underbrace{\left(...\hat{r}_{o} \hat{r}_{s} \right)}_{p-q-1}=0.
\end{split}
\end{equation} 
where the identity $\varepsilon _{nlm} \varepsilon _{nis} =\left(\delta _{il} \delta _{ms} -\delta _{im} \delta _{ls} \right)$ is used and $q$ is an even number for \textit{q} indices in term $\delta _{ij} \delta _{..} ..$. From Eq. \eqref{ZEqnNum391500}, 
\begin{equation} 
\label{ZEqnNum243735} 
\begin{split}
&\left\langle u_{i} u_{j} ...u_{k} u_{o} u_{l}^{'} \right\rangle _{,m} \left(\delta _{il} \hat{r}_{m} \right)\underbrace{\delta _{jk} \delta _{..} ..}_{q}\underbrace{\left(...\hat{r}_{o} \right)}_{p-q-2}\\
&=\left\langle u_{i} u_{j} ...u_{k} u_{o} u_{l}^{'} \right\rangle _{,m} \left(\delta _{im} \hat{r}_{l} \right)\underbrace{\delta _{jk} \delta _{..} ..}_{q}\underbrace{\left(...\hat{r}_{o} \right)}_{p-q-2}\\
&=\left\langle u_{i} u_{j} ...u_{k} u_{o} u_{l}^{'} \right\rangle _{,i} \underbrace{\delta _{jk} \delta _{..} ..}_{q}\underbrace{\left(...\hat{r}_{o} \hat{r}_{l} \right)}_{p-q-1}.
\end{split}
\end{equation} 
Using the product rule of differentiation, the LHS (left-hand side) in Eq. \eqref{ZEqnNum243735} becomes 
\begin{equation} 
\label{eq:123}
\begin{split}
&LHS=\left(\left({}_{\left(p\right)} Q_{ij...} {}_{kl} \right)\delta _{il} \left({}_{\left(p-2\right)} \Omega _{j...k} \right)\right)_{,m} \hat{r}_{m}\\
&-\left({}_{\left(p\right)} Q_{ij...} {}_{kl} \right)\delta _{il} \left({}_{\left(p-2\right)} \Omega _{j...k} \right)_{,m} \hat{r}_{m}\\
&\textrm{or equivalently}\\
&LHS=\left\langle u^{q} u_{L}^{p-q-2} \boldsymbol{\mathrm{u}}\cdot \boldsymbol{\mathrm{u}}^{'} \right\rangle _{,m} \hat{r}_{m} -\left\langle u^{q} ...u_{o} \boldsymbol{\mathrm{u}}\cdot \boldsymbol{\mathrm{u}}^{'} \right\rangle \left(...\hat{r}_{o} \right)_{,m} \hat{r}_{m}.
\end{split}
\end{equation} 
Using identity \eqref{ZEqnNum430198}, $(...\hat{r}_{o})_{,m} (\hat{r}_{m})=0$, LHS term becomes 
\begin{equation} 
\label{ZEqnNum777708} 
\begin{split}
LHS&=\left\langle u^{q} u_{L}^{p-q-2} \boldsymbol{\mathrm{u}}\cdot \boldsymbol{\mathrm{u}}^{'} \right\rangle _{,m} \left(\hat{r}_{m} \right)\\
&=\left\langle u^{q} u_{L}^{p-q-2} \boldsymbol{\mathrm{u}}\cdot \boldsymbol{\mathrm{u}}^{'} \right\rangle _{,r} =\left(R_{p,q+1} \right)_{,r}.     
\end{split}
\end{equation} 

Now, using the product rule of differentiation, the right-hand side (RHS) in Eq. \eqref{ZEqnNum243735} reads, 
\begin{equation} 
\label{eq:125} 
RHS=\left\langle u^{q} u_{L}^{p-q-2} u_{i} u_{L}^{'} \right\rangle _{,i} -\left\langle u^{q} u_{i} ...u_{o} u_{l}^{'} \right\rangle \left(...\hat{r}_{o} \hat{r}_{l} \right)_{,i} .      
\end{equation} 
Using identity \eqref{ZEqnNum107942}, the right-hand-side term becomes
\begin{equation} 
\label{eq:126} 
\begin{split}
&RHS=\left\langle u^{q} u_{L}^{p-q-2} u_{i} u_{L}^{'} \right\rangle _{,i} \\
&-\frac{1}{r} \left\langle u^{q} u_{i} ...u_{o} u_{l}^{'} \right\rangle \left[{}_{\left(p-q\right)} \Pi _{...oli} -\left(p-q-1\right)\left(...\hat{r}_{o} \hat{r}_{l} \hat{r}_{i} \right)\right].
\end{split}
\end{equation} 
Using the identity \eqref{ZEqnNum781857}, the definition in \eqref{ZEqnNum991520}, \eqref{ZEqnNum442788}, and \eqref{ZEqnNum340773}, 
\begin{equation} 
\label{ZEqnNum311718}
\begin{split}
&RHS=\left\langle u^{q} u_{L}^{p-q-1} u_{L}^{'} \hat{r}_{i} \right\rangle _{,i}-\frac{1}{r} \left(\left\langle u^{q} u_{L}^{p-q-2} \boldsymbol{\mathrm{u}}\cdot \boldsymbol{\mathrm{u}}^{'} \right\rangle\right.\\ &\left.+\left(p-q-2\right)\left\langle u^{q+2} u_{L}^{p-q-3} u_{L}^{'} \right\rangle -\left(p-q-1\right)\left\langle u^{q} u_{L}^{p-q-1} u_{L}^{'} \right\rangle \right). 
\end{split}
\end{equation} 
Again, applying the product rule of differentiation on the first term of RHS of Eq. \eqref{ZEqnNum311718} leads to,
\begin{equation} 
\label{ZEqnNum108255} 
\begin{split}
RHS&=L_{\left(p,q\right)} +\frac{2}{r} L_{\left(p,q\right)} -\frac{1}{r} \left(R_{\left(p,q+1\right)}\right.\\ &\left.+\left(p-q-2\right)L_{\left(p,q+2\right)} -\left(p-q-1\right)L_{\left(p,q\right)} \right). 
\end{split}
\end{equation} 
Equating Eq. \eqref{ZEqnNum108255} with Eq. \eqref{ZEqnNum777708} leads to the final kinematic relations between velocity correlation functions of arbitrary order \textit{p} and even number $0\le q\le p-1$,
\begin{equation} 
\label{ZEqnNum900611} 
\begin{split}
\left(R_{\left(p,q+1\right)} r\right)_{,r} +\left(p-q-2\right)&L_{\left(p,q+2\right)}\\
&=\frac{1}{r^{p-q} } \left(r^{p-q+1} L_{\left(p,q\right)} \right)_{,r},
\end{split}
\end{equation} 
\begin{equation} 
\label{ZEqnNum602698} 
\begin{split}
\left(p-q\right)R_{\left(p,q+1\right)} -\left(p-q-2\right)&L_{\left(p,q+2\right)}\\
&=\frac{2}{r^{p-q} } \left(r^{p-q+1} T_{\left(p,q\right)} \right)_{,r},
\end{split}
\end{equation} 
\begin{equation} 
\label{ZEqnNum433592} 
\left(p-q\right)L_{\left(p,q\right)} -\left(p-q-2\right)L_{\left(p,q+2\right)} =2\left(rT_{\left(p,q\right)} \right)_{,r} .      
\end{equation} 
In Eqs. \eqref{ZEqnNum900611}-\eqref{ZEqnNum433592}, terms involving correlation function $L_{\left(p,q+2\right)} $ should vanish if $q\ge p-2$.

\label{lastpage}
\end{document}